\documentclass[11pt,a4paper]{article}
\pdfoutput=1
\usepackage{cancel}
\usepackage{enumitem}
\usepackage{etex}
\usepackage{amsthm}
\usepackage{geometry}
\usepackage{dcolumn}
\usepackage{epsf}
\usepackage{mathrsfs}
\usepackage{multirow}
\usepackage{booktabs}
\usepackage{tabularx}
\usepackage{array}
\usepackage{slashed}
\usepackage{float}
\usepackage{leftidx}
\usepackage{setspace}
\usepackage{verbatim}
\usepackage{adjustbox}
\usepackage{tikz}
\usepackage[caption=false]{subfig}
\usepackage{arydshln}
\usepackage{jheppub}
\usepackage{shuffle}

\numberwithin{equation}{section}
\usetikzlibrary{arrows,decorations.markings,shapes.arrows,patterns,positioning}

\newcommand{\bea}{\begin{eqnarray}}
\newcommand{\eea}{\end{eqnarray}}
\newcommand{\bean}{\begin{eqnarray*}}
\newcommand{\eean}{\end{eqnarray*}}
\newcommand{\nn}{\nonumber\\}
\newcommand{\Sl}{\sum\limits}

\def\W #1{\widetilde{#1}}

\def\Label#1{\label{#1}%
  \smash{\hbox to0pt{\raise1ex\hbox{\tiny[#1]}\hss}}}
\def\Label#1{\label{#1}}
\renewcommand{\eqref}[1]{eq.~(\ref{#1})}

\newcommand{\figref}[1]{Fig.~\ref{#1}}
\newcommand{\tabref}[1]{table~\ref{#1}}
\newcommand{\secref}[1]{section~\ref{#1}}

\newcommand{\tabincell}[2]{\begin{tabular}{@{}#1@{}}#2\end{tabular}}

\def\Sl{\sum\limits}

\newcommand{\ctobedelete}[1]{}
\allowdisplaybreaks

\title{A graphic approach to gauge invariance induced identity}

\author[a]{Linghui Hou \footnote{The unusual ordering of authors is just to let authors get proper recognition of contributions under
outdated practice in China.}}\author[a,b]{Yi-Jian Du\footnote{Corresponding author}}

\affiliation[a]{Center for Theoretical Physics, School of Physics and Technology,
Wuhan University, \\
No.299 Bayi Road, Wuhan 430072, China}
\affiliation[b]{Suzhou Institute of Wuhan University,\\
No.377 Linquan Street, Suzhou, 215123, China}

\emailAdd{hlh@whu.edu.cn,yijian.du@whu.edu.cn}

\date{\today}
\abstract{All tree-level amplitudes in Einstein-Yang-Mills  (EYM) theory and gravity (GR) can be expanded in terms of color ordered Yang-Mills (YM) ones whose coefficients are polynomial functions of Lorentz inner products and are constructed by a graphic rule. Once the gauge invariance condition of any graviton is imposed, the expansion of a tree level EYM or gravity amplitude induces a nontrivial identity between color ordered YM amplitudes. Being different from traditional Kleiss-Kuijf (KK) and Bern-Carrasco-Johansson (BCJ) relations, the gauge invariance induced identity includes polarizations in the coefficients. In this paper, we investigate the relationship between the gauge invariance induced identity and traditional BCJ relations. By proposing a refined graphic rule, we prove that all the gauge invariance induced identities for single trace tree-level EYM amplitudes can be precisely expanded in terms of traditional BCJ relations, without referring any property of polarizations. When further considering the transversality of polarizations and momentum conservation, we prove that the gauge invariance induced identity for tree-level GR (or pure YM) amplitudes can also be expanded in terms of traditional BCJ relations for YM (or bi-scalar) amplitudes. As a byproduct, a graph-based BCJ relation is proposed and proved.
}
\keywords{Amplitude Relation, Gauge invariance}

\begin{document}
\maketitle \flushbottom

%%%%%%%%%%%%%%%%%%%%%%
\section{Introduction}
%%%%%%%%%%%%%%%%%%%%%%%%%
%{\blue 1.Inducing the recursive expansion of EYM amplitudes by gauge invariance}
Gauge invariance has been shown to provide a strong constraint on scattering amplitudes in recent years.
It was not only used to solve  amplitudes explicitly (see e.g., \cite{Boels:2016xhc,Arkani-Hamed:2016rak,Boels:2017gyc,Boels:2018nrr})  but also applied in understanding hidden structures of amplitudes. One interesting application is that a recursive expansion of single-trace Einstein-Yang-Mills (EYM) amplitudes can be determined by  gauge invariance conditions of external gravitons in addition with a proper ansatz \cite{Fu:2017uzt,Chiodaroli:2017ngp} (the approach based on Cachazo-He-Yuan (CHY) formula \cite{Cachazo:2013gna,Cachazo:2013hca,Cachazo:2013iea,Cachazo:2014xea} is provided in \cite{Teng:2017tbo}). Applying the recursive expansion repeatedly, one expresses an arbitrary tree level single-trace EYM amplitude in terms of pure Yang-Mills (YM) ones whose coefficients are conveniently constructed by a graphic rule \cite{Du:2017kpo}. This expansion can be regarded as a generalization of the earlier studies on EYM amplitudes with few gravitons \cite{Stieberger:2016lng,Nandan:2016pya,delaCruz:2016gnm,Schlotterer:2016cxa}. When the relation between gravity amplitudes and  single-trace EYM amplitudes are further considered \cite{Fu:2017uzt}, we immediately expand any tree level gravity amplitude as a combination of Yang-Mills ones. It was shown that the recursive expansion for single-trace EYM amplitudes can be
extended to all tree level multi-trace amplitudes via replacing some gravitons by gluon traces appropriately \cite{Du:2017gnh}\footnote{Discussions on multi-trace EYM amplitudes can also be found in \cite{Nandan:2016pya,Chiodaroli:2017ngp}.}.

%{\blue 2 gauge invariance identity as nontrivial relations for YM amplitudes}

In the explicit form of the recursive expansion proposed in \cite{Fu:2017uzt}, gauge invariance conditions for all but one graviton (so-called `fiducial graviton') are naturally encoded into the strength tensors as well as amplitudes with fewer gravitons. Nevertheless, the gauge invariance for the fiducial graviton does not manifest and further implies an nontrivial identity between EYM amplitudes. This identity plays as a consistency condition, which guarantees locality, in the Britto-Cachazo-Feng-Witten (BCFW) recursion approach \cite{Britto:2004ap,Britto:2005fq} to the expansion of EYM amplitudes \cite{Fu:2017uzt}. Equivalently, if we take the gauge invariance condition of the fiducial graviton in the pure YM expansion, we immediately obtain a nontrivial identity for color-ordered YM amplitudes whose coefficients are polynomial functions of Lorentz inner products and depend on both polarizations of gravitons and external momenta. Similar discussions on generating relations for YM amplitudes by imposing gauge invariance can be found in \cite{Barreiro:2013dpa,Boels:2016xhc,Nandan:2016pya,Chiodaroli:2017ngp,Du:2017gnh,Boels:2017gyc,Du:2018khm}. An application of the gauge invariance induced identity is the proof of  the equivalence between distinct approaches \cite{Du:2016tbc,Carrasco:2016ldy,Du:2017kpo} to scattering amplitudes in nonlinear sigma model \cite{Du:2018khm}.

%{\blue 3.BCJ relations}
Apart from the gauge invariance induced identity, color-ordered Yang-Mills amplitudes have also been shown to satisfy Kleiss-Kuijf (KK) \cite{Kleiss:1988ne} and Bern-Carrasco-Johansson (BCJ) relations \cite{Bern:2008qj}, which reduce the number of independent amplitudes to $(n-3)!$ (proofs can be found in \cite{Stieberger:2009hq,BjerrumBohr:2009rd,Feng:2010my,Chen:2011jxa}). A notable difference from the gauge invariance induced identity  is that  KK and BCJ relations do not include external polarizations in the coefficients. Thus it seems that the gauge invariance induced identity provides a new relation beyond KK and BCJ relations. However, direct evaluations of several examples \cite{Fu:2017uzt,Chiodaroli:2017ngp,Du:2017gnh} imply that the gauge invariance induced identity is just a combination of BCJ relations. Due to the complexity of the coefficients in the expansion of EYM amplitudes, these examples provided in \cite{Fu:2017uzt,Chiodaroli:2017ngp,Du:2017gnh} cannot be straightforwardly generalized to arbitrary case. Hence there is still lack of a  systematical study on the full connection between the gauge invariance induced identity and BCJ relations.

%{\blue 5.In this paper, we ....}

In this paper, we investigate the relationship between an arbitrary gauge invariance induced identity, which is derived from the expansion of tree level single-trace EYM amplitudes or pure GR amplitudes, and general BCJ relations. We propose a `refined graphic rule' in which different types of Lorentz inner products $\epsilon\cdot\epsilon$, $\epsilon\cdot k$ and $k\cdot k$, constructed by half polarizations of gravitons $\epsilon^{\mu}$ and external momenta $k^{\mu}$, are distinguished by different types of lines.
Similar graphic rules have already been applied in the study of CHY formula (see \cite{Huang:2017ydz,Lam:2018tgm}). By expressing the coefficients in the gauge invariance induced identity according to the refined graphic rule and collecting those terms with the same structure of the lines corresponding to coefficients $\epsilon\cdot\epsilon$ and $\epsilon\cdot k$ (such a structure is called a \emph{skeleton}) in particular examples, we find that the gauge invariance induced identity can always be expressed by a combination of BCJ relations.

To generalize our observations to arbitrary case, more hidden structures of graphs should be revealed. We show that a skeleton of a graph corresponding to the gauge invariance induced identity always consists of no less than two maximally connected subgraphs (\emph{components}) which are mutually disjoint with each other. Any physical graph, which agrees with the refined graphic rule, can be reconstructed by (i) first connecting these components (via $k\cdot k$ lines) into a graph with only two disjoint (called \emph{final upper and lower}) blocks properly, (ii) then connecting the two disjoint blocks by a $k\cdot k$ line. For any given configuration of the final upper and lower blocks, \emph{spurious graphs} are also introduced for convenience. We prove that the spurious graphs belonging to different configurations of the two disjoint blocks always cancel in pairs. Hence the sum over all physical graphs containing a same skeleton can be given by summing over both physical and spurious graphs with the same skeleton. A critical observation is that the total contribution of all physical and spurious graphs containing a same configuration of the final upper and lower blocks induces a \emph{graph-based BCJ relation} which can be expanded in terms of traditional BCJ relations. As a consequence, the general gauge invariance induced identity from the expansion of tree level single-trace EYM amplitude is precisely expanded in terms of (traditional) BCJ relations. Similar discussions are applied to the gauge invariance induced identity of tree level GR amplitudes.  The only notable difference is that the transversality of polarizations and momentum conservation in the pure gravity case should be taken into account. By the help of the language of CHY formulas \cite{Cachazo:2013gna,Cachazo:2013hca,Cachazo:2013iea,Cachazo:2014xea}, we further conclude that
the gauge invariance induced identities of Yang-Mills-scalar (YMS) and pure YM amplitudes are combinations of BCJ relations for bi-scalar amplitudes.

%{\blue 6.The structure of this paper...}
The structure of this paper is following. In \secref{sec:Background}, we review the expansion of tree level single-trace EYM amplitudes, the gauge invariance induced identity as well as traditional BCJ relations, which will be useful in the remaining sections. The refined graphic rule for the gauge invariance induced identity, which is derived from single-trace EYM amplitudes, as well as the main idea of this paper are provided in \secref{sec:RefinedRule}. In \secref{sec:DirectEvalutations}, direct evaluations of examples are presented. The construction rules of all physical and spurious graphs containing a same skeleton are given in \secref{section:GraphsForASkeleton}. We show that the total contribution of physical and spurious graphs corresponding to a same configuration of the two disjoint blocks induces the graph-based-BCJ relation which is further proven to be a combination of traditional BCJ relations in \secref{sec:GraphBasedBCJ}. Discussions on identities induced by the gauge invariance of pure GR and YM amplitudes are displayed in \secref{sec:GRandYM}. We summarize this work in \secref{sec:Conclusions}. Conventions and complicated graphs are included in the appendix.

%%%%%%%%%%%%%%%%%
\section{Expansion of tree level single-trace EYM amplitudes, gauge invariance induced identity and traditional BCJ relation}\label{sec:Background}
%%%%%%%%%%%%%%%%%%%
In this section, we review the recursive expansion of single-trace EYM amplitudes, the graphic rule, the gauge invariance induced identity from single-trace EYM amplitudes as well as the BCJ relations for Yang-Mills amplitudes.
%%%%%%%%%%
\subsection{Recursive expansion of tree level single-trace EYM amplitudes}
%%%%%%%%%
Tree level single-trace EYM amplitude $A(1,2,\dots,r\Vert\,\mathsf{H})$ with $r$ gluons $1$, $2$, ... , $r$ and $s$ gravitons $h_1$, $h_2$, ..., $h_s$ can be recursively expanded as (see \cite{Fu:2017uzt})
\bea
A(1,2,\dots,r\Vert\,\mathsf{H})&=&\Sl_{\pmb{h}\vert\,\W{\mathsf{h}}}\,\Sl_{\pmb{\sigma}\in\{2,\dots,r-1\}\shuffle\{\pmb{h},h_a\}}C_{h_a}(\pmb{h}) A\bigl(1,\pmb{\sigma},r\big\Vert\,\W{\mathsf{h}}\bigr),\Label{Eq:RecursiveExpansion}
\eea
where the full graviton set $\{h_1,\dots,h_s\}$ is denoted by $\mathsf{H}$ and an arbitrarily chosen $h_a\in \mathsf{H}$ is called  \emph{fiducial graviton}.
The first summation on the RHS is taken over all possible splittings of the set $\mathsf{H}\setminus {h_a}\to \pmb{h}\vert\,\W{\mathsf{h}}$ and all permutations of elements in $\pmb{h}$ for a given splitting. For a fixed splitting $\mathsf{H}\setminus {h_a}\to \pmb{h}\vert\,\W{\mathsf{h}}$ and a fixed permutation of elements in $\pmb{h}$, the second summation is taken over all the possible shuffle permutations
$\pmb{\sigma}\in\{2,\dots,r-1\}\shuffle\{\pmb{h},h_a\}$ (i.e. permutations in which the relative orders of elements in each set are kept). The coefficient $C_{h_a}(\pmb{h})$ for a given splitting $\mathsf{H}\setminus {h_a}\to \pmb{h}\vert\,\W{\mathsf{h}}$, a given permutation $\{i_1,i_{2},\dots,i_j\}$ of elements in $\pmb{h}$ and a given shuffle permutation $\pmb{\sigma}$ is defined by
\bea
C_{h_a}(\pmb{h})\equiv \epsilon_{h_a}\cdot F_{i_j}\cdot\dots \cdot F_{i_2}\cdot F_{i_1}\cdot Y_{i_1}(\pmb{\sigma}),\Label{Eq:RecExpCoefficient}
\eea
where the strength tensor $F_i^{\mu\nu}$ is
\bea
F_i^{\mu\nu}\equiv k_i^{\mu}\epsilon_i^{\nu}-k_i^{\nu}\epsilon_i^{\mu}~\Label{Eq:StrengthTensor}
\eea
and $Y_{i_1}(\pmb{\sigma})$ denotes the sum of all momenta of gluons $l\in\{1,2,\dots,r-1\}$ s.t. $\pmb{\sigma}^{-1}(l)<\pmb{\sigma}^{-1}(i_1)$.
%%%%%%%%%%%%%%%%%%%%%%%%%%%%%%%%%%%%%%%%%%%%%%%%%%%%%
\subsection{Graphic rule for the pure YM expansion of tree level single-trace  EYM amplitudes}
%%%%%%%%%%%%%%%%%%%%%%%%%%%%%%%%%%%%%%%%%%%%%%%%%%%%%%%
Applying the recursive expansion (\ref{Eq:RecursiveExpansion}) repeatedly,  we finally express the single-trace EYM amplitude $A(1,2,\dots,r\Vert\,\mathsf{H})$ in terms of pure YM ones
\bea
A(1,2,\dots,r\Vert{\mathsf{H}})&=&\Sl_{\pmb{\sigma}\in\{2,\dots,r-1\}\shuffle\,\text{perms}\,{\mathsf{H}}} \mathcal{C}(1,\pmb{\sigma},r)A(1,\pmb{\sigma},r).\Label{Eq:PureYMExpansion}
\eea
In the above equation, we summed over all possible permutations $\pmb{\sigma}\in\{2,\dots,r-1\}\shuffle\,\text{perms}\,{\mathsf{H}}$ in which the relative order of elements in $\{2,\dots,r-1\}$ is preserved and $\text{perms}\,{\mathsf{H}}$ are all possible permutations of elements in $\mathsf{H}$.
The coefficient $\mathcal{C}(1,\pmb{\sigma},r)$ for any permutation $\pmb{\sigma}$ is given by
    \bea
    \mathcal{C}(1,\pmb{\sigma},r)=\Sl_{\mathcal{F}\in\mathcal{G}(1,\pmb{\sigma},r)}\mathcal{C}^{[\mathcal{F}]},\Label{Eq:Coefficients}
    \eea
    where $\{\mathcal{G}(1,\pmb{\sigma},r)\}$ is used to denote the set of graphs $\mathcal{F}$ constructed by the following \emph{graphic rule}:
\begin{itemize}
\item [(1)] Define a reference order $\pmb{\rho}$ of gravitons, then all $s$ gravitons are arranged into an ordered set
\bea
\mathsf{R}=\{h_{\rho(1)},h_{\rho(2)},\dots,h_{\rho(s)}\}.~~\Label{Eq:ReferenceOrder}
\eea
The position $\rho^{-1}(l)$ of $l$ ($l\in \mathsf{H}$) in the ordered set $\mathsf{R}$ is called \emph{the weight of $l$}. Apparently, $h_{\rho(s)}$ is the highest-weight node in $\mathsf{R}$.

\item [(2)] Pick the highest-weight element $h_{\rho(s)}$ (the fiducial graviton for the first step recursive expansion) from the ordered set $\mathsf{R}$, an arbitrary gluon (\emph{root}) $l\in \{1,2,\dots,r-1\}$ as well as gravitons $i_1, {i_2}, \dots,i_j\in \mathsf{H}$ s.t. $\sigma^{-1}(l)< \sigma^{-1}(i_1)< \sigma^{-1}(i_2)<\dots \sigma^{-1}(i_j)< \sigma^{-1}(h_{\rho(s)})$. Here the position of $i$ in the permutation $\sigma$ is denoted by $\sigma^{-1}(i)$. By considering each particle in the set $\{l,i_1,i_2, \dots,i_j,h_{\rho(s)}\}$ as a node, we construct a \emph{chain} $\mathbb{CH}=\left[h_{\rho(s)},i_j,\dots,i_1,l\right]$ which starts from the node $h_{\rho(s)}$ towards the node $l$. The graviton $h_{\rho(s)}$, the gluon $l$ and gravitons $i_1$, $i_2$, ... , $i_j$ are correspondingly mentioned as \emph{the starting node}, \emph{the ending node} and \emph{the internal nodes} of this chain. We defined \emph{the weight of a chain} by the weight of the starting node of the chain.  The factor associated to this chain is
\bea
\epsilon_{h_{\rho(s)}}\cdot F_{i_j}\cdot F_{i_{j-1}}\cdot \dots \cdot F_{{i_1}}\cdot k_l.
\eea
Redefine $\mathsf{R}$ by removing ${i_1}$, ${i_2}$, ..., ${i_j}$, $h_{\rho(s)}$: $\mathsf{R}\to\mathsf{R}\,'=\mathsf{R}\setminus \{{i_1},{i_2}, ...,{i_j},h_{\rho(s)}\}$.

\item [(3)] Picking $l'\in \{1,2,\dots,r-1\}\cup\{{i_1},{i_2}, ...,{i_j},h_{\rho(s)}\}$, the highest-weight element $h_{\rho'(s')}$ (which is the fiducial graviton for the second step recursive expansion) in $\mathsf{R}\,'$ as well as gravitons ${i'_1}$, ${i'_2}$, ..., ${i'_{j'}}\in\mathsf{R}\,'$ s.t. $\sigma^{-1}(l')<\sigma^{-1}({i_1'})<\sigma^{-1}({i_2'})<\dots<\sigma^{-1}({i_{j'}'})<\sigma^{-1}(h_{\rho'(s')})$, we define a chain $\mathbb{CH}=\{h_{\rho'(s')},i'_{j'},i'_{j'-1},\dots,i'_{1},l'\}$ starting from $h_{\rho(s')}$ and ending at $l'$. This chain is associated with a factor
    \bea
    \epsilon_{h_{\rho'(s')}}\cdot F_{{i'_{j'}}}\cdot F_{{i'_{j'-1}}}\cdot \dots \cdot F_{{i_{1}'}}\cdot k_{l'}.
    \eea
    Remove ${i_1'}$,${i_2'}$, ..., ${i_{j'}'}$, $h_{\rho'(s')}$ from the ordered set $\mathsf{R}\,'$ and redefine $\mathsf{R}\to\mathsf{R}\,''=\mathsf{R}'\setminus\{{i_1'},{i_2'},\dots, {i_{j'}'},h_{\rho'(s')}\}$.
\item [(4)] Repeating the above steps until the ordered set $\mathsf{R}$ is empty, we obtain a graph $\mathcal{F}$ in which  graviton trees are planted at gluons (roots) $\{1,\dots,r-1\}$. For any given graph $\mathcal{F}$, the product of  the factors accompanied to all chains produces a term $\mathcal{C}^{[\mathcal{F}]}$ in the coefficient $\mathcal{C}(1,\pmb{\sigma},r)$ in \eqref{Eq:PureYMExpansion}. Thus the final expression of $\mathcal{C}(1,\pmb{\sigma},r)$ is given by summing over all possible graphs constructed by the above steps, i.e. \eqref{Eq:Coefficients}.
\end{itemize}

%%%%%%%%%%%%%%%%%%%%%%%%%%%%%%%%%%%%%%%%%%%%%%
\subsection{Gauge invariance induced identity from tree level  single-trace EYM amplitudes}
%%%%%%%%%%%%%%%%%%%%%%%%%%%%%%%%%%%%%%%%%%%%%%%
%
Gauge invariance requires that an EYM amplitude has to vanish when the `half' graviton polarization $\epsilon_{h_i}$ is replaced by momenta $k_{h_i}$ for any $h_i\in \mathsf{H}$. Consequently, the pure-YM expansion \eqref{Eq:PureYMExpansion} should become zero under the replacement $\epsilon_{h_i}\to k_{h_i}$. Assuming that the half polarization $\epsilon_{h_i}$ is included in the coefficients in \eqref{Eq:PureYMExpansion}, our discussion can be classified into the following two cases:
\begin{itemize}

\item If $h_i$ is the highest-weight element $h_{\rho(s)}$ (the fiducial graviton for the first-step expansion) in the reference order $\mathsf{R}$, it must be a starting node of some chain but cannot be an internal node of any chain. The gauge invariance condition for $h_i$ is not manifest and implies the following nontrivial relation for pure Yang-Mills amplitudes
\bea
\Sl_{\pmb{\sigma}\in\{2,\dots,r-1\}\shuffle\,\text{perms}\,{\mathsf{H}}} \mathcal{C}(1,\pmb{\sigma},r)\big |_{\scriptsize\epsilon_{h_{\rho(s)}}\to k_{h_{\rho(s)}}}A(1,\pmb{\sigma},r)=0,\Label{Eq:GaugeInv2}
\eea
in which the coefficient $\mathcal{C}(1,\pmb{\sigma},r)\big |_{\scriptsize\epsilon_{h_{\rho(s)}}\to k_{h_{\rho(s)}}}$ is obtained from \eqref{Eq:Coefficients} via replacing $\epsilon_{h_{\rho(s)}}$ by $k_{h_{\rho(s)}}$. In other words, the chains led by $h_{\rho(s)}$ are of the form $k_{h_{\rho(s)}}\cdot F_{i_j}\cdot\dots\cdot F_{i_2}\cdot F_{i_1}\cdot k_l$.

\item If $h_i$ is some graviton other than $h_{\rho(s)}$, it can be either an internal node or a starting node of some chain. The former case vanishes naturally because of the antisymmetry of the strength tensor $F^{\mu\nu}_{h_i}$. The latter case is achieved if the gauge invariance condition \eqref{Eq:GaugeInv2} is already satisfied by amplitudes with fewer gravitons because in this case: (i) $h_i$ plays as the fiducial graviton for some intermediate-step recursive expansion in the graphic rule, and (ii) the sum over all the graphs, which contain the same chain structure produced by the preceding steps, is proportional to the LHS of the gauge invariance condition (\ref{Eq:GaugeInv2}) (with $h_{\rho(s)}\to h_i$) for fewer-graviton EYM amplitudes (Elements on chains produced by the preceding steps are considered as gluons).
\end{itemize}
Since the gauge invariance condition for  $h_i\neq h_{\rho(s)}$ is always achieved when the identity \eqref{Eq:GaugeInv2} for fewer gravitons holds, our discussion can just be focused on  the case with $h_i=h_{\rho(s)}$.

%%%%%%%%%%%%%%%%%%%%%%%%%%%%%%%%%%%%
\subsection{BCJ relation}
%%%%%%%%%%%%%%%%%%%%%%%%%%%%%%%%%%%%
Tree level color-ordered YM amplitudes have been proven to satisfy the following general BCJ relation (this general BCJ relation was introduced in \cite{BjerrumBohr:2009rd,Chen:2011jxa}):
\bea
\Sl_{\pmb{\sigma}\in\pmb{\beta}\,\shuffle\,\pmb{\alpha}}\,\Sl_{l\in\pmb{\beta}}\left(k_{l}\cdot X_l(\pmb{\sigma})\right)A(1,\pmb{\sigma},r)=0,\Label{Eq:BCJRelation}
\eea
where $\pmb{\beta}$ and $\pmb{\alpha}$ are two ordered sets of external gluons, $X_l(\pmb{\sigma})$ denotes the sum of all  momenta of gluons $a\in\{1,\pmb{\sigma}\}\cup\pmb{\beta}$ satisfying $\sigma^{-1}(a)<\sigma^{-1}(l)$ (the gluon $1$ is always considered as the first one in the permutation $\pmb{\sigma}$).
In this paper, we  the expression on the LHS of the \eqref{Eq:BCJRelation} is denoted as
\bea
\mathcal{B}(1\,|\,\pmb{\beta},\pmb{\alpha}\,|\,r)=\Sl_{\pmb{\sigma}\in\pmb{\beta}\,\shuffle\,\pmb{\alpha}}\,\Sl_{l\in\pmb{\beta}}\left(k_{l}\cdot X_l(\pmb{\sigma})\right)A(1,\pmb{\sigma},r).\Label{Eq:LHSBCJ}
\eea
%

%%%%%%%%%%%%%%%%%
%\section{Expansion of EYM amplitudes and gauge invariance induced relations}
%%%%%%%%%%%%%%%%%%

%%%%%%%%%%%%%%%%%
%\subsection{Recursive expansion of tree level EYM amplitudes}
%%%%%%%%%%%%%%%%%

%%%%%%%%%%%%%%%%%%
%\subsection{Graphic expansion of tree level EYM amplitudes}
%%%%%%%%%%%%%%%%%%

%%%%%%%%%%%%%%%%%%%%%%
%\subsection{Gauge invariance induced identities}
%%%%%%%%%%%%%%%%%%%%%%

%%%%%%%%%%%%%%%%%%
\section{Refined graphic rule and the main idea}\label{sec:RefinedRule}
%%%%%%%%%%%%%%%%%%
In the previous section, we have shown that the gauge invariance condition for a single-trace EYM amplitude induces a nontrivial identity  \eqref{Eq:GaugeInv2} for pure Yang-Mills ones. The difference from BCJ relation is that coefficients in  \eqref{Eq:GaugeInv2} contain not only Mandelstam variables $k_a\cdot k_b$ but also other two types of Lorentz inner products $\epsilon_a\cdot \epsilon_b$ and $\epsilon_a\cdot k_b$ which involve half polarizations $\epsilon^{\mu}$.
Such feature can be straightforwardly understood when all strength tensors $F_i^{\mu\nu}$ on both types of chains $\epsilon_a\cdot F_{i_j}\cdot\dots\cdot F_{i_2}\cdot F_{i_1}\cdot k_l$ and $k_{\rho(s)}\cdot F_{i_j}\cdot\dots\cdot F_{i_2}\cdot F_{i_1}\cdot k_l$ are expanded according to the definition (\ref{Eq:StrengthTensor}). As shown by explicit examples in \cite{Fu:2017uzt}, the identity \eqref{Eq:GaugeInv2} in fact can be expanded in terms of BCJ relations \eqref{Eq:BCJRelation} without referring any property of external polarizations $\epsilon_i^{\mu}$.  Nevertheless, this observation cannot be trivially extended to the general identity \eqref{Eq:GaugeInv2} because of the complexity of coefficients in \eqref{Eq:GaugeInv2}. Thus the relationship between the gauge invariance induced identity \eqref{Eq:GaugeInv2} and BCJ relation \eqref{Eq:BCJRelation} is still unclear.
In this section, we propose a refined graphic rule and show the main idea for studying the relationship between the identity \eqref{Eq:GaugeInv2} and BCJ relation \eqref{Eq:BCJRelation}, which will be helpful for our generic study in the coming sections.

%%%%%%%%%%%%%%%%%%%%%%%%%%%%%%%%%%%%%%%%%%%%%%%%%%%
\subsection{Refined graphic rule for single trace tree-level EYM amplitudes}
%%%%%%%%%%%%%%%%%%%%%%%%%%%%%%%%%%%%%%%%%%%%%%%%%%%

 In the graphs constructed by the rule in section \ref{sec:Background}, the three types of Lorentz inner products $\epsilon_a\cdot\epsilon_b$, $\epsilon_a\cdot k_b$ and $k_a\cdot k_b$ cannot be distinguished. To investigate the general relationship between the gauge invariance induced identity (\ref{Eq:GaugeInv2}) and BCJ relation \eqref{Eq:BCJRelation}, we propose the following graphic rule by expanding all strength tensors $F_{a}^{\mu\nu}$  s.t. the three types of Lorentz inner products are represented by three distinct types of lines:
\begin{figure}
\centering
\includegraphics[width=0.32\textwidth]{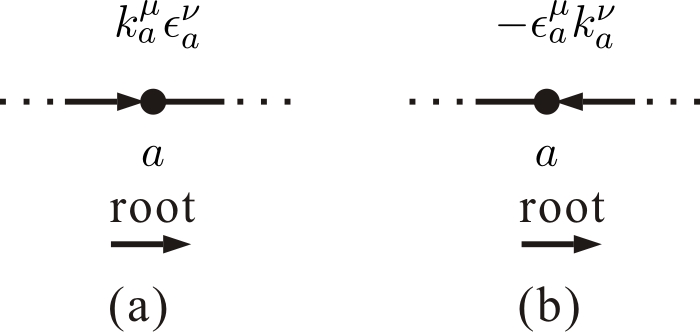}
\caption{Two types of internal nodes of chains}\label{InternalNode}
\end{figure}
 \begin{figure}
\centering
\includegraphics[width=0.28\textwidth]{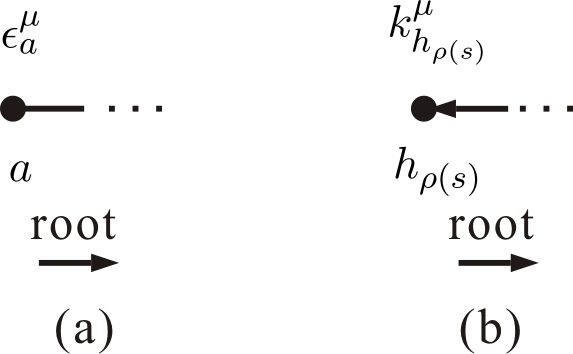}
\caption{Two types of starting points of chains.}\label{StartingPoints}
\end{figure}
\begin{figure}
\centering
\includegraphics[width=0.38\textwidth]{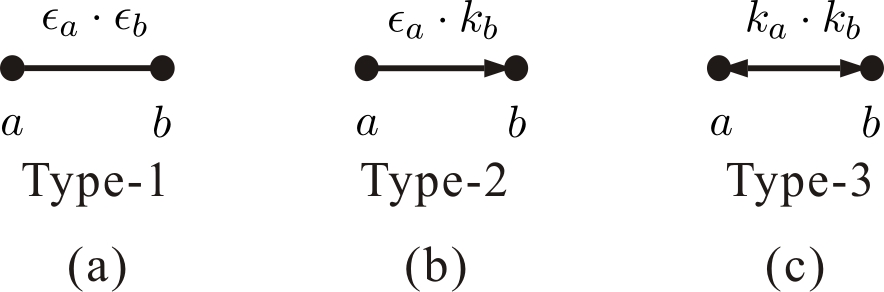}
\caption{Three types of lines between nodes.}\label{Lines}
\end{figure}
\begin{itemize}
\item [(1)] \emph{Internal nodes} In the original graphic rule, each internal node stands for a strength tensor $F_{a}^{\mu\nu}$. When $F_{a}^{\mu\nu}$ is expanded into $k_a^{\mu}\epsilon_a^{\nu}-k_a^{\nu}\epsilon_a^{\mu}$, the corresponding internal node represents either $k_a^{\mu}\epsilon_a^{\nu}$ or $-k_a^{\nu}\epsilon_a^{\mu}$. Here the momentum $k_a^{\mu}$ and the `half' polarization $\epsilon_a^{\nu}$ are respectively presented by an ingoing arrow line and an outcoming solid line. Then the strength tensor $F_a^{\mu\nu}$ becomes the sum of the two graphs in \figref{InternalNode}. As shown in \figref{InternalNode} (a), we associate a plus with an arrow pointing to the direction of root. An arrow pointing deviate from root is associated with a minus (see \figref{InternalNode} (b)).

\item[(2)] \emph{Starting nodes of chains} In the gauge invariance induced identity (\ref{Eq:GaugeInv2}),
each starting node of a chain is associated with either a `half' polarization $\epsilon_a^{\mu}$ of some element other than $h_{\rho(s)}$ or the momentum $k_{h_{\rho(s)}}^{\mu}$ of the highest-weight element $h_{\rho(s)}$ in the ordered set $\mathsf{R}$. Thus two distinct types of starting nodes are required. Noting that all chains are directed to roots, we introduce these two types of starting nodes \figref{StartingPoints} (a) and (b) by removing $(...\cdot k)$ and $(-1)(...\cdot \epsilon)$ from the internal nodes shown by \figref{InternalNode} (a) and (b) respectively.

%In addition, we will also encounter other two types of starting nodes which are shown by  \figref{StartingPoints} (b) and (d) when separating the Mandelstam variables from the other two types of Lorentz inner products in the coming discussions.

\item [{(3)}]\emph{Ending nodes of chains} Each internal/starting node of a chain or a root (the element in the original gluon set $\{1,\dots,r\}$) can also be the ending node of another chain. The contraction of an ending node $b$ of some chain with its neighbor on the same chain always has the form $(...\cdot k_b)$. Therefore, the ending node of a chain should be attached by a line of form $k_a\cdot k_b$ or $\epsilon_a\cdot k_b$.

\item [{(4)}] \emph{Three types of lines between nodes} Contractions of Lorentz indices are represented by connecting lines associated to nodes together.  There are three distinct types of lines as shown by figure \ref{Lines} (a) (type-1), (b) (type-2) and (c) (type-3) corresponding to the Lorentz inner products $\epsilon_a\cdot\epsilon_b$, $\epsilon_a\cdot k_b$ and $k_a\cdot k_b$.

\end{itemize}
With the above improvement, the coefficients $\mathcal{C}(1,\pmb{\sigma},r)\Big |_{\epsilon_{h_{\rho(s)}}\to k_{h_{\rho(s)}}}$ of the gauge invariance induced identity \eqref{Eq:GaugeInv2} is then written by summing over all refined graphs. Thus \eqref{Eq:GaugeInv2} becomes
\bea
\Sl_{\pmb{\sigma}\in\{2,\dots,r-1\}\shuffle\,\text{perms}\,{\mathsf{H}}}\left[\Sl_{\mathcal{F}\in \mathbb{G}(1,\pmb{\sigma},r)}\mathcal{D}^{[\mathcal{F}]}\right]\,A(1,\pmb{\sigma},r)=0,\Label{Eq:GaugeInv3}
\eea
where $\mathbb{G}(1,\pmb{\sigma},r)$ denotes the set of all refined graphs which are allowed by the permutation $\{1,\pmb{\sigma},r\}$ according to the refined graphic rule. \emph{It is worth noting that a given graph $\mathcal{F}$ can be allowed by various permutations.}

\begin{figure}
\centering
\includegraphics[width=0.85\textwidth]{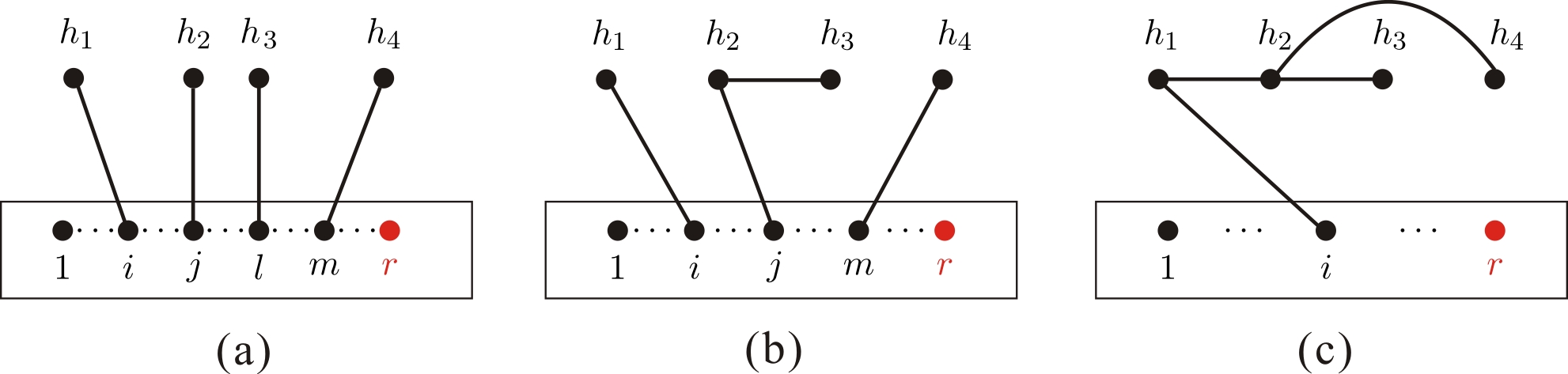}
\caption{Examples of old-version graphs for the gauge invariance induced identity \eqref{Eq:GaugeInv2}. Here, the node $h_4$ denotes $k^{\mu}_{h_4}$, while the nodes $h_1$, $h_2$ and $h_3$ denote $\epsilon^{\mu}_{h_1}$, $\epsilon^{\mu}_{h_2}$ and $\epsilon^{\mu}_{h_3}$, respectively.  We color the element $r$ by red to remind that the element $r$ always plays as the rightmost element in any permutation corresponding to a given graph. Only $1,\dots,r-1$ can be roots. }\label{OldVersionGraph}
\end{figure}
\begin{figure}
\centering
\includegraphics[width=1\textwidth]{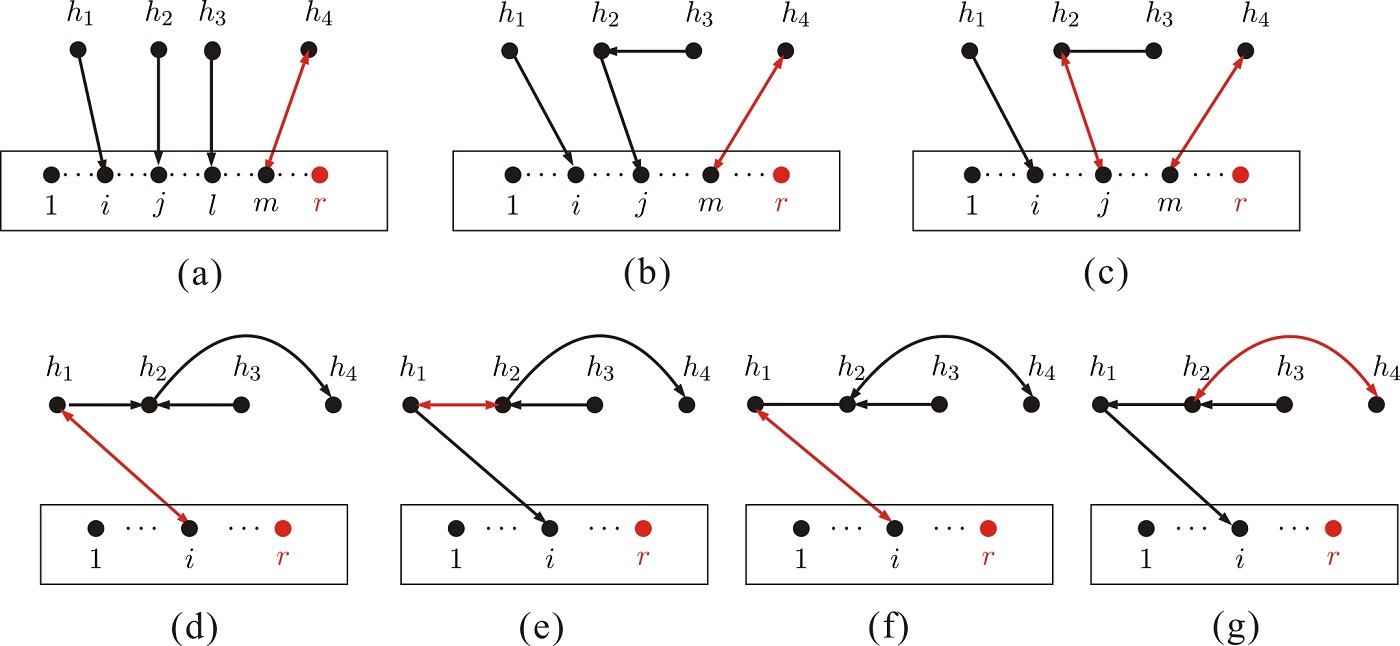}
\caption{Graphs constructed by the refined graphic rule. The graph (a) corresponds to the old-version graph \figref{OldVersionGraph} (a), the graphs (b) and (c) correspond to the old-version graph  \figref{OldVersionGraph} (b), while the graphs (d)-(g) correspond to  the old-version graph \figref{OldVersionGraph} (c). }\label{ImprovedGraph}
\end{figure}

\subsubsection{Examples for the refined graphic rule}

Now we take the identity induced by the gauge invariance of the amplitude $A(1,\dots,r\Vert\,h_1,h_2,h_3,h_4)$ as an example. We assume the reference order is  $\mathsf{R}=\{h_1,h_2,h_3,h_4\}$ and consider the gauge invariance induced identity \eqref{Eq:GaugeInv3} with $\epsilon_{h_4}\to k_{h_4}$.

{\bf Example-1}~~For any given permutation $\pmb{\sigma}\in\{2,\dots,r-1\}\shuffle\{h_1\}\shuffle\{h_2\}\shuffle\{h_3\}\shuffle\{h_4\}$, there must be old-version graphs \figref{OldVersionGraph} (a) contributing terms of the form $(k_{h_4}\cdot k_m)(\epsilon_{h_3}\cdot k_l) (\epsilon_{h_2}\cdot k_j)(\epsilon_{h_1}\cdot k_i)$ ($i,j,l,m\in\{1,\dots,r-1\}$) to the coefficient $\mathcal{C}(1,\pmb{\sigma},r)|_{\tiny\epsilon_{h_{\rho(s)}}\to k_{h_{\rho(s)}}}$.
The refined graph for  such a coefficient is shown by \figref{ImprovedGraph} (a), in which each chain led by $h_1$, $h_2$ or $h_3$ consists of only one type-2 line, while the chain led by $h_4$ is a type-3 line.

{\bf Example-2}~~For any permutation $\pmb{\sigma}\in\{2,\dots,r-1\}\shuffle\{h_4\}\shuffle\{h_2,h_3\}\shuffle\{h_1\}$, there exist old-version graphs \figref{OldVersionGraph} (b)  consisting of three chains  $(k_{h_4}\cdot k_m)$, $(\epsilon_{h_3}\cdot F_{h_2}\cdot k_j)$ and $(\epsilon_{h_1}\cdot k_i)$ $(i,j,m\in\{1,\dots,r-1\})$.
For given $i,j,m\in\{1,\dots,r-1\}$, the graph \figref{OldVersionGraph} (b) contributes a term $(k_{h_4}\cdot k_m)(\epsilon_{h_3}\cdot F_{h_2}\cdot k_j) (\epsilon_{h_1}\cdot k_i)$ to $\mathcal{C}(1,\pmb{\sigma},r)|_{\tiny\epsilon_{h_{\rho(s)}}\to k_{h_{\rho(s)}}}$. According to the refined graphic rule, this coefficient is given by summing the two graphs \figref{ImprovedGraph} (b) and (c) together:
\bea
&&\text{(b)}:(k_{h_4}\cdot k_m)(\epsilon_{h_3}\cdot k_{h_2})(\epsilon_{h_2}\cdot k_j)(k_{h_1}\cdot k_i),~~
\text{(c)}:(-1)(k_{h_4}\cdot k_m)(\epsilon_{h_3}\cdot \epsilon_{h_2})(k_{h_2}\cdot k_j)(k_{h_1}\cdot k_i).
\eea

{\bf Example-3}~~For any permutation $\pmb{\sigma}\in\{2,\dots,r-1\}\shuffle\{h_1,h_2,\{h_3\}\shuffle\{h_4\}\}$, there exist old-version graphs \figref{OldVersionGraph} (c) (for $i\in\{1,\dots,r-1\}$) each of which contains two distinct chains $(k_{h_4}\cdot F_{h_2}\cdot F_{h_1}\cdot k_i)$ and $(\epsilon_{h_3}\cdot k_2)$. Thus the total contribution of such a graph is $(k_{h_4}\cdot F_{h_2}\cdot F_{h_1}\cdot k_i)(\epsilon_3\cdot k_2)$.
According to the refined graphic rule, the coefficient for a given $i\in\{1,\dots,r-1\}$ is provided by the sum of the four graphs \figref{ImprovedGraph} (d), (e), (f) and (g):
\bea
&&\text{(d)}:(-1)^2(k_{h_4}\cdot\epsilon_{h_2})(k_{h_2}\cdot\epsilon_{h_1})(k_{h_1}\cdot k_i)(\epsilon_{h_3}\cdot k_{h_2})~\,\text{(e)}:(-1)(k_{h_4}\cdot\epsilon_{h_2})(k_{h_2}\cdot k_{h_1})(\epsilon_{h_1}\cdot k_i)(\epsilon_{h_3}\cdot k_{h_2})\nn
&&\text{(f)}:(-1)(k_{h_4}\cdot k_{h_2})(\epsilon_{h_2}\cdot\epsilon_{h_1})(k_{h_1}\cdot k_i)(\epsilon_{h_3}\cdot k_{h_2})~~~\,\text{(g)}:~(k_{h_4}\cdot k_{h_2})(\epsilon_{h_2}\cdot k_{h_1})(\epsilon_{h_1}\cdot k_i)(\epsilon_{h_3}\cdot k_{h_2}).
\eea

%%%%%%%%%%%%%%%%%%%%%%%%%%%
\subsection{The main idea}
%%%%%%%%%%%%%%%%%%%%%%%%%%%%
Having established the refined graphic rule, we are now ready for studying the relationship between the gauge invariance induced identity \eqref{Eq:GaugeInv3} and BCJ relation \eqref{Eq:BCJRelation}. In the coming sections, we will prove that the gauge invariance induced identity \eqref{Eq:GaugeInv3} can be expanded into a combination of BCJ relations.
The main idea is following:

{\bf Step-1} \emph{The factorization of coefficients}~~For any graph $\mathcal{F}$ constructed by the refined graphic rule, the coefficient $\mathcal{D}^{[\mathcal{F}]}$  can be factorized as a product of two coefficients $\mathcal{P}^{[\mathcal{F}']}$ and $\mathcal{K}^{[{\mathcal{F}\setminus\mathcal{F}'}]}$ associated with a total factor $(-1)^{\mathcal{N}(\mathcal{F})}$. Here, \emph{the skeleton} $\mathcal{F}'$ is the subgraph which is obtained by deleting all type-3 lines from $\mathcal{F}$. The factor $\mathcal{P}^{[\mathcal{F}']}$ associated to the skeleton $\mathcal{F}'$ contains only factors of forms $(\epsilon_a\cdot\epsilon_b)$ and $(\epsilon_a\cdot k_b)$. We use $\mathcal{F}\setminus \mathcal{F}'$ to stand for the subgraph which is obtained by deleting all type-1 and 2 lines from $\mathcal{F}$ (i.e. the complement of the skeleton)\footnote{In this paper, `$\mathsf{A}\setminus \mathsf{B}$' for a set $\mathsf{A}$ and its subset $\mathsf{B}$, if the sets are not considered as graphs, means we remove all elements of $\mathsf{B}$ from $\mathsf{A}$.  For a graph $\mathcal{A}$ and its subgraph $\mathcal{B}$, the expression `$\mathcal{A}\setminus \mathcal{B}$' means we remove all lines that are attached to nodes in $\mathcal{B}$ from $\mathcal{A}$ but keep the nodes. The expression `$\mathcal{A}-\mathcal{B}$' is defined by removing all nodes in $\mathcal{B}$ and lines attached to these nodes from $\mathcal{A}$.}. The  factor $\mathcal{K}^{[{\mathcal{F}\setminus\mathcal{F}'}]}$  corresponding to ${\mathcal{F}\setminus\mathcal{F}'}$ contains only Mandelstam variables $(k_a\cdot k_b)$. The total factor $(-1)^{\mathcal{N}(\mathcal{F})}$ depends on the number $\mathcal{N}(\mathcal{F})$ of arrows  pointing deviate from the direction of roots ( except the one connected to the highest-weight node $h_{\rho(s)}$ because we do not associate a minus to the arrow \figref{StartingPoints} (b) ). For instance, the factor $\mathcal{D}^{[\mathcal{F}']}$ for the graph \figref{ImprovedGraph} (d) is $(-1)^2(\epsilon_{h_1}\cdot k_{h_2})(\epsilon_{h_2}\cdot k_{h_4})(\epsilon_{h_3}\cdot k_{h_3})(k_{h_1}\cdot k_i)$ which can be factorized into $\mathcal{P}^{[\mathcal{F}']}=(\epsilon_{h_1}\cdot k_{h_2})(\epsilon_{h_2}\cdot k_{h_4})(\epsilon_{h_3}\cdot k_{h_3})$ and the factor  $\mathcal{K}^{[{\mathcal{F}\setminus\mathcal{F}'}]}=(k_{h_1}\cdot k_i)$ with a total sign $(-1)^{\mathcal{N}(\mathcal{F})}=(-1)^2$. With this factorization, the LHS of the gauge invariance induced identity \eqref{Eq:GaugeInv3} is expressed by
\bea
T\equiv\Sl_{\pmb{\sigma}\in\{2,\dots,r-1\}\shuffle\,\text{perms}\,{\mathsf{H}}}\biggl[\Sl_{\mathcal{F}\in \mathbb{G}(1,\pmb{\sigma},r)}  (-1)^{\mathcal{N}(\mathcal{F}) }\mathcal{P}^{[\mathcal{F}']}\mathcal{K}^{[{\mathcal{F}\setminus\mathcal{F}'}]}\biggr]\,A(1,\pmb{\sigma},r).\Label{Eq:GaugeInv4}
\eea
Here, we emphasize that a given skeleton $\mathcal{F}'$ can belong to different permutations $\pmb{\sigma}\in\{2,\dots,r-1\}\shuffle\,\text{perms}\,{\mathsf{H}}$.

{\bf Step-2} \emph{Collecting all terms corresponding to the graphs  $\mathcal{F}\supset\mathcal{F}'$ for any given skeleton $\mathcal{F}'$ } When all graphs containing the same skeleton are collected together, the expression \eqref{Eq:GaugeInv4} becomes
    \bea
T=\Sl_{\mathcal{F}'}\mathcal{P}^{[\mathcal{F}']}\biggl[\Sl_{\substack{\mathcal{F}\text{~s.t.}\\ \mathcal{F}\supset\mathcal{F}'}} (-1)^{\mathcal{N}(\mathcal{F})}\mathcal{K}^{[{\mathcal{F}\setminus\mathcal{F}'}]}\Sl_{\pmb{\sigma}^{\mathcal{F}}}\,A(1,\pmb{\sigma}^{\mathcal{F}},r)\biggr],\Label{Eq:GaugeInv5}
\eea
where we summed over (i) all possible skeletons $\mathcal{F}'$, (ii) all possible graphs $\mathcal{F}$ which are constructed by the refined graphic rule and satisfy $\mathcal{F}\supset\mathcal{F}'$ for a given skeleton $\mathcal{F}'$, (iii) all permutations $\pmb{\sigma}^{\mathcal{F}}$ for a given $\mathcal{F}$.

 {\bf Step-3} \emph{Finding out the relationship between terms associated with a skeleton and the LHS of BCJ relations }
For any given skeleton $\mathcal{F}'$ in \eqref{Eq:GaugeInv5}, we will prove that the expression in the square brackets can be written in terms of the LHS of BCJ relations.

We first present direct evaluations of simple examples in the next section and then prove the general cases in \secref{section:GraphsForASkeleton} and \secref{sec:GraphBasedBCJ}.

%%%%%%%%%%%%%%%%%%%%%%%%%%%%%
\section{Direct evaluations}\label{sec:DirectEvalutations}
%%%%%%%%%%%%%%%%%%%%%%%%%%%%%

In this section, we show that the expression in the square brackets in \eqref{Eq:GaugeInv5} can be expanded in terms of the LHS of BCJ relations by direct evaluation of  simple examples.

\subsection{The identity with $\mathsf{H}=\{h_1\}$}
\begin{figure}
\centering
\includegraphics[width=3.6in]{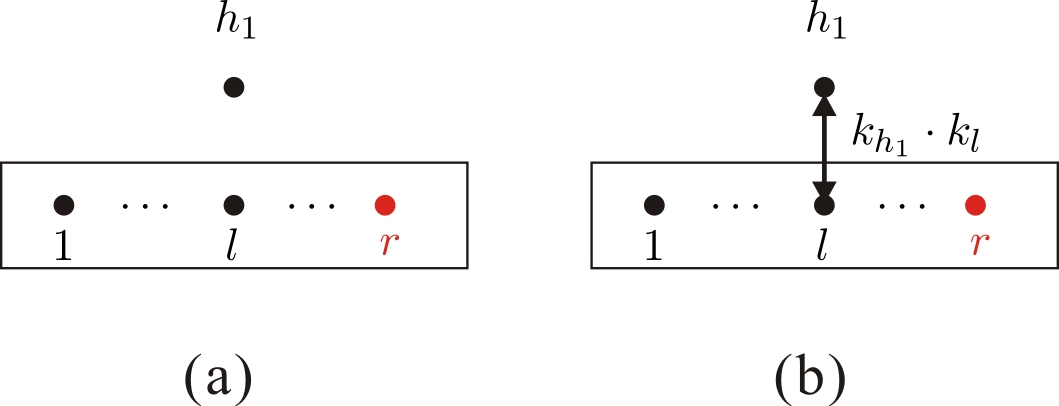}
\caption{The graph (a) is the skeleton $\mathcal{F}\,'$ for the gauge invariance induced identity (\ref{Eq:GaugeInv3}) with $\mathsf{H}=\{h_1\}$. The graph (b) for $l\in\{1,\dots,r-1\}$ is a typical graph $\mathcal{F}$ containing the skeleton (a).}\label{Fig:Figure1}
\end{figure}
When the set $\mathsf{H}$ contains only one element $h_1$, the LHS of the identity \eqref{Eq:GaugeInv2} has the form
\bea
T=\Sl_{\pmb{\sigma}}(k_{h_1}\cdot Y_{h_1}(\pmb{\sigma}))A(1,\pmb{\sigma}\in\{2,\dots,r-1\}\shuffle\{h_1\},r), \Label{Eq:SingleID1}
\eea
which is apparently  $\mathcal{B}(1\,\vert\,\{h_1\}, \{2,\dots,r-1\}\,\vert\, r)$ (see \eqref{Eq:LHSBCJ}). Eq. (\ref{Eq:SingleID1}) can be understood as \eqref{Eq:GaugeInv5} for the specific case $\mathsf{H}=\{h_1\}$: (i) The skeleton $\mathcal{F}'$ is  shown by \figref{Fig:Figure1} (a) and contributes a trivial factor $\mathcal{P}^{[\mathcal{F}']}=1$.
(ii) Graphs $\mathcal{F}\supset\mathcal{F}'$ (for all $l\in\{1,2,\dots,r-1\}$) are given by \figref{Fig:Figure1} (b) and the kinematic factors $\mathcal{K}^{[\mathcal{F}\setminus\mathcal{F}']}$ are $k_{h_1}\cdot k_l$.
(iii) Permutations allowing the graph \figref{Fig:Figure1} (b) for a given $l\in\{1,2,\dots,r-1\}$ are $\pmb{\sigma}^{l}\in\{2,\dots,l,\{h_1\}\shuffle\{l+1,\dots,r-1\}\}$.
All together, \eqref{Eq:GaugeInv5} for this example reads
  \bea
T= 1\cdot \biggl[\Sl_{l\in\{1,2,\dots,r-1\}}(k_{h_1}\cdot k_l)\Sl_{\pmb{\sigma}^l}A(1,\pmb{\sigma}^l\in\{2,\dots,l,\{h_1\}\shuffle\{l+1,\dots,r-1\}\},r)\biggr].\Label{Eq:SingleID1-1}
\eea
We can collect the coefficients for a given permutation $\pmb{\sigma}\in\{1,\{h_1\}\shuffle\{2,\dots,r-1\},r\}$ in the above equation. Noting that each $l$ satisfying $\pmb{\sigma}^{-1}(l)<\pmb{\sigma}^{-1}(h_1)$ contributes a factor $k_{h_1}\cdot k_{l}$, we find that the total factor for $\pmb{\sigma}$ is $(k_{h_1}\cdot Y_{h_1}(\pmb{\sigma}))$.
When all permutations $\pmb{\sigma}\in\{2,\dots,r-1\}\shuffle\{h_1\}$ are considered, the expression \eqref{Eq:SingleID1-1} becomes  \eqref{Eq:SingleID1}.

\subsection{The identity with $\mathsf{H}=\{h_1,h_2\}$}

\begin{figure}
\centering
\includegraphics[width=0.5\textwidth]{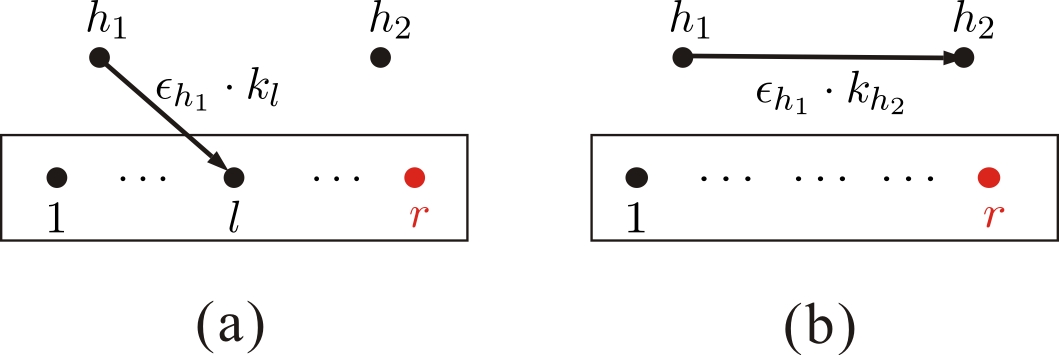}
\caption{Graphs (a) and (b) are the skeletons $\mathcal{F}'$ for the gauge invariance induced identity (\ref{Eq:GaugeInv3}) with $\mathsf{H}=\{h_1,h_2\}$.}\label{Fig:Figure2}
\end{figure}
\begin{figure}
\centering
\includegraphics[width=1\textwidth]{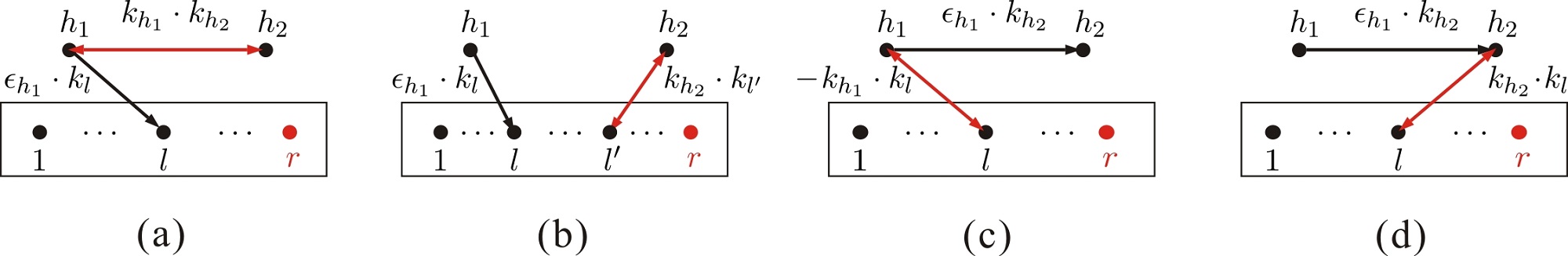}
\caption{Graphs (a) and (b) are those graphs $\mathcal{F}$ containing the skeleton $\mathcal{F}'=\text{\figref{Fig:Figure2} (a)}$. Graphs containing the skeleton $\mathcal{F}'=\text{\figref{Fig:Figure2} (b)}$ are given by (c) and (d). }\label{Fig:Figure3}
\end{figure}
The next to simplest example is the identity (\ref{Eq:GaugeInv2}) with two gravitons $h_1$ and $h_2$. Here, we assume that the reference order is $\mathsf{R}=\{h_1,h_2\}$ and  consider the gauge invariance condition for $h_2$.  According to the old-version graphic rule,  the LHS of the gauge invariance induced identity \eqref{Eq:GaugeInv2} is written as
\bea
T&=&\Bigl[k_{h_2}\cdot F_{h_1}\cdot Y_{h_1}+(k_{h_2}\cdot Y_{h_2})(\epsilon_{h_1}\cdot Y_{h_1})\Bigr]A(1,\{2,\dots,r-1\}\shuffle\{h_1,h_2\},r)\nn
&+&\Bigl[(k_{h_2}\cdot Y_{h_2})(\epsilon_{h_1}\cdot Y_{h_1})+(k_{h_2}\cdot Y_{h_2})(\epsilon_{h_1}\cdot k_{h_2})\Bigr]A(1,\{2,\dots,r-1\}\shuffle\{h_2,h_1\},r).\Label{Eq:DoubleID1}
\eea
When the strength tensor $F^{\mu\nu}$ is expanded, we arrive the following expression
\bea
T&=&\Bigl[(k_{h_2}\cdot k_{h_1})(\epsilon_{h_1}\cdot Y_{h_1})-(k_{h_2}\cdot \epsilon_{h_1})(k_{h_1}\cdot Y_{h_1})+(k_{h_2}\cdot Y_{h_2})(\epsilon_{h_1}\cdot Y_{h_1})\Bigr]A(1,\{2,\dots,r-1\}\shuffle\{h_1,h_2\},r)\nn
&+&\Bigl[(k_{h_2}\cdot Y_{h_2})(\epsilon_{h_1}\cdot Y_{h_1})+(k_{h_2}\cdot Y_{h_2})(\epsilon_{h_1}\cdot k_{h_2})\Bigr]A(1,\{2,\dots,r-1\}\shuffle\{h_2,h_1\},r),\Label{Eq:DoubleID2}
\eea
which can be regarded as a result of the refined graphic rule. In \eqref{Eq:DoubleID2}, only one polarization $\epsilon_{h_1}$ appears in the coefficients. Since $\epsilon_{h_1}$ can be contracted with the momentum of either $l\in\{1,\dots,r-1\}$ or $h_2$, the skeletons $\mathcal{F}'$ can be classified into two categories: (i) \figref{Fig:Figure2}.(a) (for $l\in\{1,2,\dots,r-1\}$)  and (ii) \figref{Fig:Figure2}.(b).

\begin{figure}
\centering
\includegraphics[width=0.75\textwidth]{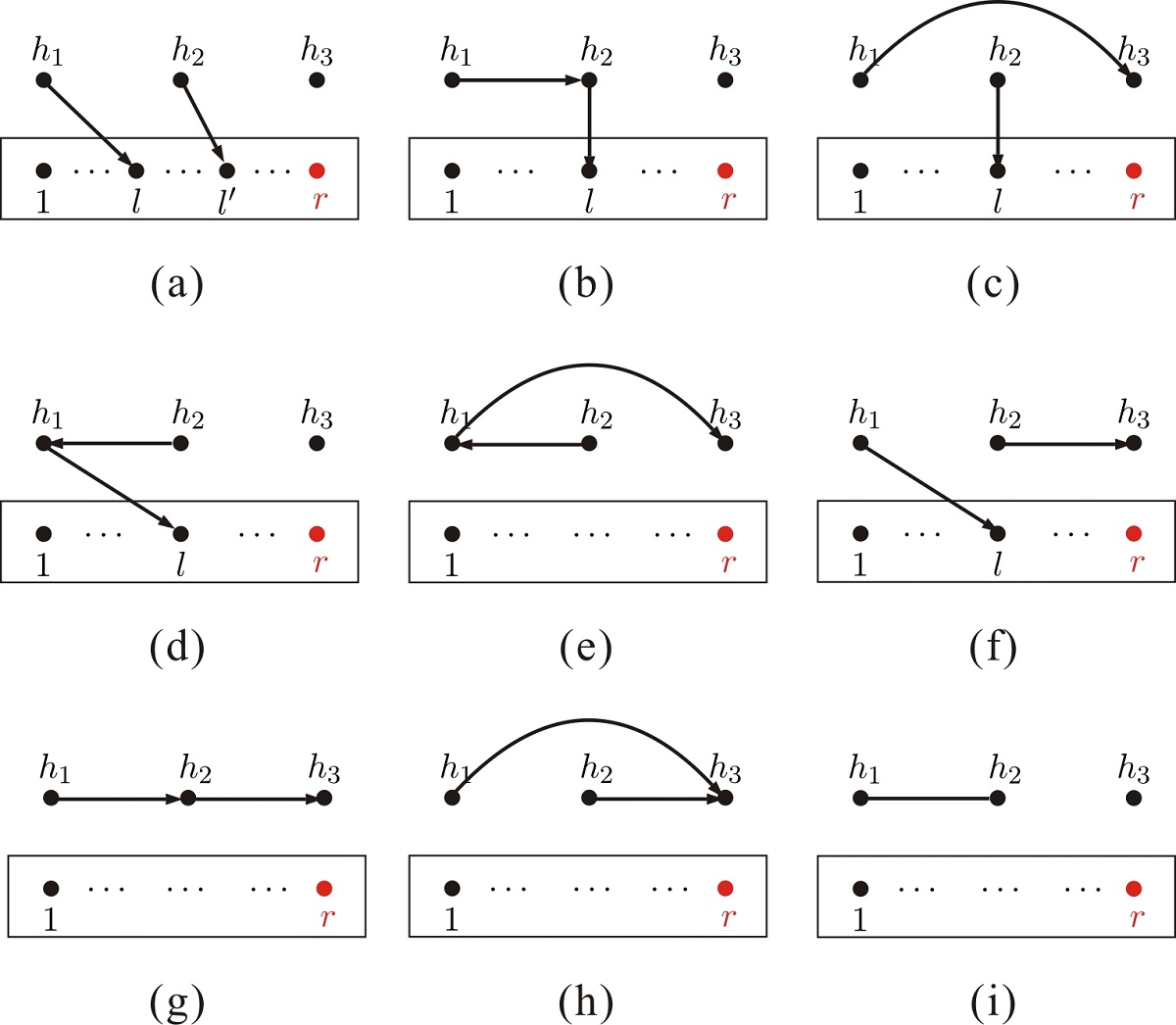}
\caption{All skeletons for the gauge invariance induced identity with $\mathsf{H}=\{h_1,h_2,h_3\}$.}\label{Fig:Figure4}
\end{figure}

The skeleton \figref{Fig:Figure2} (a) with a given $l\in\{1,2,\dots,r-1\}$ contributes a factor $\mathcal{P}^{[\mathcal{F}']}=\epsilon_{h_1}\cdot k_l$.
Graphs $\mathcal{F}$ containing the skeleton \figref{Fig:Figure2} (a) are displayed by \figref{Fig:Figure3} (a) and (b). The factor $\mathcal{K}^{[\mathcal{F}\setminus\mathcal{F}']}$ for $\mathcal{F}=\text{\figref{Fig:Figure3} (a)}$ is $k_{h_2}\cdot k_{h_1}$ and the corresponding permutations $\pmb{\sigma}^{\mathcal{F}}$ are
\bea
\pmb{\sigma}^{\mathcal{F}}\in\{2,\dots,l,\{h_1,h_2\}\shuffle\{l+1,\dots,r-1\}\}.
\eea
The factor $\mathcal{K}^{[\mathcal{F}\setminus\mathcal{F}']}$ for $\mathcal{F}=\text{\figref{Fig:Figure3} (b)}$ is $k_{h_2}\cdot k_{l'}$ $(l'=\{1,2,\dots,r-1\})$ and the corresponding permutations $\pmb{\sigma}^{\mathcal{F}}$ satisfy
\bea
&&\pmb{\sigma}^{\mathcal{F}}\in\{2,\dots,l,\{h_1\}\shuffle\{l+1,\dots,l',\{h_2\}\shuffle\{l'+1,\dots,r-1\}\}\}~~(1\leq l\leq l'\leq r-1)\nn
&&\pmb{\sigma}^{\mathcal{F}}\in\{2,\dots,l',\{h_2\}\shuffle\{l'+1,\dots,l,\{h_1\}\shuffle\{l+1,\dots,r-1\}\}\}~~(1\leq l'< l\leq r-1).
\eea
Then the expression in the square brackets of \eqref{Eq:GaugeInv4}, which is associated to the skeleton $\mathcal{F}\,'=\text{\figref{Fig:Figure2} (a)}$ for a given $l\in\{1,2,\dots,r-1\}$, is explicitly written as
\bea
&&(k_{h_2}\cdot k_{h_1})\Sl_{\pmb{\sigma}}A(1,\pmb{\sigma}\in\{2,\dots,l,\{h_1,h_2\}\shuffle\{l+1,\dots,r-1\}\},r)\Label{Eq:DoubleID3}\\
&+&\Sl_{\substack{l\,'\in\{1,2,\dots,r-1\}\\\text{s.t.~}l\,'\geq l}}(k_{h_2}\cdot k_{l\,'})\Sl_{\pmb{\sigma}}A(1,\pmb{\sigma}\in\{2,\dots,l,\{h_1\}\shuffle\{l+1,\dots,l',\{h_2\}\shuffle\{l'+1,\dots,r-1\}\}\},r)\nn
&+&\Sl_{\substack{l\,'\in\{1,2,\dots,r-1\}\\\text{s.t.~}l\,'< l}}(k_{h_2}\cdot k_{l\,'})\Sl_{\pmb{\sigma}}A(1,\pmb{\sigma}\in\{2,\dots,l',\{h_2\}\shuffle\{l'+1,\dots,l,\{h_1\}\shuffle\{l+1,\dots,r-1\}\}\},r).\nonumber
\eea
The sum over all permutations $\pmb{\sigma}$ in the above expression can be obtained by summing over all possible $\pmb{\sigma}\in\{h_2\}\shuffle\pmb{\beta}$ for a fixed $\pmb{\beta}\in\{2,\dots,l,\{h_1\}\shuffle\{l+1,\dots,r-1\}\}$ and then summing over all possible $\pmb{\beta}\in\{2,\dots,l,\{h_1\}\shuffle\{l+1,\dots,r-1\}\}$.
For a fixed $\pmb{\beta}\in\{2,\dots,l,\{h_1\}\shuffle\{l+1,\dots,r-1\}\}$, the coefficients for $\pmb{\sigma}\in\{h_2\}\shuffle\,\pmb{\beta}$ are collected as
\bea
&\Sl_{\substack{l\,'\in\{1,\dots,r-1\}\\\sigma^{-1}(l\,')<\sigma^{-1}(h_2)}}k_{h_2}\cdot k_{l\,'}&~~~~~\text{if }\sigma^{-1}(h_2)<\sigma^{-1}(l) ,\\
&\Sl_{\substack{l\,'\in\{1,\dots,r-1\}\\\sigma^{-1}(l\,')<\sigma^{-1}(l)}}k_{h_2}\cdot k_{l\,'}+\Sl_{\substack{l\,'\in\{1,\dots,r-1\}\\\sigma^{-1}(l)<\sigma^{-1}(l\,')<\sigma^{-1}(h_2)}}k_{h_2}\cdot k_{l\,'} &~~~~~\text{if }\sigma^{-1}(l)<\sigma^{-1}(h_2)<\sigma^{-1}(h_1),\nn
&\Sl_{\substack{l\,'\in\{1,\dots,r-1\}\\\sigma^{-1}(l\,')<\sigma^{-1}(l)}}k_{h_2}\cdot k_{l\,'}+\Sl_{\substack{l\,'\in\{1,\dots,r-1\}\\\sigma^{-1}(l)<\sigma^{-1}(l\,')<\sigma^{-1}(h_2)}}k_{h_2}\cdot k_{l\,'}+k_{h_2}\cdot k_{h_1}& ~~~~~\text{if }\sigma^{-1}(h_2)>\sigma^{-1}(h_1),\nonumber
\eea
where the first line in the above expression comes from the last term of \eqref{Eq:DoubleID3}, the second line gets contributions from both the second  and the last terms of \eqref{Eq:DoubleID3}, the third line gets contributions of all the three terms of \eqref{Eq:DoubleID3}. Apparently, all the three cases can be uniformly written as $k_{h_2}\cdot X_{h_2}(\pmb{\sigma})$ where $X_{h_2}(\pmb{\sigma})$ is the sum of all elements (including the element $1$) appearing on the LHS of $h_2$ in the permutation $\pmb{\sigma}$. Therefore \eqref{Eq:DoubleID3} can be reorganized as
\bea
\Sl_{\sigma\in\{h_2\}\shuffle\,\pmb{\beta}}\big(k_{h_2}\cdot X_{h_2}(\pmb{\sigma})\big)A(1,\pmb{\sigma},r)=\mathcal{B}(1\,\vert\,\{h_2\},\pmb{\beta}\,\vert\,r),~(\text{for } \pmb{\beta}\in\{2,\dots,l,\{h_1\}\shuffle\{l+1,\dots,r-1\}\} )
\eea
which is a combination of the LHS of fundamental BCJ relations.

The skeleton $\mathcal{F}'=\text{\figref{Fig:Figure2} (b)}$ contributes a factor $P^{[\mathcal{F}']}=\epsilon_{h_1}\cdot k_{h_2}$. Graphs $\mathcal{F}\supset \mathcal{F}'$  containing the skeleton \figref{Fig:Figure2} (b) are presented by \figref{Fig:Figure3} (c) and (d) which are associated with the factors $\mathcal{K}^{[\mathcal{F}\setminus\mathcal{F}\,']}=-k_{h_1}\cdot k_l$ and $\mathcal{K}^{[\mathcal{F}\setminus\mathcal{F}\,']}=k_{h_2}\cdot k_l$, respectively. Permutations $\pmb{\sigma}^{\mathcal{F}}$ corresponding to \figref{Fig:Figure3} (c) and (d) are
\bea
\pmb{\sigma}^{\mathcal{F}}\in\{2,\dots,l,\{h_1,h_2\}\shuffle\{l+1,\dots,r-1\}\} \text{ and } \pmb{\sigma}^{\mathcal{F}}\in\{2,\dots,l,\{h_2,h_1\}\shuffle\{l+1,\dots,r-1\}\},
\eea
where $1\leq l\leq r-1$. Thus the expression in the square brackets of \eqref{Eq:GaugeInv4} for the skeleton \figref{Fig:Figure2} (b) is given by
\bea
&-&\Sl_{l\in\{1,2,\dots,r-1\}}(k_{h_1}\cdot k_l)\Sl_{\pmb{\sigma}}A(1,\dots,l,\pmb{\sigma}\in\{h_1,h_2\}\shuffle\{l+1,\dots,r-1\},r)\nn
&+&\Sl_{l\in\{1,2,\dots,r-1\}}(k_{h_2}\cdot k_l)\Sl_{\pmb{\sigma}}A(1,\dots,l,\pmb{\sigma}\in\{h_2,h_1\}\shuffle\{l+1,\dots,r-1\},r).
\eea
All possible permutations $\pmb{\sigma}$ in the above expression belong to $\{2,\dots,r-1\}\shuffle\{h_1\}\shuffle\{h_2\}$, thus one can pick out $\pmb{\sigma}$ and collect all coefficients accompanying with it.
The coefficient  is $(k_{h_1}\cdot Y_{h_1}(\pmb{\sigma}))$ for any $\pmb{\sigma}\in\{h_1,h_2\}\shuffle\{2,\dots,r-1\}$ and $(k_{h_2}\cdot Y_{h_2}(\pmb{\sigma}))$ for any $\pmb{\sigma}\in\{h_2,h_1\}\shuffle\{2,\dots,r-1\}$.
Hence \figref{Fig:Figure2} (b) becomes
\bea
-\Sl_{\pmb{\sigma}\in\{h_1,h_2\}\shuffle\{2,\dots,r-1\}}(k_{h_1}\cdot Y_{h_1}(\pmb{\sigma}))A(1,\pmb{\sigma},r)+\Sl_{\pmb{\sigma}\in\{h_2,h_1\}\shuffle\{2,\dots,r-1\}}(k_{h_2}\cdot Y_{h_2}(\pmb{\sigma}))A(1,\pmb{\sigma},r),
\eea
which can be further arranged as
\bea
&&-\Sl_{\pmb{\beta}\in\{h_2\}\shuffle\{2,\dots,r-1\}}\,\Sl_{\pmb{\sigma}\in\{h_1\}\shuffle\,\pmb{\beta}}\,(k_{h_1}\cdot X_{h_1}(\pmb{\sigma}))\,A(1,\pmb{\sigma},r)\nn
&&+\Sl_{\pmb{\sigma}\in\{h_2,h_1\}\shuffle\{2,\dots,r-1\}}(k_{h_2}\cdot X_{h_2}(\pmb{\sigma})+k_{h_1}\cdot X_{h_1}(\pmb{\sigma}))A(1,\pmb{\sigma},r)\nn
&=&-\Sl_{\pmb{\beta}\in\{h_2\}\shuffle\{2,\dots,r-1\}}\mathcal{B}(1\,\vert\,\{h_2\},\pmb{\beta}\,\vert\,r)+\mathcal{B}(1\,\vert\,\{h_2,h_1\},\{2,\dots,r-1\}\,\vert\,r).
\eea
Finally, the expression in the square brackets of \eqref{Eq:GaugeInv4} for the skeleton \figref{Fig:Figure2} (b) has been expanded into the LHS of BCJ relations.

\begin{figure}
\centering
\includegraphics[width=0.75\textwidth]{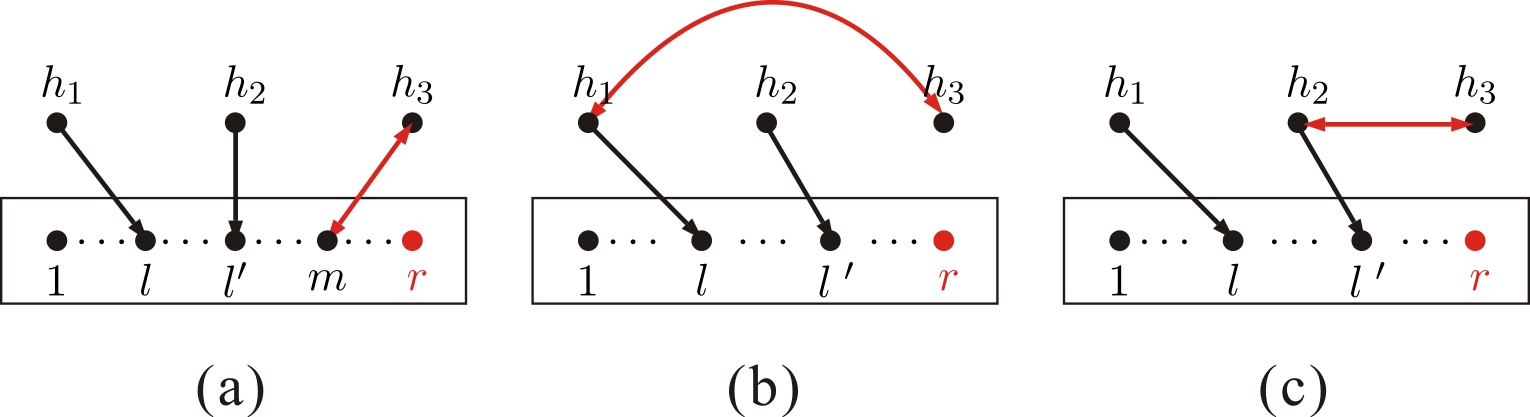}
\caption{All graphs containing the skeleton $\mathcal{F}\,'= \text{\figref{Fig:Figure4} (a)}$}\label{Fig:Figure5}
\end{figure}

\subsection{The identity with $\mathsf{H}=\{h_1,h_2,h_3\}$}\label{sec:|H|=3Example}

\begin{table}
\centering
\begin{tabular}{c|c|c}
  \hline
 Relative orders of $h_1$, $h_2$ and $h_3$ & Graphs $\mathcal{F}\supset \mathcal{F}'$ & Sum of coefficients $(-1)^{\mathcal{N}(\mathcal{F})}\mathcal{K}^{[\mathcal{F}\setminus\mathcal{F}']}$\\\hline
  $\{h_1,h_2,h_3\}$ &  \figref{Fig:Figure5} (a),(b),(c) &$k_{h_3}\cdot Y_{h_3}(\pmb{\sigma})+k_{h_3}\cdot k_{h_1}+k_{h_3}\cdot k_{h_2}$  \\
  $\{h_1,h_3,h_2\}$  & \figref{Fig:Figure5} (a),(b) & $k_{h_3}\cdot Y_{h_3}(\pmb{\sigma})+k_{h_3}\cdot k_{h_1}$\\
   $\{h_3,h_1,h_2\}$ & \figref{Fig:Figure5} (a) & $k_{h_3}\cdot Y_{h_3}(\pmb{\sigma})$ \\
  $\{h_2,h_1,h_3\}$ & \figref{Fig:Figure5} (a),(b),(c) & $k_{h_3}\cdot Y_{h_3}(\pmb{\sigma})+k_{h_3}\cdot k_{h_1}+k_{h_3}\cdot k_{h_2}$  \\
  $\{h_2,h_3,h_1\}$ &  \figref{Fig:Figure5} (a),(c) & $k_{h_3}\cdot Y_{h_3}(\pmb{\sigma})+k_{h_3}\cdot k_{h_2}$\\
  $\{h_3,h_2,h_1\}$ & \figref{Fig:Figure5} (a) & $k_{h_3}\cdot Y_{h_3}(\pmb{\sigma})$\\
  \hline
\end{tabular}
\caption{ Consider the graphs in  \figref{Fig:Figure5}, which contain the skeleton $\mathcal{F}\,'= \text{\figref{Fig:Figure4} (a)}$. According to the refined graphic rule, the factor $(-1)^{\mathcal{N}(\mathcal{F})}$ for all the three graphs in \figref{Fig:Figure5} is $(-1)^{0}=1$. For fixed $l$, $l'$ and a fixed permutation $\pmb{\beta}$ satisfying \eqref{Eq:TripleIDRelative1}, the coefficients  $(-1)^{\mathcal{N}(\mathcal{F})}\mathcal{K}^{[\mathcal{F}\setminus\mathcal{F}']}$ for each permutation $\pmb{\sigma}\in\pmb{\beta}\shuffle\{h_3\}$ are collected. For example, if the relative order of elements of $\{h_1,h_2,h_3\}$ in $\pmb{\sigma}\in\pmb{\beta}\shuffle\{h_3\}$ is $h_1$, $h_2$, $h_3$, all graphs in \figref{Fig:Figure5} contribute to this permutation. The sum of \figref{Fig:Figure5} (a) for all $m$ satisfying $\pmb{\sigma}^{-1}(m)<\pmb{\sigma}^{-1}(h_3)$ provides $k_{h_3}\cdot Y_{h_3}(\pmb{\sigma})$, while graphs \figref{Fig:Figure5} (b) and (c) contribute $k_{h_3}\cdot k_{h_1}$ and $k_{h_3}\cdot k_{h_2}$, respectively. Thus the total factor is  $k_{h_3}\cdot Y_{h_3}(\pmb{\sigma})+k_{h_3}\cdot k_{h_1}+k_{h_3}\cdot k_{h_2}$, as shown by the first row.}\label{table:triple-1}
\end{table}
Now we study the gauge invariance induced identity \eqref{Eq:GaugeInv2} with three elements in $\mathsf{H}$. The reference order is chosen as $\mathsf{R}=\{h_1,h_2,h_3\}$.
 There are two external polarizations $\epsilon_{h_1}$ and $\epsilon_{h_2}$ in the coefficients. If both of them are contracted with momenta, we have a factor of the form $\mathcal{P}^{\mathcal{F}'}=(\epsilon_{h_1}\cdot k_a)(\epsilon_{h_2}\cdot k_b)$ accompanied to the possible skeletons (a)-(h) in figure \ref{Fig:Figure4}. Else, if the two polarizations are contracted with each other, the factor $\mathcal{P}^{\mathcal{F}'}$ is  $\epsilon_{h_1}\cdot\epsilon_{h_2}$ corresponding to the skeleton $\mathcal{F}'=\text{\figref{Fig:Figure4} (i)}$. Apparently, all skeletons in \figref{Fig:Figure4} are disconnected graphs. For a given skeleton $\mathcal{F}'$, we call each maximally connected subgraph a \emph{component} (the set $\{1,\dots,r\}$ is also considered as a component). According to the number of components and the number of nodes in those components which do not contain $\{1,\dots,r\}$, we classify the skeletons into four categories {(1)}. \figref{Fig:Figure4} (a), (b) and (d), {(2)}. \figref{Fig:Figure4} (c), (f), {(3)}. \figref{Fig:Figure4} (e), (g) and (h), {(4)}. \figref{Fig:Figure4} (i).

(1) All the skeletons \figref{Fig:Figure4} (a), (b) and (d) contain two mutually disjoint components. In each skeleton, one component is the single node $h_3$ while the other one is a connected subgraph containing all elements in $\{1,\dots,r\}\cup\{h_1,h_2\}$. We take the skeleton $\mathcal{\mathcal{F}}'= \text{\figref{Fig:Figure4} (a)}$, which is associated with a factor $\mathcal{P}^{[\mathcal{F}\,']}=(\epsilon_{h_1}\cdot l)(\epsilon_{h_2}\cdot l\,')$ (for given $l,l\,'\in\{1,2,\dots,r-1\}$), for example.  The possible relative orders $\pmb{\beta}$ for elements in $\{1,\dots,r\}\cup\{h_1,h_2\}$ are
\bea
&&\pmb{\beta}\in\{2,\dots,l,\{h_1\}\shuffle\{l+1,\dots,l',\{h_2\}\shuffle\{l'+1,\dots,r-1\}\}~~(1\leq l\leq l'\leq r-1)\nn
&&\pmb{\beta}\in\{2,\dots,l',\{h_2\}\shuffle\{l'+1,\dots,l,\{h_1\}\shuffle\{l+1,\dots,r-1\}\}~~(1\leq l'< l\leq r-1).\Label{Eq:TripleIDRelative1}
\eea
All graphs $\mathcal{F}$ containing the skeleton $\mathcal{F}\,'= \text{\figref{Fig:Figure4} (a)}$ are provided by \figref{Fig:Figure5} (a), (b) and (c). For any fixed $l$, $l'$ and a corresponding $\pmb{\beta}$, all possible permutations $\pmb{\sigma}$ corresponding to graphs \figref{Fig:Figure5} (a), (b) and (c) belong to $\pmb{\beta}\shuffle\{h_3\}$.
We can collect coefficients  $(-1)^{\mathcal{N}(\mathcal{F})}\mathcal{K}^{[\mathcal{F}\setminus\mathcal{F}']}$ for each permutation $\pmb{\sigma}\in\pmb{\beta}\shuffle\{h_3\}$.
As shown by \tabref{table:triple-1}, coefficients for any permutation $\pmb{\sigma}\in\pmb{\beta}\shuffle\{h_3\}$ can be uniformly written as $k_{h_3}\cdot X_{h_3}({\pmb{\sigma}})$. Thus the expression in the square brackets of \eqref{Eq:GaugeInv5} for the skeleton $\mathcal{\mathcal{F}}'= \text{\figref{Fig:Figure4} (a)}$ (with given $l$ and $l'$) is expressed by
    \bea
    \Sl_{\pmb{\beta}}\biggl[\Sl_{\pmb{\sigma}\in\{h_3\}\shuffle\,\pmb{\beta}}(k_{h_3}\cdot X_{h_3}(\pmb{\sigma}))A(1,\pmb{\sigma},r)\biggr]=\Sl_{\pmb{\beta}}\mathcal{B}(1\,\vert\,\{h_3\},\pmb{\beta}\,\vert\,r),\Label{Eq:TripleID-1}
    \eea
    where we summed over all $\pmb{\beta}$ satisfy  \eqref{Eq:TripleIDRelative1}. Apparently, the expression in the square brackets in the above equation is the LHS of a fundamental BCJ relation.

\begin{figure}
\centering
\includegraphics[width=7in]{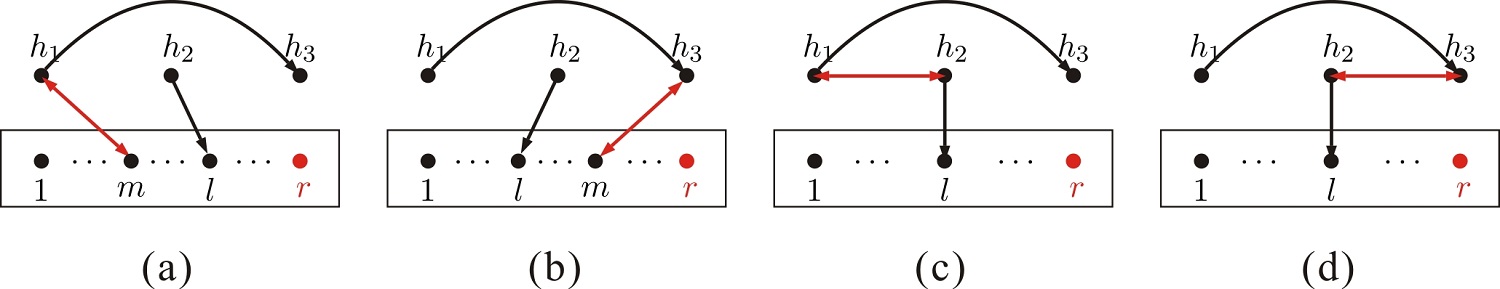}
\caption{All graphs $\mathcal{F}$ containing the skeleton $\mathcal{F}'=\text{\figref{Fig:Figure4} (c)}$.}\label{Fig:Figure6}
\end{figure}

\begin{table}
\centering
\begin{tabular}{c|c|c}
  \hline
 Relative orders of $h_1$, $h_2$ and $h_3$ & Graphs $\mathcal{F}\supset \mathcal{F}'$ & Sum of coefficients  $(-1)^{\mathcal{N}(\mathcal{F})}\mathcal{K}^{[\mathcal{F}\setminus\mathcal{F}']}$\\\hline
  $\{h_1,h_2,h_3\}$ &   \figref{Fig:Figure6} (a) &$-k_{h_1}\cdot Y_{h_1}(\pmb{\sigma})$  \\
  $\{h_1,h_3,h_2\}$  &\figref{Fig:Figure6} (a)& $-k_{h_1}\cdot Y_{h_1}(\pmb{\sigma})$ \\
   $\{h_2,h_1,h_3\}$ & \figref{Fig:Figure6} (a), (c) & $-(k_{h_1}\cdot Y_{h_1}(\pmb{\sigma})+k_{h_1}\cdot k_{h_2})$ \\
  $\{h_2,h_3,h_1\}$ & \figref{Fig:Figure6} (b), (d) & $k_{h_3}\cdot Y_{h_3}(\pmb{\sigma})+k_{h_3}\cdot k_{h_2}$ \\
  $\{h_3,h_1,h_2\}$ & \figref{Fig:Figure6} (b) & $k_{h_3}\cdot Y_{h_3}(\pmb{\sigma})$\\
  $\{h_3,h_2,h_1\}$ & \figref{Fig:Figure6} (b) & $k_{h_3}\cdot Y_{h_3}(\pmb{\sigma})$\\
  \hline
\end{tabular}
\caption{ Consider the graphs in  \figref{Fig:Figure6}, which contain the skeleton $\mathcal{F}\,'= \text{\figref{Fig:Figure4} (c)}$. According to the refined graphic rule, the factor $(-1)^{\mathcal{N}(\mathcal{F})}$ is $-1$ for \figref{Fig:Figure6} (a), (c), $1$ for \figref{Fig:Figure6} (b), (d). For a given $l$ and a given $\pmb{\beta}$ satisfying \eqref{Eq:TripleIDRelative2}, permutations $\pmb{\sigma}$ corresponding to the skeleton $\mathcal{F}'=\text{\figref{Fig:Figure4} (c)}$ belong to $\{h_1\}\shuffle\{h_3\}\shuffle\pmb{\beta}$ for $\pmb{\beta}$. For any $\pmb{\sigma}\in\{h_1\}\shuffle\{h_3\}\shuffle\pmb{\beta}$ we collect the coefficients $(-1)^{\mathcal{N}(\mathcal{F})}\mathcal{K}^{[\mathcal{F}\setminus\mathcal{F}']}$ together as we have done in the previous case.  Each row shows the expression of the sum of $(-1)^{\mathcal{N}(\mathcal{F})}\mathcal{K}^{[\mathcal{F}\setminus\mathcal{F}']}$ for a given relative order of $h_1$, $h_2$ and $h_3$. }\label{table:triple-2}
\end{table}
(2) Both skeletons \figref{Fig:Figure4} (c) and (f) contain two mutually disconnected components. In each skeleton, the component which does not contain $\{1,\dots,r\}$ has two nodes with a type-2 line between them. We take the graph  \figref{Fig:Figure4} (c) as an example. The factor $\mathcal{P}^{\mathcal{F}'}$ for the skeleton \figref{Fig:Figure4} (c) is $(\epsilon_{h_1}\cdot k_{h_3})(\epsilon_{h_2}\cdot k_l)$. Relative permutations for the elements in the component containing $\{1,\dots,r\}$ are
\bea
\pmb{\beta}\in\{2,\dots,l,\{h_2\}\shuffle\{l+1,\dots,r-1\}\}~~~~~(1\leq l\leq r-1). \Label{Eq:TripleIDRelative2}
\eea
All graphs $\mathcal{F}$ satisfying $\mathcal{F}\supset\mathcal{F}\,'=\text{\figref{Fig:Figure4} (c)}$ are presented by \figref{Fig:Figure6} and all possible permutations $\pmb{\sigma}$ allowed by the graphs in \figref{Fig:Figure6} belong to $\pmb{\beta}\shuffle\{h_1\}\shuffle\{h_3\}$ (for a given $\pmb{\beta}$ in \eqref{Eq:TripleIDRelative2}). To obtain the expression in the brackets in \eqref{Eq:GaugeInv5} corresponding to the skeleton \text{\figref{Fig:Figure4} (c)} for a given $l\in\{1,\dots,r-1\}$, we can collect the coefficients for any given permutation $\pmb{\sigma}\in \pmb{\beta}\shuffle\{h_1\}\shuffle\{h_3\}$ (see \tabref{table:triple-2}). Then sum over all possible $\pmb{\sigma}\in \pmb{\beta}\shuffle\{h_1\}\shuffle\{h_3\}$ and all possible $\pmb{\beta}$ satisfying \eqref{Eq:TripleIDRelative2}. From \tabref{table:triple-2}, we find that the coefficients in the first three rows and  the last three rows can be respectively written as $-k_{h_1}\cdot X_{h_1}(\pmb{\sigma})$ and $k_{h_3}\cdot X_{h_3}(\pmb{\sigma})$.
Thus the expression in the square brackets of \eqref{Eq:GaugeInv5} for the skeleton \text{\figref{Fig:Figure4} (c)} ($l\in\{1,\dots,r-1\}$) is given by
\bea
-\Sl_{\pmb{\beta}}\biggl[\Sl_{\pmb{\sigma}\in\{h_1,h_3\}\shuffle\,\pmb{\beta}}(k_{h_1}\cdot X_{h_1}(\pmb{\sigma}))A(1,\pmb{\sigma},r)\biggr]&+&\Sl_{\pmb{\beta}}\biggl[\Sl_{\pmb{\sigma}\in\{h_3,h_1\}\shuffle\,\pmb{\beta}} (k_{h_3}\cdot X_{h_3}(\pmb{\sigma}))A(1,\pmb{\sigma},r)\biggr],\Label{Eq:TripleID-2-1}
\eea
in which the first summation is taken over all $\pmb{\beta}$ satisfying \eqref{Eq:TripleIDRelative2}. The first term in the above expression can be rewritten as
\bea
&-&\Sl_{\pmb{\beta}}\biggl[\Sl_{\pmb{\sigma}\in\{h_1\}\shuffle\{h_3\}\shuffle\,\pmb{\beta}}(k_{h_1}\cdot X_{h_1}(\pmb{\sigma}))A(1,\pmb{\sigma},r)-\Sl_{\pmb{\sigma}\in\{h_3,h_1\}\shuffle\,\pmb{\beta}}(k_{h_1}\cdot X_{h_1}(\pmb{\sigma}))A(1,\pmb{\sigma},r)\biggr]\nn
&=&-\Sl_{\pmb{\beta}}\Sl_{\pmb{\rho}}\mathcal{B}(1\,\vert\,\{h_1\},\pmb{\rho}\in\{h_3\}\shuffle\pmb{\beta}\,\vert\,r)+\Sl_{\pmb{\beta}}\biggl[\Sl_{\pmb{\sigma}\in\{h_3,h_1\}\shuffle\,\pmb{\beta}}(k_{h_1}\cdot X_{h_1}(\pmb{\sigma}))A(1,\pmb{\sigma},r)\biggr].
\eea
Since the sum of the last term on the second line of the above equation and the second term of \eqref{Eq:TripleID-2-1} is just a combination of LHS of BCJ relations $\Sl_{\pmb{\beta}}\mathcal{B}(1\,\vert\,\{h_3,h_1\},\pmb{\beta}\,\vert\,r)$, \eqref{Eq:TripleID-2-1} is finally arranged as
\bea
-\Sl_{\pmb{\beta}}\Sl_{\pmb{\rho}}\mathcal{B}(1\,\vert\,\{h_1\},\pmb{\rho}\in\{h_3\}\shuffle\pmb{\beta}\,\vert\,r)+\Sl_{\pmb{\beta}}\mathcal{B}(1\,\vert\,\{h_3,h_1\},\pmb{\beta}\,\vert\,r),\Label{Eq:TripleID-2-2}
\eea
the first summation in each term is taken over all $\pmb{\beta}$ satisfying \eqref{Eq:TripleIDRelative2}. Thus  the expression in the brackets in \eqref{Eq:GaugeInv5} for the skeleton \figref{Fig:Figure4} (c) is written as a combination of the LHS of BCJ relations.

\begin{figure}
\centering
\includegraphics[width=5.5in]{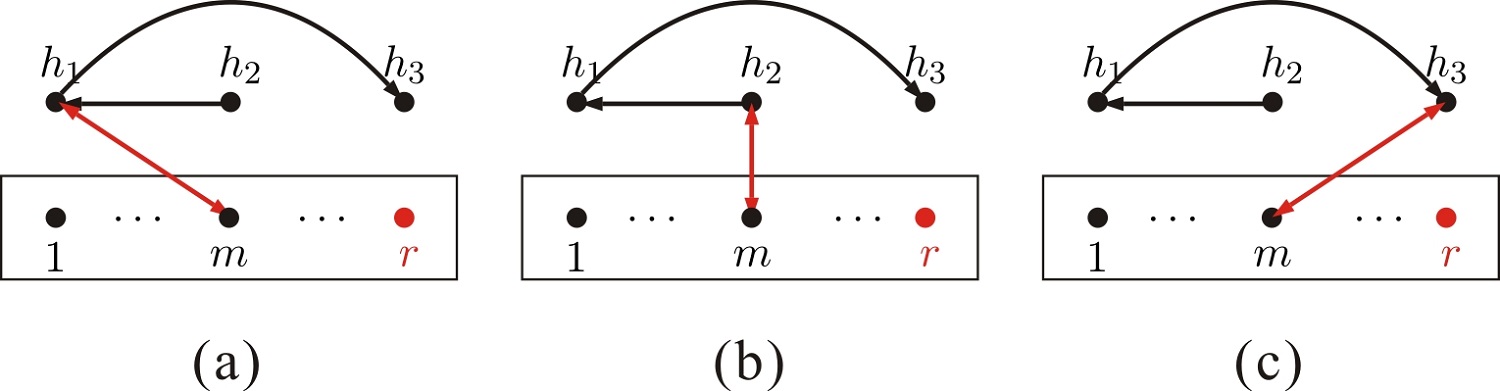}
\caption{All graphs $\mathcal{F}$ containing the skeleton $\mathcal{F}\,'=\text{\figref{Fig:Figure4} (e)}$.}\label{Fig:Figure7}
\end{figure}

\begin{figure}
\centering
\includegraphics[width=0.8\textwidth]{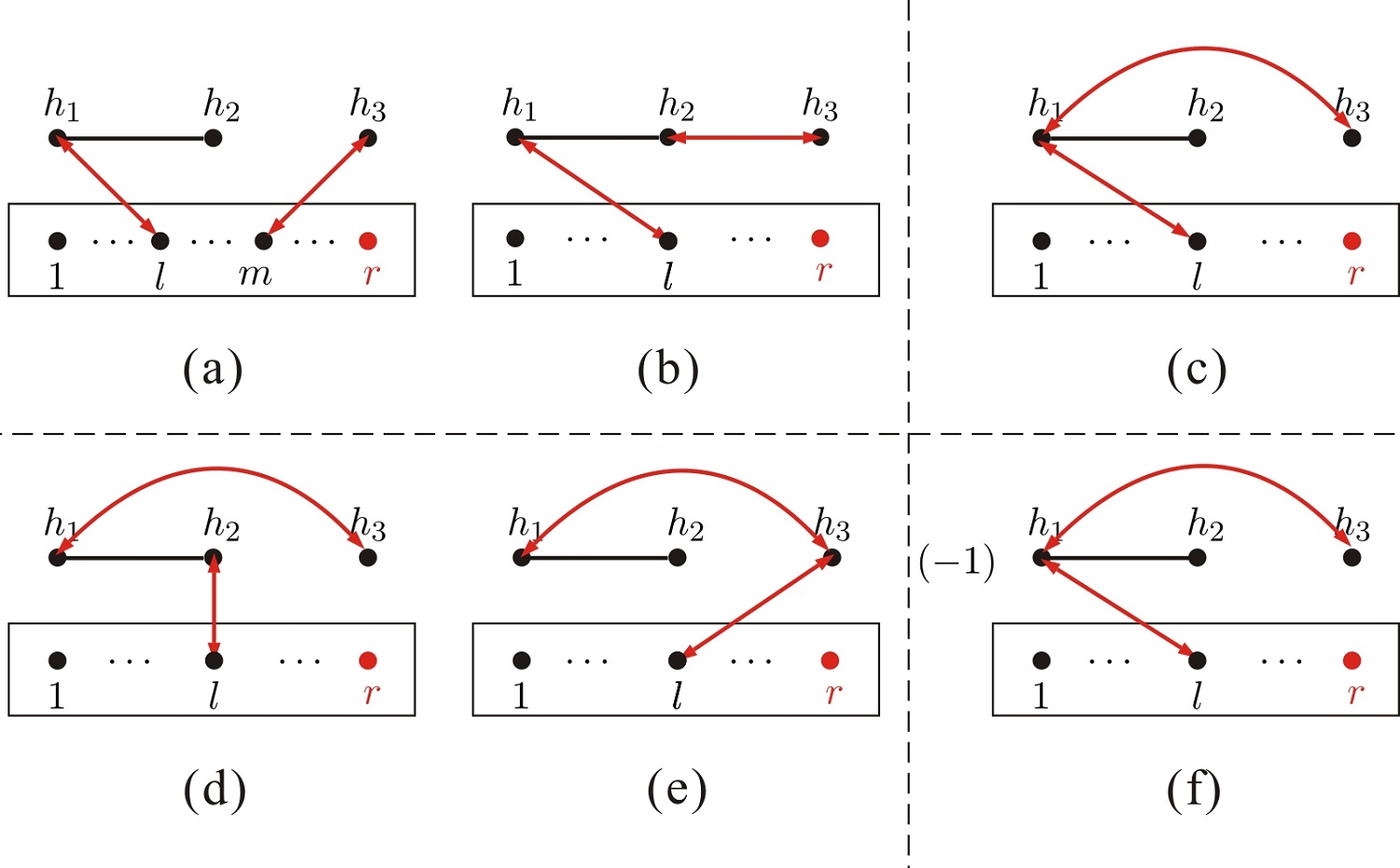}
\caption{Graphs (a), (b), (d) and (e) are physical graphs, which are constructed by the refined graphic rule, for the skeleton  $\mathcal{F}\,'=\text{\figref{Fig:Figure4} (i)}$. Graphs (d) and (e) are spurious graphs for the skeleton $\mathcal{F}\,'=\text{\figref{Fig:Figure4} (i)}$. All spurious graphs cancel with each other.}\label{Fig:Figure8}
\end{figure}
\begin{table}
\centering
    \begin{tabular}{c|c|c}
      \hline
      Relative orders of $h_1$, $h_2$ and $h_3$ & Graphs $\mathcal{F}\supset\mathcal{F}\,'$ & Coefficients $(-1)^{\mathcal{N}(\mathcal{F})}\mathcal{K}^{[\mathcal{F}\setminus\mathcal{F}']}$ \\\hline
     $\{h_1,h_2,h_3\}$ &\figref{Fig:Figure7} (a)  & $-k_{h_1}\cdot Y_{h_1}(\pmb{\sigma})$ \\
      $\{h_1,h_3,h_2\}$ & \figref{Fig:Figure7} (a)  &$-k_{h_1}\cdot Y_{h_1}(\pmb{\sigma})$  \\
      $\{h_2,h_1,h_3\}$ & \figref{Fig:Figure7} (b) & $k_{h_2}\cdot Y_{h_2}(\pmb{\sigma})$\\
      $\{h_3,h_1,h_2\}$&\figref{Fig:Figure7} (c)&$k_{h_3}\cdot Y_{h_3}(\pmb{\sigma})$\\
      \hline
    \end{tabular}
  \caption{Total coefficients  $(-1)^{\mathcal{N}(\mathcal{F})}\mathcal{K}^{[\mathcal{F}\setminus\mathcal{F}']}$ for permutations allowed by the graphs in \figref{Fig:Figure7}.}\label{table:triple-3}
  \end{table}

 (3) Each of the skeletons \figref{Fig:Figure4} (e), (g) and (h) consists of two mutually disjoint components. One of them contains all elements of the set $\mathsf{H}=\{h_1,h_2,h_3\}$ and two type-2 lines between them. The other contains all elements in $\{1,\dots,r\}$.
We now study the skeleton $\mathcal{F}\,'=\text{\figref{Fig:Figure4} (e)}$, which gives a factor $\mathcal{P}^{[\mathcal{F}\,']}=(\epsilon_{h_1}\cdot k_{h_3})(\epsilon_{h_2}\cdot k_{h_1})$, for instance. All possible graphs satisfying $\mathcal{F}\supset \mathcal{F}\,'$ are displayed in \figref{Fig:Figure7}. Relative orders of $h_1$, $h_2$ and $h_3$, allowed by the graphs in  \figref{Fig:Figure7} are $\{h_1,\{h_2\}\shuffle\{h_3\}\}$, $\{h_2,h_1,h_3\}$ and $\{h_3,h_1,h_2\}$, whose coefficients   $\mathcal{K}^{[\mathcal{F}\setminus \mathcal{F}\,']}$ are collected by \tabref{table:triple-3}. Thus the total contribution from the skeleton reads
  \bea
  &&\Sl_{\pmb{\sigma}\in \{2,\dots,r-1\}\shuffle\{h_1,\{h_2\}\shuffle\{h_3\}\}}(-k_{h_1}\cdot Y_{h_1}(\pmb{\sigma}))A(1,\pmb{\sigma},r)\Label{Eq:TripleID-3-1}\\
  &+&\Sl_{\pmb{\sigma}\in\{2,\dots,r-1\}\shuffle\{h_2,h_1,h_3\}}(k_{h_2}\cdot Y_{h_2}(\pmb{\sigma}))A(1,\pmb{\sigma},r)+\Sl_{\pmb{\sigma}\in\{2,\dots,r-1\}\shuffle\{h_3,h_1,h_2\}}(k_{h_3}\cdot Y_{h_3}(\pmb{\sigma}))A(1,\pmb{\sigma},r).\nonumber
  \eea
When expressing the last two terms respectively by
\bea
\Sl_{\pmb{\beta}\in\{2,\dots,r-1\}\shuffle\{h_1,h_3\}} \mathcal{B}(1\,\vert\,\{h_2\},\pmb{\beta}\,\vert\,r)-\biggl[\Sl_{\pmb{\sigma}\in\{2,\dots,r-1\}\shuffle\{h_1,\{h_2\}\shuffle\{h_3\}\}}(k_{h_2}\cdot X_{h_2}(\pmb{\sigma}))A(1,\pmb{\sigma},r)\biggr]
\eea
and
\bea
\Sl_{\pmb{\beta}\in\{2,\dots,r-1\}\shuffle\{h_1,h_2\}} \mathcal{B}(1\,\vert\,\{h_3\},\pmb{\beta}\,\vert\,r)-\biggl[\Sl_{\pmb{\sigma}\in\{2,\dots,r-1\}\shuffle\{h_1,\{h_2\}\shuffle\{h_3\}\}}(k_{h_3}\cdot X_{h_3}(\pmb{\sigma}))A(1,\pmb{\sigma},r)\biggr],
\eea
\eqref{Eq:TripleID-3-1} becomes
\bea
&&\Sl_{\pmb{\beta}\in\{2,\dots,r-1\}\shuffle\{h_1,h_3\}} \mathcal{B}(1\,\vert\,\{h_2\},\pmb{\beta}\,\vert\,r)+\Sl_{\pmb{\beta}\in\{2,\dots,r-1\}\shuffle\{h_1,h_2\}} \mathcal{B}(1\,\vert\,\{h_3\},\pmb{\beta}\,\vert\,r)\nn
&-&\Sl_{\pmb{\rho}\in\{h_1,\{h_2\}\shuffle\{h_3\}\}}\mathcal{B}(1\,\vert\,\pmb{\rho},\{2,\dots,r-1\}\,\vert\,r),\Label{Eq:TripleID-2-2}
\eea
which is a combination of the LHS of BCJ relations.

\begin{table}
\centering
    \begin{tabular}{c|c|c|c|c}
      \hline
       \tabincell{c}{Relative orders \\of $h_1$, $h_2$ and $h_3$} & \tabincell{c}{Physical graphs\\$\mathcal{F}\supset\mathcal{F}\,'$}& \tabincell{c}{ $(-1)^{\mathcal{N}(\mathcal{F})}\mathcal{K}^{[\mathcal{F}\setminus\mathcal{F}']}$ from\\ physical graphs} & \tabincell{c}{Spurious graphs\\$\mathcal{F}\supset\mathcal{F}\,'$}&\tabincell{c}{Factors from\\ spurious graphs}\\\hline
   $\{h_1,h_2,h_3\}$&\small\figref{Fig:Figure8} (a),(b) & {\small$-(k_{h_1}\cdot k_l)(k_{h_3}\cdot Y_{h_3}+k_{h_3}\cdot k_{h_2})$}&\small\figref{Fig:Figure8} (c)& \small$-(k_{h_3}\cdot k_{h_1})(k_{h_1}\cdot k_l)$\\
       $\{h_1,h_3,h_2\}$
   &\small\figref{Fig:Figure8} (a) & \small$-(k_{h_1}\cdot k_l)(k_{h_3}\cdot Y_{h_3})$&\small\figref{Fig:Figure8} (c) &\small$-(k_{h_3}\cdot k_{h_1})(k_{h_1}\cdot k_l)$\\
  $\{h_3,h_1,h_2\}$ & \small\figref{Fig:Figure8} (a)&\small$-(k_{h_1}\cdot k_l)(k_{h_3}\cdot Y_{h_3})$&no&0\\ \hline
 $\{h_1,h_2,h_3\}$ &no &0&\small\figref{Fig:Figure8} (f)&\small$(k_{h_3}\cdot k_{h_1})(k_{h_1}\cdot Y_{h_1})$\\
$\{h_1,h_3,h_2\}$ &no &0&\small\figref{Fig:Figure8} (f)&\small$(k_{h_3}\cdot k_{h_1})(k_{h_1}\cdot Y_{h_1})$\\
      $\{h_2,h_1,h_3\}$ & \small\figref{Fig:Figure8} (d) &\small $-(k_{h_2}\cdot Y_{h_2})(k_{h_3}\cdot k_{h_1})$&no &0\\
    $\{h_3,h_1,h_2\}$ & \small\figref{Fig:Figure8} (e)&\small$-(k_{h_1}\cdot k_{h_3})(k_{h_3}\cdot Y_{h_3})$&no&0\\
      \hline
    \end{tabular}
  \caption{For given $l$ and given $\pmb{\beta}$ satisfying \eqref{Eq:TripleIDRelative3} in the upper part of \figref{Fig:Figure8}, relative orders and corresponding factors are collected on the second, third and fourth rows. Relative orders for graphs in the lower part of \figref{Fig:Figure8} and their factors are collected on the last four rows.}\label{table:triple-4}
  \end{table}
(4) The skeleton  \figref{Fig:Figure4} (i) is a much more nontrivial example.
There are three mutually disjoint components in \figref{Fig:Figure4} (i) which contain  the elements $\{1,\dots,r\}$, the single node $h_3$ and the subgraph with nodes $h_1$ and $h_2$, correspondingly.  The factor $\mathcal{P}^{[\mathcal{F}\,']}$ for \figref{Fig:Figure4} (i) is $\epsilon_{h_1}\cdot\epsilon_{h_2}$ and the possible graphs $\mathcal{F}\supset\mathcal{F}\,'=\text{\figref{Fig:Figure4} (i)}$ are shown by \figref{Fig:Figure8} (a), (b), (d) and (e). For a given graph \figref{Fig:Figure8} (a) or (b), in which $h_1$ is contracted with an arbitrary $l\in\{1,\dots,r-1\}$, the possible relative orders of elements in $\{h_1,h_2\}\cup\{1,\dots,r-1\}$ satisfy
\bea
\pmb{\beta}\in\{1,\dots,l,\{h_1,h_2\}\shuffle\{l+1,\dots,r-1\}\}. \Label{Eq:TripleIDRelative3}
\eea
Hence all permutations contributing to graphs \figref{Fig:Figure8} (a) and (b) have the form $\pmb{\sigma}\in\{h_3\}\shuffle\pmb{\beta}$. To get the total contribution of all graphs \figref{Fig:Figure8} (a) and (b), we should collect coefficients  $(-1)^{\mathcal{N}(\mathcal{F})}\mathcal{K}^{[\mathcal{F}\setminus\mathcal{F}']}$ together for a given permutation $\pmb{\sigma}\in\{h_3\}\shuffle\pmb{\beta}$ (see \tabref{table:triple-4}), then sum over all $\pmb{\beta}$ satisfying \eqref{Eq:TripleIDRelative3} and all $l\in\{1,\dots,r-1\}$. The total contributions of all graphs \figref{Fig:Figure8} (a) and (b) thus is written as
\bea
T_1&=&-\Sl_{l\in\{1,\dots,r-1\}}(k_{h_1}\cdot k_l)\Sl_{\substack{\small\pmb{\beta}\in\{2,\dots,l,\{h_1,h_2\}\\ \small\shuffle\{l+1,\dots,r-1\}\}}}\biggl[\Sl_{\small\substack{{\pmb{\sigma}\in\{h_3\}\shuffle\pmb{\beta}}\\{\pmb{\sigma}^{-1}(h_2)<\pmb{\sigma}^{-1}(h_3)}}}(k_{h_3}\cdot Y_{h_3}(\pmb{\sigma})+k_{h_3}\cdot k_{h_2})A(1,\pmb{\sigma},r)\nn
&&+\Sl_{\small\substack{{\pmb{\sigma}\in\{h_3\}\shuffle\pmb{\beta}}\\{\pmb{\sigma}^{-1}(h_1)<\pmb{\sigma}^{-1}(h_3)<\pmb{\sigma}^{-1}(h_2)}}}(k_{h_3}\cdot Y_{h_3}(\pmb{\sigma}))A(1,\pmb{\sigma},r)+\Sl_{\small\substack{{\pmb{\sigma}\in\{h_3\}\shuffle\pmb{\beta}}\\{\scriptsize\pmb{\sigma}^{-1}(h_3)<\pmb{\sigma}^{-1}(h_1)}}}(k_{h_3}\cdot Y_{h_3}(\pmb{\sigma}))A(1,\pmb{\sigma},r)\biggr].\Label{Eq:TripleID-4-1}
\eea

For a given graph  \figref{Fig:Figure8} (d) or (e), where $h_2$ is contracted with any $l\in\{1,\dots,r-1\}$, all the possible permutations  $\pmb{\sigma}$ contributing to  \figref{Fig:Figure8} (d) and (e) satisfy
\bea
&&\pmb{\sigma}\in\{1,\dots,r-1\}\shuffle\{h_2,h_1,h_3\}, \text{~for \figref{Fig:Figure8} (d)},\nn
&&\pmb{\sigma}\in\{1,\dots,r-1\}\shuffle\{h_3,h_1,h_2\}, \text{~for \figref{Fig:Figure8} (e)}.
\eea
When collecting the coefficients $(-1)^{\mathcal{N}(\mathcal{F})}\mathcal{K}^{[\mathcal{F}\setminus\mathcal{F}']}$ for any given $\pmb{\sigma}$ (see \tabref{table:triple-4}), we arrive the total contribution of the graphs \figref{Fig:Figure8} (d) and (e)
\bea
T_2&=&-(k_{h_1}\cdot k_{h_3})\biggl[\,\Sl_{\substack{\scriptsize\pmb{\sigma}\in\{2,\dots,r-1\}\\ \scriptsize~\shuffle\{h_3,h_1,h_2\}}}(k_{h_3}\cdot Y_{h_3}(\pmb{\sigma}))A(1,\pmb{\sigma},r)+\Sl_{\substack{\scriptsize\pmb{\sigma}\in\{2,\dots,r-1\}\\ \scriptsize~\shuffle\{h_2,h_1,h_3\}}}(k_{h_2}\cdot Y_{h_2}(\pmb{\sigma}))A(1,\pmb{\sigma},r)\biggr].\Label{Eq:TripleID-4-2}
\eea

In order to reorganize $T_1+T_2$ in terms of the LHS of BCJ relations,
we introduce so-called \emph{spurious graphs} \figref{Fig:Figure8} (c) and (f) which also contains the skeleton \figref{Fig:Figure4} (i) but are not real physical graphs constructed by the refined graphic rule. The two spurious graphs \figref{Fig:Figure8} (c) and (f) have the same structure with opposite signs, thus they must cancel with one another. Relative orders associated with each spurious graph are $\{h_1,h_2,h_3\}$ and $\{h_1,h_3,h_2\}$. All the spurious graphs corresponding to \figref{Fig:Figure8} (c) contribute (see \tabref{table:triple-4})
\bea
&&-(k_{h_3}\cdot k_{h_1})\biggl[\,\Sl_{l\in\{1,\dots,r-1\}}(k_{h_1}\cdot k_l)\Sl_{\substack{\scriptsize\pmb{\sigma}\in\{2,\dots,l,\{h_1,\{h_2\}\shuffle\{h_3\}\}\\ \scriptsize\shuffle\{l+1,\dots,r-1\}\}}}A(1,\pmb{\sigma},r)\biggr]\nn
&=&-(k_{h_3}\cdot k_{h_1})\biggl[\,\Sl_{\substack{\scriptsize\pmb{\sigma}\in\{h_1,\{h_2\}\shuffle\{h_3\}\}\\ \scriptsize\shuffle\{2,\dots,r-1\}}}(k_{h_1}\cdot Y_{h_1})A(1,\pmb{\sigma},r)\biggr],\Label{Eq:TripleIDSpurious}
\eea
to the skeleton $\mathcal{F}\,'=\text{\figref{Fig:Figure4} (i)}$, while the spurious graphs corresponding to \figref{Fig:Figure8} (f) have the same contribution but with an opposite sign. For any given $l$ on the LHS of \eqref{Eq:TripleIDSpurious}, the sum over $\pmb{\sigma}$ can be achieved by first summing over all permutations $\pmb{\sigma}\in\{h_3\}\shuffle\pmb{\beta}$ which satisfy $\pmb{\sigma}^{-1}(h_2)<\pmb{\sigma}^{-1}(h_3)$ or $\pmb{\sigma}^{-1}(h_1)<\pmb{\sigma}^{-1}(h_3)<\pmb{\sigma}^{-1}(h_2)$ for any given $\pmb{\beta}$ satisfying \eqref{Eq:TripleIDRelative3}, then summing over all possible $\pmb{\beta}$ satisfying \eqref{Eq:TripleIDRelative3}.
Therefore, the sum of all contributions from the spurious graph \figref{Fig:Figure8} (c)  and $T_1$ reads
\bea
&&-\Sl_{l\in\{1,\dots,r-1\}}(k_{h_1}\cdot k_l)\Sl_{\substack{\scriptsize\pmb{\beta}\in\{2,\dots,l,\{h_1,h_2\}\\ \scriptsize\shuffle\{l+1,\dots,r-1\}\}}}\biggl[\,\Sl_{\scriptsize\substack{{\pmb{\sigma}\in\{h_3\}\shuffle\pmb{\beta}}\\ \scriptsize{\pmb{\sigma}^{-1}(h_2)<\pmb{\sigma}^{-1}(h_3)}}}(k_{h_3}\cdot Y_{h_3}(\pmb{\sigma})+k_{h_3}\cdot k_{h_2}+k_{h_3}\cdot k_{h_1})A(1,\pmb{\sigma},r)\\
&&+\Sl_{\substack{\scriptsize{\pmb{\sigma}\in\{h_3\}\shuffle\pmb{\beta}}\\ \scriptsize{\pmb{\sigma}^{-1}(h_1)<\pmb{\sigma}^{-1}(h_3)<\pmb{\sigma}^{-1}(h_2)}}}(k_{h_3}\cdot Y_{h_3}(\pmb{\sigma})+k_{h_3}\cdot k_{h_1})A(1,\pmb{\sigma},r)+\Sl_{\substack{\scriptsize{\pmb{\sigma}\in\{h_3\}\shuffle\pmb{\beta}}\\ \scriptsize{\pmb{\sigma}^{-1}(h_3)<\pmb{\sigma}^{-1}(h_1)}}}(k_{h_3}\cdot Y_{h_3}(\pmb{\sigma}))A(1,\pmb{\sigma},r)\biggr]\nn
&=&-\Sl_{l\in\{1,\dots,r-1\}}(k_{h_1}\cdot k_l)\Sl_{\substack{\scriptsize\pmb{\beta}\in\{2,\dots,l,\{h_1,h_2\}\\ \scriptsize\shuffle\{l+1,\dots,r-1\}\}}}\mathcal{B}(1|\{h_3\},\pmb{\beta}|r).
\Label{Eq:TripleID-4-3}\nonumber
\eea
The sum of the contributions from the spurious graph \figref{Fig:Figure8} (f) (here we make use of the RHS of \eqref{Eq:TripleIDSpurious} and notice that there is an extra minus)  and $T_2$ is
\bea
&&(k_{h_1}\cdot k_{h_3})\biggl[\,-\Sl_{\substack{\scriptsize\pmb{\sigma}\in\{2,\dots,r-1\}\\ \scriptsize\shuffle\{h_3,h_1,h_2\}}}(k_{h_3}\cdot Y_{h_3}(\pmb{\sigma}))A(1,\pmb{\sigma},r)-\Sl_{\substack{\scriptsize\pmb{\sigma}\in\{2,\dots,r-1\}\\ \scriptsize\shuffle\{h_2,h_1,h_3\}}}(k_{h_2}\cdot Y_{h_2}(\pmb{\sigma}))A(1,\pmb{\sigma},r)\nn
&&+\Sl_{\substack{\scriptsize\pmb{\sigma}\in\{2,\dots,r-1\}\\ \scriptsize\shuffle\{h_1,\{h_2\}\shuffle\{h_3\}\}}}(k_{h_1}\cdot Y_{h_1}(\pmb{\sigma}))A(1,\pmb{\sigma},r)\biggr]\nn
&=&(k_{h_1}\cdot k_{h_3})\biggl[\,-\Sl_{\substack{\scriptsize\pmb{\beta}\in\{2,\dots,r-1\}\\ \scriptsize\shuffle\{h_1,h_2\}}}\mathcal{B}(1\,\vert\,\{h_3\},\pmb{\beta}\,\vert\,r)
-\Sl_{\substack{\scriptsize\pmb{\beta}\in\{2,\dots,r-1\}\\ \scriptsize\shuffle\{h_1,h_3\}}}\mathcal{B}(1\,\vert\,\{h_2\},\pmb{\beta}\,\vert\,r)\nn
&&+\Sl_{{\scriptsize\pmb{\rho}\in\{h_1,\{h_2\}\shuffle\{h_3\}\}}}\mathcal{B}(1\,\vert\,\pmb{\rho},\{2,\dots,r-1\}\,\vert\,r)\biggr].\Label{Eq:TripleID-4-4}
\eea
Finally, the sum of $T_1$ and $T_2$ becomes the sum of \eqref{Eq:TripleID-4-3} and \eqref{Eq:TripleID-4-4} which are already expanded by the LHS of BCJ relations.

%%%%%%%%%%%%%%%%%%%%%%%%%%%%%%%%%%%%%%%%%%%%%%
\subsection{Comments on direct evaluations}
%%%%%%%%%%%%%%%%%%%%%%%%%%%%%%%%%%%%%%%%%%%
Let us close this section by sketching some helpful observations on the direct evaluations:
\begin{itemize}
\item {\bf\emph{Structure of skeletons $\mathcal{F}'$}}~~Each skeleton consists of at least two mutually disjoint components, one contains the highest-weight graviton $h_{\rho(s)}$ (\emph{type-II component}), the other contains all elements in $\{1,\dots,r\}$ (\emph{type-III component}). Other components (\emph{type-I components}), each of which  always has a type-1 line, may also appear in a skeleton.
\item {\bf\emph{Physical graphs  for a given skeleton}}~~Physical graphs for a given skeleton are constructed by connecting disjoint components via type-3 lines into a fully connected graph where all chains are allowed by the (refined) graphic rule.
\item {\bf\emph{Spurious graphs  for a given skeleton}}~~Spurious graphs, which contain structures not allowed by the (refined) graphic rule, are introduced for skeletons consisting of at least three components. All spurious graphs for a given skeleton have to cancel with each other.
\item {\bf \emph{The sum of physical graphs and spurious graphs}} Since the total contribution of all spurious graphs for a given skeleton must vanish, the sum of contributions of the physical graphs equals to the sum of contributions of both physical and spurious graphs. With the help of spurious graphs, we find that all physical and spurious terms together can be expanded as  a combination of the LHS of BCJ relations.
\end{itemize}
In the coming section, we will see all these features arise and play a critical role in the study of general gauge invariance induced identity \eqref{Eq:GaugeInv2}.

%%%%%%%%%%%%%%%%%%%%%%%%%%%%%%%%%%%%%%%%%%%%%%%%%%%%%%%%%%%%%%
\section{Constructing all physical and spurious graphs for a given skeleton}\label{section:GraphsForASkeleton}
%%%%%%%%%%%%%%%%%%%%%%%%%%%%%%%%%%%%%%%%%%%%%%%%%%%%%%%%%%%%%%
Through direct evaluations of simple examples in \secref{sec:DirectEvalutations}, we have shown that the expression in \eqref{Eq:GaugeInv5} for a given skeleton $\mathcal{F}\,'$ can be expanded in terms of the LHS of BCJ relations. To investigate the general identity \eqref{Eq:GaugeInv2}, we need to find out all possible graphs containing any given skeleton $\mathcal{F}\,'$. In this section, we prove that every skeleton $\mathcal{F}\,'$ consists of at least two disjoint components, as we have already seen via examples. All \emph{physical graphs} (graphs allowed by the refined graphic rule) with a given skeleton $\mathcal{F}\,'$ can be reconstructed by connecting type-3 lines between the components of  $\mathcal{F}\,'$  according to two distinct versions of rules which are in fact equivalent with each other. \emph{Spurious graphs} (graphs   not allowed by the refined graphic rule) associated with a given skeleton $\mathcal{F}\,'$ are also introduced. We show that the spurious graphs for any skeleton have to cancel out. When both physical and spurious graphs are considered, the expression in the square brackets of \eqref{Eq:GaugeInv5} can be reexpressed in a more convenient form which is a combination of the LHS of so-called \emph{graph-based BCJ relations}. In the next section, we prove that the LHS of graph-based-BCJ relations can always be written as a combination of the LHS of traditional BCJ relations \eqref{Eq:BCJRelation}. Hence we conclude that the gauge invariance induced identity \eqref{Eq:GaugeInv2} (or equivalently \eqref{Eq:GaugeInv3}) in general can be expanded in terms of BCJ relations.

%%%%%%%%%%%%%%%%%%%%%%%%%%%%%%%%%%%%%%
\subsection{General structure of skeletons}\label{section:GeneralStructure}
%%%%%%%%%%%%%%%%%%%%%%%%%%%%%%%%%%%%%%%

To study the general pattern of a skeleton, we recall that there are two kinds of chains in a graph (defined by the old-version graphic rule): (i) a chain led by any node $a$ in $\mathsf{H}\setminus\{h_{\rho(s)}\}$ has the form $\epsilon_{a}\cdot F_{i_j}\cdot...\cdot F_{i_1}\cdot k_b$ (here $b$ can be either a node on a higher-weight chain or an element of $\{1,\dots,r-1\}$),
(ii) the chain led by the highest weight element $h_{\rho(s)}$ has the form $k_{h_{\rho(s)}}\cdot F_{i_j}\cdot...\cdot F_{i_1}\cdot k_b$ ($b\in\{1,\dots,r-1\}$) because we have replaced $\epsilon_{h_{\rho(s)}}$ by  $k_{h_{\rho(s)}}$ in the expansion of EYM amplitude. We shall study the structures of these two types of chains in turn.

\begin{figure}
\centering
\includegraphics[width=1\textwidth]{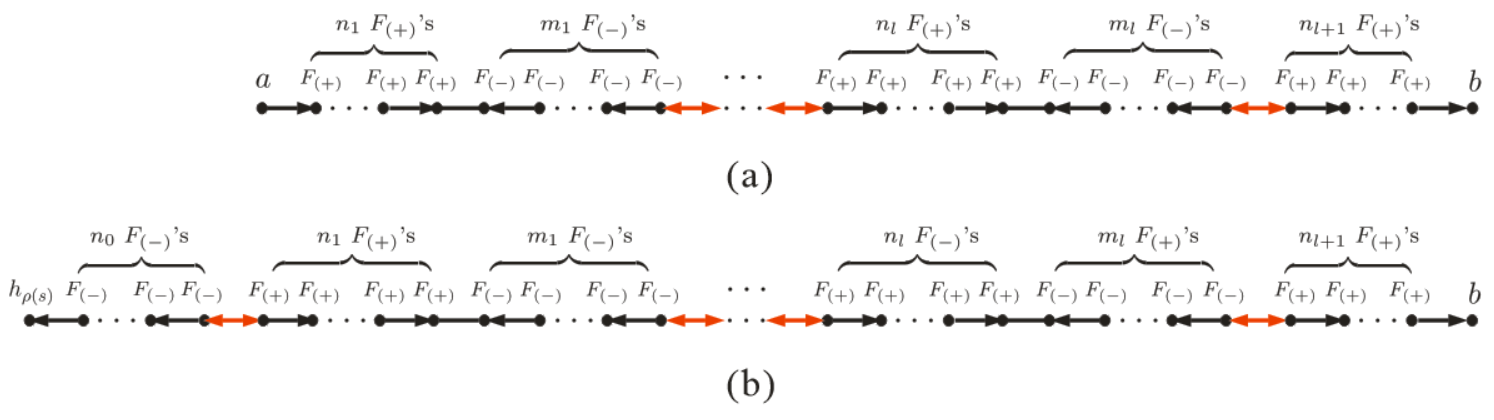}
\caption{Graphs (a) and (b) correspond to chains led by $a\in \mathsf{H}\setminus\{h_{\rho(s)}\}$ and $h_{\rho(s)}$ defined by the refined graphic rule.  A contraction of the form $(...\cdot F_{(+)})\cdot (F_{(-)}\cdot...)$ (in (a) and (b)) or $\epsilon_a\cdot (F_{(-)}\cdot...)$ (in (a)) provides a type-1 line, while a contraction of the form $(...\cdot F_{(-)})\cdot (F_{(+)}\cdot...)$ (in (a) and (b)), $k_{h_{\rho(s)}}\cdot F_{(+)}$ (in (b)), $F_{(-)}\cdot k_b$ (in (a) and (b)) or $k_{h_{\rho(s)}}\cdot k_b$ (in (b)) provides a type-3 line. }\label{Fig:SectorsOfChains}
\end{figure}

(i) When all strength tensors $F^{\mu\nu}$ are expressed by its definition \eqref{Eq:StrengthTensor}, a chain  $\epsilon_{a}\cdot F_{i_j}\cdot ... \cdot F_{i_2}\cdot F_{i_1}\cdot k_b$ ( $a\in \mathsf{H}\setminus\{h_{\rho(s)}\}$, $b$ belongs to a higher-weight chain or an element in $\{1,\dots,r-1\}$) is expanded as a sum of chains defined by the refined graphic rule. Each in the sum has the general form
\bea
\epsilon_a\mathop{\underbrace{\cdot(\mathop{\overbrace{F_{(+)}\cdot...\cdot F_{(+)}}}^{\text{$n_1$ $F_{(+)}$'s}}\cdot \mathop{\overbrace{F_{(-)}\cdot...\cdot F_{(-)}}}^{\text{$m_1$ $F_{(-)}$'s}})\cdot...\cdot(\mathop{\overbrace{F_{(+)}\cdot...\cdot F_{(+)}}}^{\text{$n_l$ $F_{(+)}$'s}}\cdot \mathop{\overbrace{F_{(-)}\cdot...\cdot F_{(-)}}}^{{\text{$m_l$ $F_{(-)}$'s}}})}}_{\text{$l$ sectors}}\cdot(\mathop{\overbrace{F_{(+)}\cdot...\cdot F_{(+)}}}^{\text{$n_{l+1}$ $F_{(+)}$'s}})\cdot k_b, \Label{Eq:Chain-(i)}
\eea
where $F^{\mu\nu}_{(+)}\equiv k^{\mu}\epsilon^{\nu}$,  $F^{\mu\nu}_{(-)}\equiv-\epsilon^{\mu}k^{\nu}$ correspond to the nodes \figref{InternalNode} (a) and (b), $l\geq 0$. In the language of graphs, such a chain is characterized by \figref{Fig:SectorsOfChains} (a). If $l=0$,
$n_{l+1=1}\geq 0$, the chain \eqref{Eq:Chain-(i)} in this case has no type-3 line. If $l>0$, the value of integers $n_i$, $m_i$ satisfy $n_1,n_{l+1}\geq 0$, $n_{i=2,\dots,l}>0$, $m_{i=1,\dots,l}>0$ and the chain in this case at least has one type-3 line.

\begin{figure}
\centering
\includegraphics[width=0.8\textwidth]{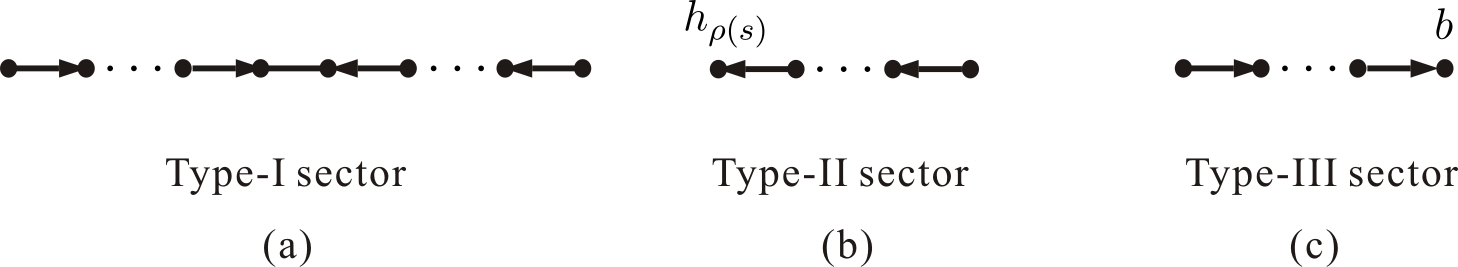}
\caption{Three types of sectors of chains when all type-3 lines are removed}\label{Fig:TypeOfSectors}
\end{figure}
(ii) Similarly, the chain $k_{h_{\rho(s)}}\cdot F_{i_j}\cdot...\cdot F_{i_1}\cdot k_b$ ($b\in\{1,\dots,r-1\}$) can also be expanded as a sum of chains defined by the refined graphic rule. Each chain in the sum has the general form
\bea
{\small k_{h_{\rho(s)}}\cdot(\mathop{\overbrace{F_{(-)}\cdot...\cdot F_{(-)}}}^{\text{$n_0$ $F_{(-)}$'s}})\mathop{\underbrace{\cdot(\mathop{\overbrace{F_{(+)}\cdot...\cdot F_{(+)}}}^{\text{$n_1$ $F_{(+)}$'s}}\cdot \mathop{\overbrace{F_{(-)}\cdot...\cdot F_{(-)}}}^{\text{$m_1$ $F_{(-)}$'s}})\cdot...\cdot(\mathop{\overbrace{F_{(+)}\cdot...\cdot F_{(+)}}}^{\text{$n_l$ $F_{(+)}$'s}}\cdot \mathop{\overbrace{F_{(-)}\cdot...\cdot F_{(-)}}}^{{\text{$m_l$ $F_{(-)}$'s}}})}}_{\text{$l$ sectors}}\cdot(\mathop{\overbrace{F_{(+)}\cdot...\cdot F_{(+)}}}^{\text{$n_{l+1}$ $F_{(+)}$'s}})\cdot  k_b,}\Label{Eq:Chain-(ii)}\nn
\eea
where $l\geq 0$. The value of integers $n_i$, $m_i$ satisfy $n_{0},n_{l+1}\geq 0$ (for $l\geq 0$), $n_{i=1,\dots,l}>0$, $m_{i=1,\dots,l}>0$ (for $l>0$). In the language of graphs, this chain is characterized by \figref{Fig:SectorsOfChains} (b). Since the arrow lines connected to the starting node $h_{\rho(s)}$  and the ending node $b$ in \figref{Fig:SectorsOfChains} point to opposite directions, there must be at least one type-3 line in \figref{Fig:SectorsOfChains} (b). In order to obtain the skeleton $\mathcal{F}'$ of a graph $\mathcal{F}$, we delete all type-3 lines from $\mathcal{F}$. Then chains are divided into disjoint sectors in general. All sectors can be classified into the following types.
\begin{itemize}
\item \emph{Type-I sector: A sector containing a type-1 line} (see \figref{Fig:TypeOfSectors} (a))~~This sector can also have type-2 lines whose arrows point to the direction of the two end nodes of the type-1 line. Both types of chains \figref{Fig:SectorsOfChains} (a) and (b) can contain type-I sectors.
\item \emph{Type-II sector: A sector only containing type-2 lines whose arrows point to the direction of the starting node of the chain} (see \figref{Fig:TypeOfSectors} (b))~~Only the highest-weight chain \figref{Fig:SectorsOfChains} (b) involve a type-II sector. The single node $h_{\rho(s)}$ with no line (on this chain) connected to it  (i.e. \figref{Fig:SectorsOfChains} (b) with $n_0=0$) is considered as a special type-II sector.
\item \emph{Type-III sector: A sector only containing type-2 lines whose arrows point to the direction of the ending node of a chain} (see \figref{Fig:TypeOfSectors} (c))~~Both types of chains \figref{Fig:SectorsOfChains} (a) and (b) contain type-III sectors. The single node $b$ with no line (on this chain) connected to it ( i.e. \figref{Fig:SectorsOfChains} (a) or (b) with $n_{l+1}=0$) is a special type-III sector.
\end{itemize}

%Then in general, the graph \figref{Fig:SectorsOfChains} (a) (for $a\in \mathsf{H}\setminus\{h_{\rho(s)}\}$) is divided into disjoint sectors: the sectors which do not contain $b$ correspond to factors of the form $\left[(\epsilon\cdot k)\dots(\epsilon\cdot k)(\epsilon\cdot\epsilon)(k\cdot\epsilon)\dots (k\cdot\epsilon)\right]$, while the sector containing $b$ provides a factor of the form $\left[(\epsilon\cdot k)\dots(\epsilon\cdot k_b)\right]$.
%
\begin{figure}
\centering
\includegraphics[width=0.9\textwidth]{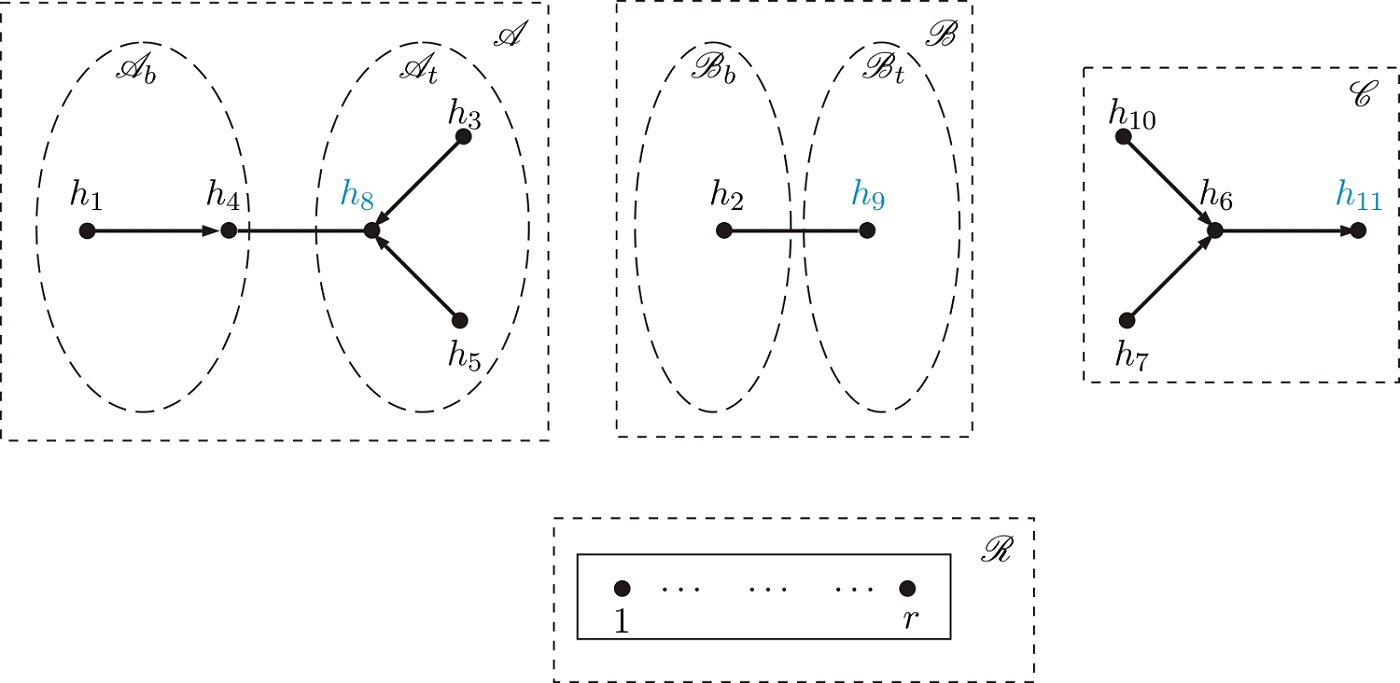}
\caption{A skeleton consisting of four mutually disjoint components for the gauge invariance induced identity \eqref{Eq:GaugeInv2} with $|\mathsf{H}|=11$. The reference order is chosen as $\mathsf{R}=\{h_1,h_2,\dots,h_{10},h_{11}\}$. Here, both $\mathscr{A}$ and $\mathscr{B}$ are type-I components. The kernel of $\mathscr{A}$ is given by the nodes $h_4$ and $h_8$ with a type-1 line between them (i.e. the factor $\epsilon_{h_4}\cdot\epsilon_{h_8}$), while the kernel of $\mathscr{B}$  is given by $h_2$ and $h_9$ with a type-1 line between them (i.e. $\epsilon_{h_2}\cdot\epsilon_{h_9}$). On each side of the component $\mathscr{A}$, the arrows of type-2 lines point to the corresponding end node of the kernel. The component  $\mathscr{C}$ is the type-II component. The set $\mathscr{R}=\{1,\dots,r\}$ itself is the type-III component (Although only the elements $1$, ..., $r-1$ can play as roots in the graphic rule, we also include the last element $r$ in the component $\mathscr{R}$ for convenience; A more general type-III componnet can involve type-II lines whose arrows pointing to roots).}\label{Fig:Figure9}
\end{figure}

According to the refined graphic rule, the highest-weight chain in a graph must be of the form \figref{Fig:SectorsOfChains} (b). This chain at least contains two mutually disjoint sectors: a type-II sector and a type-III sector. It can also have type-I sectors. The type-III sector of the highest-weight chain \figref{Fig:SectorsOfChains} (b) must end at a root $b\in\{1,\dots,r-1\}$ while all nodes on this chain can be ending nodes of type-III sectors of other chains. If we look at a chain of the form \figref{Fig:SectorsOfChains} (a), we find that the type-III sector of this chain can end at either a node of a higher-weight chain or a root $b\in\{1,\dots,r-1\}$, while any node on this chain can be the ending node of the type-III sector of a lower-weight chain. After putting all these sectors together, we conclude that a general skeleton is composed of the following types of components:
\begin{itemize}

\item  {\emph{Type-I component: component containing a type-1 line }(see the $\mathscr{A}$ and $\mathscr{B}$ components in \figref{Fig:Figure9} for example)~} Such a component consists of a type-I sector with possible type-III sectors attached to it. Each component of this type must have only one type-1 line in it and may also have type-2 lines pointing to the direction of the end nodes of the type-1 line. A type-I component can be reconstructed by connecting a type-1 line between two separate parts, each only contains type-2 lines. We define the part to which the highest-weight node of a type-I component belongs as the \emph{top side}. The other part is called the \emph{bottom side}. The type-1 line with its two end nodes together is defined as the \emph{kernel} of this type-I component.

\item {\emph{Type-II component: the component containing the highest-weight graviton $h_{\rho(s)}$} (see the $\mathscr{C}$  components in \figref{Fig:Figure9} for example)} This component consists of the type-II sector of the highest-weight chain with possible type-III sectors (belonging to other chains) attached on it. Apparently, type-II component involves only type-2 lines whose arrows point to the direction of the node $h_{\rho(s)}$.

\item{\emph{Type-III component: the component containing the set $\{1,\dots,r\}$} (see the $\mathscr{R}$ components in \figref{Fig:Figure9} for example)~} This component is obtained by connecting possible type-III sectors to roots $b\in\{1,\dots,r-1\}$. Thus type-III components contains type-2 lines whose arrows point to the root set.
\end{itemize}
All the above three types of components can be considered as connected subgraphs where tree structures with only type-2 lines pointing to (i) the kernel (for type-I components), (ii) the highest-weight node $h_{\rho(s)}$ (for the type-II component) and (iii) roots (for the type-III component).
If the highest-weight chain contains only one type-3 line in the original graph $\mathcal{F}$ while other chains do not have any type-3 line, the skeleton of $\mathcal{F}$ must consist of only two mutually disjoint components: the type-II and the type-III components (see \figref{Fig:Figure4} (a)-(h) for example).

\begin{figure}
\centering
\includegraphics[width=1\textwidth]{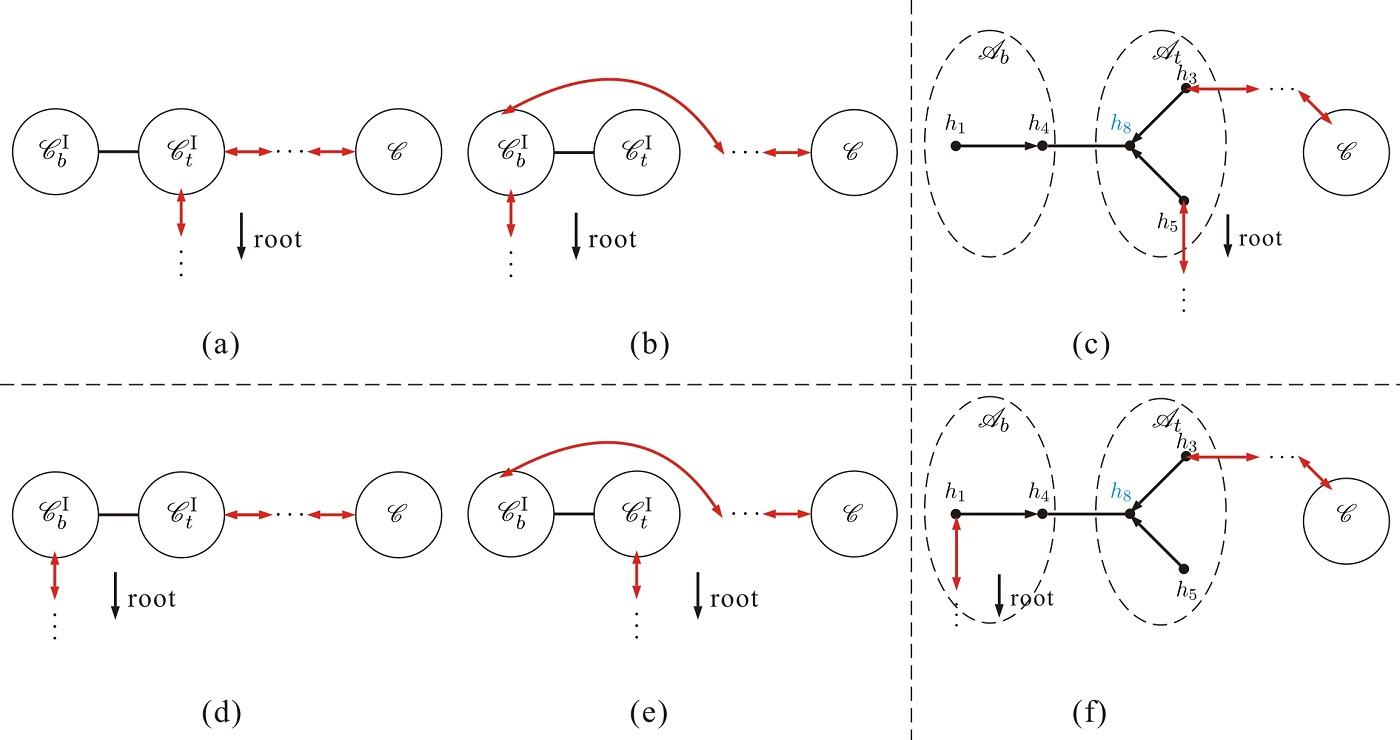}
\caption{We use a circle to stand for an arbitrary component or the top or bottom side of a type-I component; A solid line between the top and bottom sides of a type-I component $\mathscr{C}^{\text{I}}$ stands for a type-1 line connecting two arbitrary nodes on the corresponding sides; A double arrow line between two distinct components stands for a type-3 line between two arbitrary nodes in corresponding components. Assume that $\mathscr{C}$ is the type-II component or a type-I component with a higher weight than $\mathscr{C}^{\text{I}}$. The graphs (a) and (b), in which a higher-weight chain passes through only one side of  $\mathscr{C}^{\text{I}}$, are not allowed by the refined graphic rule because a chain cannot contain conflict arrow line towards root (e.g. the graph (c)). Thus if a chain passes through a component, it must pass through both sides of the component. In other words, the kernel of the component $\mathscr{C}^{\text{I}}$ must be on that chain (see the graphs (d) and (e) as well as the explicit example (f)).}\label{Fig:GraphsNotAllowed}
\end{figure}
\begin{figure}
\centering
\includegraphics[width=1\textwidth]{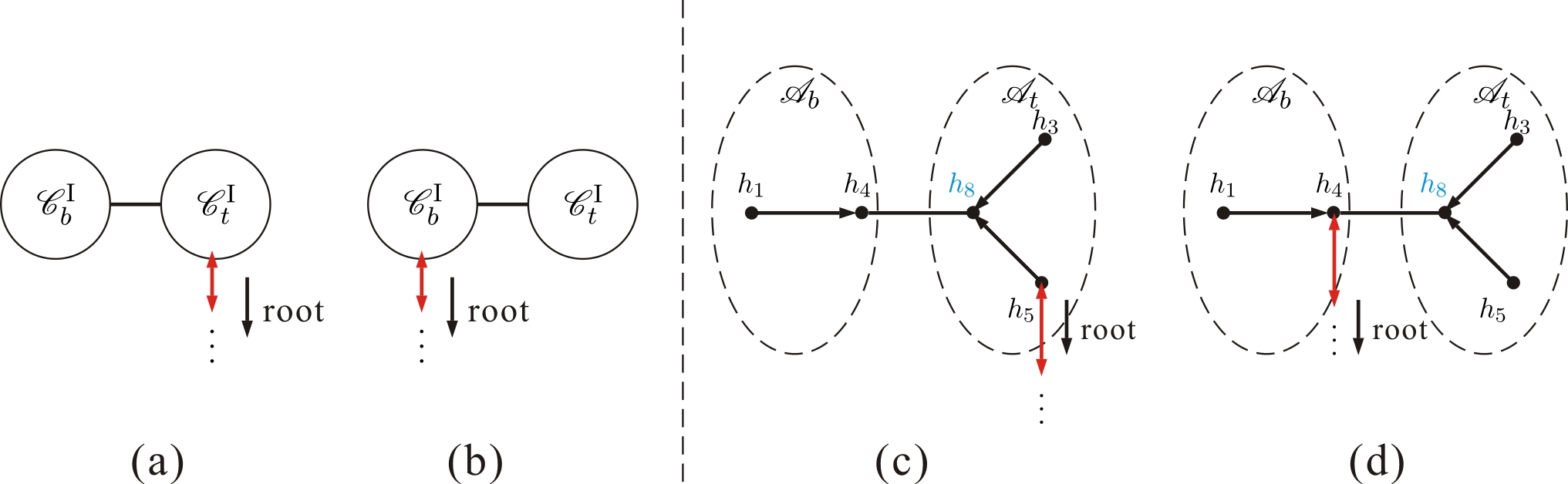}
\caption{If there is a chain starting at the highest-weight node of a type-I component $\mathscr{C}^{\text{I}}$, this chain must contain the kernel of $\mathscr{C}^{\text{I}}$. In other words, the bottom-side end node of the kernel must be nearer to root than the top-side one. Thus the graph (b) (a more concrete example is given by (d)) is  allowed but (a)  (a more concrete example is given by (c)) is not allowed.}\label{Fig:GraphsNotAllowed2}
\end{figure}

With the general structure of skeletons in hand, we will construct all graphs for an arbitrary skeleton in the rest of this section.

%%%%%%%%%%%%%%%%%%%%%%%%%%%%%%%%%%%%%%%%%%%%%%%%%%%%%%%%%%%%%%%%%%%%%%%%%%%%%%
\subsection{The construction of physical graphs for a given skeleton}\label{section:PhysicalGraphs}
%%%%%%%%%%%%%%%%%%%%%%%%%%%%%%%%%%%%%%%%%%%%%%%%%%%%%%%%%%%%%%%%%%%%%%%%%%%%
Physical graphs corresponding to a given skeleton $\mathcal{F}'$ can be obtained by connecting the disjoint components of  $\mathcal{F}'$ into a fully connected graph via type-3 lines. If the skeleton consists of $N$ disjoint components, the number of  type-3 lines must be $N-1$. For the convenience of coming discussions, we define \emph{reference order} $\mathsf{R}_{\mathscr{C}}$ of all type-I and type-II components in a skeleton by the relative order of their highest-weight nodes in  $\mathsf{R}=\{h_{\rho(1)},h_{\rho(2)},\dots,h_{\rho(s)}\}$. For instance, the reference order of the components in \figref{Fig:Figure9} is $\mathsf{R}_{\mathscr{C}}=\{\mathscr{A},\mathscr{B},\mathscr{C}\}$.  The position of a component in the ordered set  $\mathsf{R}_{\mathscr{C}}$ is called \emph{the weight of the component}. To construct all possible physical graphs containing a given skeleton $\mathcal{F}'$, we should notice the following constraints from the refined graphic rule:
\begin{itemize}
\item  Assume we have two components $\mathscr{C}^{\text{I}}$ and $\mathscr{C}$. Here, $\mathscr{C}^{\text{I}}$ is a type-I component while $\mathscr{C}$ is either the type-II or a type-I component whose weight is higher than that of $\mathscr{C}^{\text{I}}$. The structures shown in \figref{Fig:GraphsNotAllowed} (a) and (b), in which  the chain led by the highest weight node in $\mathscr{C}$ passes through only a single side (the top or the bottom side) of the component $\mathscr{C}^{\text{I}}$ (via two type-3 lines), are forbidden. This is because if only the top (or bottom) side of $\mathscr{C}^{\text{I}}$ is passed through by the chain, there must exist an internal node which is attached by two conflict arrows on that chain (for example the node $h_8$ in \figref{Fig:GraphsNotAllowed}.(c)). Such a structure is forbidden by the definition of the strength tensor $F^{\mu\nu}$. Therefore, only the graphs \figref{Fig:GraphsNotAllowed} (d) and (e) where the higher-weight chain pass through both sides of $\mathscr{C}^{\text{I}}$ are allowed (a more specific example is given by \figref{Fig:GraphsNotAllowed} (f)). In other words, if a chain of higher weight passes through a type-I component $\mathscr{C}^{\text{I}}$, the kernel of the component $\mathscr{C}^{\text{I}}$ (e.g. $\epsilon_{h_4}\cdot\epsilon_{h_8}$ for \figref{Fig:GraphsNotAllowed} (f)) must be on this chain.

\item A chain,  say $\mathbb{CH}_1$,  starting at the highest-weight node of a type-I component $\mathscr{C}^{\text{I}}$ must pass through both sides of $\mathscr{C}^{\text{I}}$ (or equivalently the kernel of  $\mathscr{C}^{\text{I}}$). If not (as shown by \figref{Fig:GraphsNotAllowed2} (a) or a more  explicit example \figref{Fig:GraphsNotAllowed2} (c)), there must exist another chain, say $\mathbb{CH}_2$ which  ends at some node $a\in\mathscr{C}^{\text{I}}_t\cap\mathbb{CH}_1$.  The node $a$ plays as the ending node of $\mathbb{CH}_2$ but can only supply a $\epsilon$ to $\mathbb{CH}_2$, which apparently conflicts to the refined graphic rule.
\end{itemize}
With the help of these two constraints, we can construct all physical graphs for a given skeleton. Let us first illustrate this procedure by the skeleton  \figref{Fig:Figure9}.

%%%%%%%%%%%%%%%%%%%%%%%%%%%%%%%%%%%%%%%%%%%%
\subsubsection{An example: physical graphs for the skeleton \figref{Fig:Figure9}}
%%%%%%%%%%%%%%%%%%%%%%%%%%%%%%%%%%%%%%%%%%%%

 To construct all physical graphs containing the skeleton \figref{Fig:Figure9}, we need to connect the four disjoint components $\mathscr{A}$, $\mathscr{B}$, $\mathscr{C}$ and $\mathscr{R}$ by three type-3 lines according to the refined graphic rule. All physical graphs for the skeleton \figref{Fig:Figure9} can be conveniently generated by constructing \emph{chains of components} which reflect the structure of chains led by the highest-weight nodes in the corresponding components. Here we present two equivalent constructions of physical graphs for the skeleton \figref{Fig:Figure9}.

{\centering\subsubsection*{Physical graphs for the skeleton \figref{Fig:Figure9}: construction-1}}

{\bf Step-1~~}According to the graphic rule, there is always a chain which starts at the highest-weight node $h_{11}\in \mathscr{C}$ in a physical graph and ends at an element in $\mathscr{R}=\{1,\dots,r-1\}$. Corresponding to distinct configurations of the highest-weight chain (the chain led by the highest weight element $h_{11}$), we construct distinct chains of components which are led by $\mathscr{C}$ and ended at $\mathscr{R}$ as follows.
\begin{itemize}
\item [{\bf{(i)}}] Each graph where the highest-weight chain neither passes through $\mathscr{A}$ nor $\mathscr{B}$ contains a substructure where two nodes $a_1\in\mathscr{C}$ and $b_1\in\mathscr{R}$  are connected by a type-3 line. Such a substructure defines a chain of components: $\mathbb{CH}=[\mathscr{C},\mathscr{R}]$ (adjacent components on this chain are connected by a type-3 line).

\item [{\bf{(ii)}}] If $\mathscr{B}$ component is passed through by the highest-weight chain while $\mathscr{A}$ is not, there should exist $a_1\in\mathscr{C}$, $b_1\in\mathscr{B}$ connected by a type-3 line and $a_2\in\mathscr{B}$, $b_2\in\mathscr{R}$ by another one on this chain.
Since a chain must pass through both sides of a type-I component, $b_1$ and $a_2$ should belong to distinct sides of $\mathscr{B}$, i.e. either $b_1\in\mathscr{B}_t$, $a_2\in\mathscr{B}_b$  or $b_1\in\mathscr{B}_b$, $a_2\in\mathscr{B}_t$ is satisfied. Correspondingly, a chain of components $\mathbb{CH}=[\mathscr{C},\mathscr{B}_t-\mathscr{B}_b,\mathscr{R}]$ or $\mathbb{CH}=[\mathscr{C},\mathscr{B}_b-\mathscr{B}_t,\mathscr{R}]$ is defined (Here a `$-$' denotes the type-1 line between $\mathscr{B}_t$ and $\mathscr{B}_b$).

\item [{\bf{(iii)}}] Similarly with {\bf{(ii)}}, if  $\mathscr{A}$ component is passed trough by the highest weight chain while $\mathscr{B}$ is not, we have $a_1\in\mathscr{C}$, $b_1\in\mathscr{A}$ connected by a type-3 line and $a_2\in\mathscr{A}$, $b_2\in\mathscr{R}$ by another. The nodes $b_1$ and $a_2$ can only belong to distinct sides of $\mathscr{A}$, thus we arrive two possible chains of components $\mathbb{CH}=[\mathscr{C},\mathscr{A}_t-\mathscr{A}_b,\mathscr{R}]$ and $\mathbb{CH}=[\mathscr{C},\mathscr{A}_b-\mathscr{A}_t,\mathscr{R}]$.

\item [{\bf{(iv)}}] If the highest-weight chain, which starts from $\mathscr{C}$ and ends at $\mathscr{R}$, passes through $\mathscr{A}$ and $\mathscr{B}$ in turn,
each pair of nodes $(a_i,b_i)$ ($i=1,2,3, a_1\in \mathscr{C},b_1,a_2\in\mathscr{A},b_2,a_3\in\mathscr{B}, b_3\in\mathscr{R}$) is connected by a type-3 line. In addition, $b_1$, $a_2$ ($b_2$, $a_3$) must belong to distinct sides of the component $\mathscr{A}$ ($\mathscr{B}$). Hence we obtain the following four chains of components
\bea
 &&\mathbb{CH}=[\mathscr{C},\mathscr{B}_{t}-\mathscr{B}_{b},\mathscr{A}_{t}-\mathscr{A}_{b},\mathscr{R}]:~\text{\figref{Fig:Figure14}} (a), \mathbb{CH}=[\mathscr{C},\mathscr{B}_{t}-\mathscr{B}_{b},\mathscr{A}_{b}-\mathscr{A}_{t},\mathscr{R}]:~\text{\figref{Fig:Figure14}} (b) \nn
 &&\mathbb{CH}=[\mathscr{C},\mathscr{B}_{b}-\mathscr{B}_{t},\mathscr{A}_{t}-\mathscr{A}_{b},\mathscr{R}]:~\text{\figref{Fig:Figure14}} (c), \mathbb{CH}=[\mathscr{C},\mathscr{B}_{b}-\mathscr{B}_{t},\mathscr{A}_{b}-\mathscr{A}_{t},\mathscr{R}]:~\text{\figref{Fig:Figure14}} (d).
\eea

\item [{\bf{(v)}}]  If the highest-weight chain passes trough $\mathscr{B}$ and $\mathscr{A}$ in turn, we just exchange the roles of components $\mathscr{A}$ and  $\mathscr{B}$ in the previous case. Then the following chains of components are obtained
\bea
 &&\mathbb{CH}=[\mathscr{C},\mathscr{A}_{t}-\mathscr{A}_{b},\mathscr{B}_{t}-\mathscr{B}_{b},\mathscr{R}]:~\text{\figref{Fig:Figure14}} (e), \mathbb{CH}=[\mathscr{C},\mathscr{A}_{t}-\mathscr{A}_{b},\mathscr{B}_{b}-\mathscr{B}_{t},\mathscr{R}]:~\text{\figref{Fig:Figure14}} (f) \nn
 &&\mathbb{CH}=[\mathscr{C},\mathscr{A}_{b}-\mathscr{A}_{t},\mathscr{B}_{t}-\mathscr{B}_{b},\mathscr{R}]:~\text{\figref{Fig:Figure14}} (g), \mathbb{CH}=[\mathscr{C},\mathscr{A}_{b}-\mathscr{A}_{t},\mathscr{B}_{b}-\mathscr{B}_{t},\mathscr{R}]:~\text{\figref{Fig:Figure14}} (h).
\eea
       \end{itemize}
Having constructed a chain of components $\mathbb{CH}$  starting from $\mathscr{C}$ and ending at $\mathscr{R}$, we redefine the reference order $\mathsf{R}_{\mathscr{C}}$  by deleting the components which have been used, i.e, $\mathsf{R}_{\mathscr{C}}\to \mathsf{R}_{\mathscr{C}}'=\mathsf{R}_{\mathscr{C}}\setminus\mathbb{CH}$. For the cases (iv) and (v), the ordered set $\mathsf{R}_{\mathscr{C}}$  have been clear up, thus the procedure is terminated. For the cases (i), (ii) and (iii), we should further construct other chains of components in the next step.

{\bf{Step-2~~}}Pick out the highest-weight component from the redefined $\mathsf{R}_{\mathscr{C}}$ (if it is not empty) and construct a new chain of components towards a component on $\mathbb{CH}$ which have been constructed previously.
\begin{itemize}
\item  [{\bf{(i)}}] Based on the chain $\mathbb{CH}=[\mathscr{C},\mathscr{R}]$,  the reference order $\mathsf{R}_{\mathscr{C}}$ is redefined as  $\mathsf{R}_{\mathscr{C}}'= \{\mathscr{A}, \mathscr{B}\}$ where the highest-weight component is $\mathscr{B}$. The chain led by the highest-weight node in $\mathscr{B}$ can directly ends at a node in $\mathscr{C}\oplus\mathscr{R}$ ($\oplus$ denotes disjoint union) or passes through the component  $\mathscr{A}$. For the former case, there exist two nodes on the chain $a'_1\in\mathscr{B}$ and $b'_1\in \mathscr{C}\oplus\mathscr{R}$ which are connected by a type-3 line. Considering that the bottom-side end node of the kernel of $\mathscr{B}$ is nearer to roots than the top-side one, possible new chains of components are constructed as
\bea
\mathbb{CH}'=[\mathscr{B}_{t}-\mathscr{B}_{b},\mathscr{C}\text{~or~}\mathscr{R}].
\eea

For the latter case, one can find two pairs of nodes $a'_1$, $b'_1$ ($a'_1\in\mathscr{B}$, $b'_1\in\mathscr{A}$) and $a'_2$, $b'_2$ ($a'_2\in\mathscr{A}$, $b'_2\in\mathscr{C}~\text{or}~\mathscr{R}$), each pair is connected by a type-3 line. Since $b'_1$ and $a'_2$ must live in distinct sides of $\mathscr{A}$, possible chains of components are constructed as
\bea
\mathbb{CH}'=[\mathscr{B}_{t}-\mathscr{B}_{b},\mathscr{A}_t-\mathscr{A}_b,\mathscr{C}],~~~\mathbb{CH}'=[\mathscr{B}_{t}-\mathscr{B}_{b},\mathscr{A}_t-\mathscr{A}_b,\mathscr{R}],\nn
\mathbb{CH}'=[\mathscr{B}_{t}-\mathscr{B}_{b},\mathscr{A}_b-\mathscr{A}_t,\mathscr{C}],~~~\mathbb{CH}'=[\mathscr{B}_{t}-\mathscr{B}_{b},\mathscr{A}_b-\mathscr{A}_t,\mathscr{R}].
\eea
Each of the above four possible chains of components together with the chain $\mathbb{CH}=[\mathscr{C},\mathscr{R}]$ constructed by step-1 (i) forms a full physical graph (see \figref{Fig:Figure12} (e), (g), (f) or (h)).

\item [{\bf{(ii)}}]The  reference order of components defined in step-1 (ii) is
     $\mathsf{R}_{\mathscr{C}}'= \{\mathscr{A}\}$. The chain led by the highest-weight node in $\{\mathscr{A}\}$ must ends at a node in $\mathscr{C}\oplus\mathscr{B}\oplus\mathscr{R}$. Thus this chain contains a node $a'_1\in\mathscr{A}$ and a node $b'_1\in\mathscr{C}\oplus\mathscr{B}\oplus\mathscr{R}$ which are connected with each other by a type-3 line. Since the chain led by the highest-weight node $h_8$ in $\mathscr{A}$ have to pass through both top and bottom sides of $\mathscr{A}$,  $a'_1$ must be in  $\mathscr{A}_t$. Possible chains of components are defined as
    \bea
  &&\mathbb{CH}'=[\mathscr{A}_t-\mathscr{A}_b,\mathscr{C}\text{~or~}\mathscr{B}\text{~or~}\mathscr{R}].
    \eea
This chain together with the chains  $\mathbb{CH}=[\mathscr{C},\mathscr{B}_t-\mathscr{B}_b,\mathscr{R}]$ and $\mathbb{CH}=[\mathscr{C},\mathscr{B}_b-\mathscr{B}_t,\mathscr{R}]$  constructed in step-1 (ii) provides full physical graphs \figref{Fig:Figure13} (a), (b) and \figref{Fig:Figure13} (c), (d), respectively.

  \item [{\bf(iii)}] The reference order of components defined in step-1 (iii) is $\mathsf{R}_{\mathscr{C}}'= \{\mathscr{B}\}$. Similarly with (ii) in this step, one can construct a chain of component led by $\mathscr{B}$ towards the components that have been used in  step-1 (iii): %
    \bea
  &&\mathbb{CH}'=[\mathscr{B}_t-\mathscr{B}_b,\mathscr{C}\text{~or~}\mathscr{A}\text{~or~}\mathscr{R}],
    \eea
which reflects that a chain led by the highest-weight node $h_9$  ends at a node in  $\mathscr{C}\oplus\mathscr{A}\oplus\mathscr{R}$.
Together with the chains $\mathbb{CH}=[\mathscr{C},\mathscr{A}_t-\mathscr{A}_b,\mathscr{R}]$ and $\mathbb{CH}=[\mathscr{C},\mathscr{A}_b-\mathscr{A}_t,\mathscr{R}]$ produced in step-1 (iii), this chain of components provides the full graphs \figref{Fig:Figure13} (e), (f), (g), (h) and \figref{Fig:Figure13} (i), (j), (k), (l), respectively.
\end{itemize}
Having constructed all chains for the second step, we redefine the reference order $\mathsf{R}_{\mathscr{C}}$ of the components again by removing the components which have been used in step-2. This procedure is terminated for the cases with empty redefined $\mathsf{R}_{\mathscr{C}}$ and all graphs completed in this step are given by \figref{Fig:Figure12}.(e)-(h) and \figref{Fig:Figure13}.(a)-(l). Then only the $\mathsf{R}_{\mathscr{C}}=\{\mathscr{A}\}$  which is redefined after the construction of the chain $\mathbb{CH}'=[\mathscr{B}_{t}-\mathscr{B}_{b},\mathscr{C}\text{~or~}\mathscr{R}]$  in step-2 (i) is nonempty and requires a further step.

{\bf Step-3~~}The highest-weight component in the redefined reference order $\mathsf{R}_{\mathscr{C}}=\{\mathscr{A}\}$, which is based on the chain $\mathbb{CH}'=[\mathscr{B}_{t}-\mathscr{B}_{b},\mathscr{C}\text{~or~}\mathscr{R}]$ constructed in step-2 (i), is $\mathscr{A}$. There must be a node $a''_1\in \mathscr{A}$ and  a node $b''_1\in \mathscr{B}\oplus\mathscr{C}\oplus\mathscr{R}$ which are connected with each other by a type-3 line on this chain.
This structure corresponds to a chain of components
\bea
\mathbb{CH}''=[\mathscr{A},\mathscr{B}\text{~or~}\mathscr{C}\text{~or~}\mathscr{R}].
\eea
The chains of components  $\mathbb{CH}=[\mathscr{C},\mathscr{R}]$ and $\mathbb{CH}'=[\mathscr{B}_{t}-\mathscr{B}_{b},\mathscr{C}\text{~or~}\mathscr{R}]$ respectively constructed by step-1 (i) and step-2 (i) together with the above chain produce full graphs \figref{Fig:Figure12} (a)-(d).

All graphs constructed by the above steps with all possible choices of end nodes of type-3 lines together form the full set of physical graphs (graphs constructed by the refined graphic rule) containing the skeleton \figref{Fig:Figure9}.

%%%%%%%%%%%%%%%%%%%%%%%%%%%%%%%%%%%%%%%%%%%%%%%%%%%%%%%%%
{\centering\subsubsection*{Physical graphs for the skeleton \figref{Fig:Figure9}: construction-2}}
%%%%%%%%%%%%%%%%%%%%%%%%%%%%%%%%%%%%%%%%%%%%%%%%%%%%%%%%%%%
Physical graphs for a given skeleton can be constructed in a distinct way. We define the reference order of all type-I components by removing the type-II component from $\mathsf{R}_{\mathscr{C}}$: $\mathsf{R}^{\,\text{I}}_{\mathscr{C}}\equiv\mathsf{R}_{\mathscr{C}}\setminus\mathscr{C}=\{\mathscr{A},\mathscr{B}\}$. We also define the upper block $\mathscr{U}$ and lower block $\mathscr{L}$ of a graph by the maximally connected graphs (constructed in an intermediate step) that contain the type-II component $\mathscr{C}$ and the type-III component $\mathscr{R}$, respectively. All physical graphs can be obtained by the following steps.

{\bf Step-1~~} At the beginning $\mathscr{U}=\mathscr{C}$ and $\mathscr{L}=\mathscr{R}$. We consider the nodes in $\mathscr{U}$ and in $\mathscr{L}$ (except the node $r$) as two distinct sets of roots. Pick out the highest-weight component $\mathscr{B}$ from $\mathsf{R}^{\,\text{I}}_{\mathscr{C}}$ and construct a chain of component which starts from $\mathscr{B}$ towards either $\mathscr{U}$ or $\mathscr{L}$. Note that the bottom-side end node of the kernel of $\mathscr{B}$  must be nearer to root than the top-side one. According to whether $\mathscr{A}$ component is on this chain and whether the chain ends at the upper block $\mathscr{U}$ or lower block $\mathscr{L}$, we construct the following possible configurations:
    \bea
&&\text{(i)}.~~~\mathscr{U}\to\mathscr{U}'=[\mathscr{B}_{t}-\mathscr{B}_{b},\mathscr{C}],~\mathscr{L}\to\mathscr{L}'=\mathscr{R}; ~~~~\text{(ii)}.~\mathscr{U}\to\mathscr{U}'=\mathscr{C},~\mathscr{L}\to\mathscr{L}'=[\mathscr{B}_{t}-\mathscr{B}_{b},\mathscr{R}];~~\Label{Eq:CMPTClassification}\\
&&\text{(iii)}.~\,\mathscr{U}\to\mathscr{U}'=[\mathscr{B}_{t}-\mathscr{B}_{b},\mathscr{A}_{t}-\mathscr{A}_{b},\mathscr{C}],~\mathscr{L}\to\mathscr{L}'=\mathscr{R};\nn &&\text{(iv)}.~\,\mathscr{U}\to\mathscr{U}'=[\mathscr{B}_{t}-\mathscr{B}_{b},\mathscr{A}_{b}-\mathscr{A}_{t},\mathscr{C}],~\mathscr{L}\to\mathscr{L}'=\mathscr{R};\nn
&&\text{(v)}.~~\,\mathscr{U}\to\mathscr{U}'=\mathscr{C},~\mathscr{L}\to[\mathscr{B}_{t}-\mathscr{B}_{b},\mathscr{A}_{t}-\mathscr{A}_{b},\mathscr{R}]; \text{(vi)}.~\mathscr{U}\to\mathscr{U}'=\mathscr{C},~\mathscr{L}\to[\mathscr{B}_{t}-\mathscr{B}_{b},\mathscr{A}_{b}-\mathscr{A}_{t},\mathscr{R}],\nonumber
    \eea
   where we have redefined the upper and lower blocks by the new maximally connected graphs $\mathscr{U}'$ and $\mathscr{L}'$ that contain $\mathscr{C}$ and $\mathscr{R}$ respectively.
Redefine $\mathsf{R}^{\,\text{I}}_{\mathscr{C}}$ by removing the components which have been used on the chain led by $\mathscr{B}$. In (iii), (iv), (v) and (vi) of \eqref{Eq:CMPTClassification}, the chain led by $\mathscr{B}$ passes through $\mathscr{A}$ (the last three lines of \eqref{Eq:CMPTClassification}) and the redefined $\mathsf{R}^{\,\text{I}}_{\mathscr{C}}$ becomes empty (thus this process is terminated). Contrarily, in (i) and (ii), the component $\mathscr{A}$ is not on the chain led by $\mathscr{B}$ and the redefined $\mathsf{R}^{\,\text{I}}_{\mathscr{C}}$ is a nonempty set $\{\mathscr{A}\}$. Thus we have to turn to the next step for (i) and (ii).

 {\bf Step-2~~} Since the ordered set $\mathsf{R}^{\,\text{I}}_{\mathscr{C}}$ redefined after (i) and (ii) in the previous step is an nonempty one $\{\mathscr{A}\}$, we need to construct chains,  which start from the component $\mathscr{A}$, towards either the upper block $\mathscr{U}'$ or the lower block $\mathscr{L}'$ (which is defined in (i) and (ii) (see \eqref{Eq:CMPTClassification})). Based on \eqref{Eq:CMPTClassification} (i) and (ii), possible structures are further constructed as follows
\bea
&&\text{(i)}.~~\mathscr{U}\to\mathscr{U}''=[\mathscr{A}_{t}-\mathscr{A}_{b}, \mathscr{U}'],~~\mathscr{L}\to\mathscr{L}''=\mathscr{L}';\nn
&&\text{(ii)}.~\mathscr{U}\to\mathscr{U}''=\mathscr{U}',~~~~~~~~~~~~~~~\,\mathscr{L}\to\mathscr{L}''=[\mathscr{A}_{t}-\mathscr{A}_{b},\mathscr{L}'],
\eea
where $\mathscr{U}'$ and $\mathscr{L}'$ are   the upper and lower blocks defined in either \eqref{Eq:CMPTClassification} (i) or \eqref{Eq:CMPTClassification}  (ii).
Redefine the ordered set $\mathsf{R}^{\,\text{I}}_{\mathscr{C}'}=\{\mathscr{A}\}$ obtained in the previous step by removing $\mathscr{A}$. Then $\mathsf{R}^{\,\text{I}}_{\mathscr{C}'}$ becomes empty.

{\bf Step-3~~} After the previous steps, the ordered set $\mathsf{R}^{\text{I}}_{\mathscr{C}'}$ becomes empty and all graphs consists of only two disjoint blocks, the redefined $\mathscr{U}$ and $\mathscr{L}$ which will be called  \emph{the final upper and lower blocks}, respectively. Now we connect two nodes correspondingly belonging to the final blocks $\mathscr{U}$ and $\mathscr{L}-\{r\}$\footnote{ Since $r$ cannot be a root, it cannot be attached by the type-3 line between the final upper and lower blocks $\mathscr{U}$ and $\mathscr{L}$. Thus the end of this type-3 line in $\mathscr{L}$ can only belong to $\mathscr{L}-\{r\}$.},  by a type-3 line and make sure the chain led by the component $\mathscr{C}$ cannot only pass through a single side of any type-I component (i.e. graphs with structures \figref{Fig:GraphsNotAllowed} (a) and (b) are excluded). The graph then becomes a connected one. Since the highest-weight node $h_{11}$ belongs to $\mathscr{U}$ while all roots $\{1,\dots,r-1\}$ are contained by $\mathscr{L}-\{r\}$, the highest-weight chain that starts from $h_{11}$ and ends at an element in $\{1,\dots,r-1\}$ is constructed in this step. Finally, all graphs constructed in this way are presented by  \figref{Fig:Figure15-1}.(a1), (a2), (a3), (a4), \figref{Fig:Figure15}.(a1), (b1), (c1), (d1) and \figref{Fig:Figure16}.(a1), (b1), (c1), (d1).
%\footnote{Here, a graph with chosen an arbitrary dashed arrow line and fix other structure stands for a possibility. For example, \figref{Fig:Figure15}.(a1) stands for three possible graphs.}.

{\centering\subsubsection*{The equivalence between the two constructions}}
The first version of construction naturally inherits the refined graphic rule because a chain of components is just constructed by keeping track of the chain led by the highest weight node in the starting component. We need to verify that the second construction also provides the same physical graphs with the first approach.
The explicit correspondence between the second and the first construction is following
\bea
\text{\figref{Fig:Figure15-1} (a1)-(c1)}&\Leftrightarrow&\text{\figref{Fig:Figure12} (a)},~\text{\figref{Fig:Figure13} (a),(e)},~\text{\figref{Fig:Figure14} (e)};\nn
\text{\figref{Fig:Figure15-1} (d1)}&\Leftrightarrow&\text{\figref{Fig:Figure12} (b)},~\text{\figref{Fig:Figure13} (b),(i)},~\text{\figref{Fig:Figure14} (g)};\nn
\text{\figref{Fig:Figure15} (a1)}&\Leftrightarrow&\text{\figref{Fig:Figure12} (d)},~\text{\figref{Fig:Figure13} (d),(h)},~\text{\figref{Fig:Figure14} (c)};\nn
\text{\figref{Fig:Figure15} (b1)-(d1)}&\Leftrightarrow&\text{\figref{Fig:Figure12} (c)},~\text{\figref{Fig:Figure13} (c),(l)},~\text{\figref{Fig:Figure14} (d)};\nn
\text{\figref{Fig:Figure16} (a1)}&\Leftrightarrow&\text{\figref{Fig:Figure12} (e)},~\text{\figref{Fig:Figure13} (g)},~\text{\figref{Fig:Figure14} (a)};\nn
\text{\figref{Fig:Figure16} (b1)}&\Leftrightarrow&\text{\figref{Fig:Figure12} (f)},~\text{\figref{Fig:Figure13} (k)},~\text{\figref{Fig:Figure14} (b)};\nn
\text{\figref{Fig:Figure16} (c1)}&\Leftrightarrow&\text{\figref{Fig:Figure12} (g)},~\text{\figref{Fig:Figure13} (j)},~\text{\figref{Fig:Figure14} (h)};\nn
\text{\figref{Fig:Figure16} (d1)}&\Leftrightarrow&\text{\figref{Fig:Figure12} (h)},~\text{\figref{Fig:Figure13} (f)},~\text{\figref{Fig:Figure14} (f)}.
\eea
Thus the two constructions for this example precisely match with each other.

\subsubsection{General rules for the construction of physical graphs}
Inspired by the two constructions of physical graphs for the skeleton \figref{Fig:Figure9}, we propose two distinct rules for constructing all physical graphs for an arbitrary skeleton. We will prove that these two rules in fact provide the same physical graphs.

{\centering\subsubsection*{Rule-1}}

As we have already stated in \secref{section:GeneralStructure}, each skeleton $\mathcal{F}\,'$ in \eqref{Eq:GaugeInv5} at least  contains a type-II component $\mathscr{C}^{\text{II}}$ and a type-III component $\mathscr{C}^{\text{III}}$. Moreover, $N$ type-I components $\mathscr{C}_1^{\text{I}}$, $\mathscr{C}_2^{\text{I}}$, ..., $\mathscr{C}_{N}^{\text{I}}$ may also be involved. To construct all physical graphs for a skeleton with $N+2$ components, we should connect these components together via $N+1$ type-3 lines properly. This can be achieved by the following steps.

{\bf Step-1} Assuming that the reference order of all type-I and type-II components is given by $\mathsf{R}_{\mathscr{C}}=\{\mathscr{C}_1^{\text{I}},\mathscr{C}_2^{\text{I}},\dots,\mathscr{C}_{N}^{\text{I}},\mathscr{C}^{\text{II}}\}$, we pick out the type-II component $\mathscr{C}^{\text{II}}$ (the highest-weight component) as well as arbitrary type-I components $\mathscr{C}_{a_1}^{\text{I}}$, $\mathscr{C}_{a_2}^{\text{I}}$, ..., $\mathscr{C}_{a_i}^{\text{I}}$ (not necessary to preserve the relative order in $\mathsf{R}_{\mathscr{C}}$)
 and construct a chain $\mathbb{CH}$, whose starting component is $\mathscr{C}^{\text{II}}$ and internal components are $\mathscr{C}_{a_1}^{\text{I}}$, $\mathscr{C}_{a_2}^{\text{I}}$, ..., $\mathscr{C}_{a_i}^{\text{I}}$, towards $\mathscr{C}^{\text{III}}$:
\bea
\mathbb{CH}=\left[\mathscr{C}^{\text{II}},\left(\mathscr{C}_{a_i}^{\text{I}}\right)_{t\,(\text{or~}b)} -\left(\mathscr{C}_{a_i}^{\text{I}}\right)_{b\,(\text{or~}t)}
,\dots,\left(\mathscr{C}_{a_1}^{\text{I}}\right)_{t\,(\text{or~}b)}-\left(\mathscr{C}_{a_1}^{\text{I}}\right)_{b\,(\text{or~}t)},\mathscr{C}^{\text{III}}\right].
\eea
Redefine the reference order by $\mathsf{R}_{\mathscr{C}}\to \mathsf{R}_{\mathscr{C}}'\equiv\mathsf{R}_{\mathscr{C}}\setminus\left\{\mathscr{C}^{\text{II}},\mathscr{C}_{a_1}^{\text{I}},\mathscr{C}_{a_2}^{\text{I}},\dots,\mathscr{C}_{a_i}^{\text{I}}\right\}
$.

{\bf Step-2} After the redefinition in the previous step, the ordered set $\mathsf{R}_{\mathscr{C}}=\mathsf{R}_{\mathscr{C}}'$ only consists of type-I components.
Now we pick out the highest-weight component, say $\mathscr{C}_{a_{i'}'}^{\text{I}}$  as well as arbitrary type-I components $\mathscr{C}_{a_1'}^{\text{I}}$, ..., $\mathscr{C}_{a_{i'}'}^{\text{I}}$ from $\mathsf{R}_{\mathscr{C}}'$  and construct a chain $\mathbb{CH}'$ towards a component $\mathscr{C}\in\bigl\{\mathscr{C}^{\text{II}},\mathscr{C}_{a_1}^{\text{I}},\mathscr{C}_{a_2}^{\text{I}},\dots,\mathscr{C}_{a_i}^{\text{I}},\mathscr{C}^{\text{III}}\bigr\}$, which has been used in the previous step, as follows:
\bea
\mathbb{CH}'=\left[\bigl(\mathscr{C}_{a_{i'}'}^{\text{I}}\bigr)_{t\,(\text{or~}b)}-\bigl(\mathscr{C}_{a_{i'}'}^{\text{I}}\bigr)_{b\,(\text{or~}t)}
,\bigl(\mathscr{C}_{a_{i'-1}'}^{\text{I}}\bigr)_{t\,(\text{or~}b)}-\bigl(\mathscr{C}_{a_{i'-1}'}^{\text{I}}\bigr)_{b\,(\text{or~}t)}
,\dots,\bigl(\mathscr{C}_{a_1'}^{\text{I}}\bigr)_{t\,(\text{or~}b)}-\bigl(\mathscr{C}_{a_1'}^{\text{I}}\bigr)_{b\,(\text{or~}t)}\leftrightarrow\mathscr{C}\right].\nn
\eea
Redefine the reference order by $\mathsf{R}_{\mathscr{C}}\to \mathsf{R}_{\mathscr{C}}''\equiv\mathsf{R}_{\mathscr{C}}'\setminus\{\mathscr{C}_{a_1'}^{\text{I}},\mathscr{C}_{a_2'}^{\text{I}},\dots,\mathscr{C}_{a_{i'}'}^{\text{I}}\}$.

{\bf Step-3} Repeat the above steps. In each step, construct a chain, whose starting component is the highest-weight one in  $\mathsf{R}_{\mathscr{C}}$ redefined by the previous step, towards an arbitrary component which has been used in the previous steps. Then redefine $\mathsf{R}_{\mathscr{C}}$ by removing the starting and internal components which have been used in this step. This procedure is terminated till the ordered set $\mathsf{R}_{\mathscr{C}}$  becomes empty. We obtain a physical graph containing the skeleton $\mathcal{F}\,'$.

All graphs (all possible chain structures of components and all possible choices of the end nodes of each type-3 lines) together form the set of all physical graphs $\mathcal{F}$ for the skeleton $\mathcal{F}\,'$.

{\centering\subsubsection*{Rule-2}}
Inspired by the second construction of physical graphs  for the skeleton \figref{Fig:Figure9}, we propose the following rule which is essentially equivalent with rule-1.

{\bf Step-1} Define the upper and lower blocks  $\mathscr{U}$ and $\mathscr{L}$ by the maximally connected graphs $\mathscr{C}^{\text{II}}$ and $\mathscr{C}^{\text{III}}$ that contains the highest weight node $h_{\rho(h)}$ and the elements $\{1,\dots,r-1\}$ respectively and define the reference order of all type-I components by $\mathsf{R}^{\,\text{I}}_{\mathscr{C}}\equiv\mathsf{R}_{\mathscr{C}}\setminus\mathscr{C}^{\text{II}}=\{\mathscr{C}_1^{\text{I}},\mathscr{C}_2^{\text{I}},\dots,\mathscr{C}_{N}^{\text{I}}\}$.
We construct a chain, whose starting component is the highest-weight one in $\mathsf{R}^{\,\text{I}}_{\mathscr{C}}$ (i.e., $\mathscr{C}_{N}^{\text{I}}$) and internal components are arbitrarily chosen as $\mathscr{C}_{a_1}^{\text{I}}$, $\mathscr{C}_{a_2}^{\text{I}}$, ..., $\mathscr{C}_{a_i}^{\text{I}}$ (not necessary in the relative order in $\mathsf{R}^{\,\text{I}}_{\mathscr{C}}$), towards either the upper block $\mathscr{U}$ or the lower block $\mathscr{L}$. Correspondingly, we get two possible structures
\bea
\text{(i)}.&&\mathscr{U}\to\mathscr{U}'=\bigl[\big(\mathscr{C}_{N}^{\text{I}}\bigr)_{t}-\big(\mathscr{C}_{N}^{\text{I}}\bigr)_{b}
,\big(\mathscr{C}_{a_i}^{\text{I}}\bigr)_{t\,(\text{or~}b)}-\big(\mathscr{C}_{a_i}^{\text{I}}\bigr)_{b\,(\text{or~}t)}
,\dots,\big(\mathscr{C}_{a_1}^{\text{I}}\bigr)_{t\,(\text{or~}b)}-\big(\mathscr{C}_{a_1}^{\text{I}}\bigr)_{b\,(\text{or~}t)}
,\mathscr{C}^{\text{II}}\bigr],\nn
&&\mathscr{L}\to\mathscr{L}'=\mathscr{C}^{\text{III}};\nn
\text{(ii)}.&&\mathscr{U}\to\mathscr{U}'=\mathscr{C}^{\text{II}},\nn
&&\mathscr{L}\to\mathscr{L}'=\bigl[\big(\mathscr{C}_{N}^{\text{I}}\bigr)_{t}-\big(\mathscr{C}_{N}^{\text{I}}\bigr)_{b}
,\big(\mathscr{C}_{a_i}^{\text{I}}\bigr)_{t\,(\text{or~}b)}-\big(\mathscr{C}_{a_i}^{\text{I}}\bigr)_{b\,(\text{or~}t)}
,\dots,\big(\mathscr{C}_{a_1}^{\text{I}}\bigr)_{t\,(\text{or~}b)}-\big(\mathscr{C}_{a_1}^{\text{I}}\bigr)_{b\,(\text{or~}t)}
,\mathscr{C}^{\text{III}}\bigr].\Label{Eq:Rule-2Step-1}
\eea
Redefine $\mathsf{R}_{\mathscr{C}}^{\,\text{I}}$ by ${\mathsf{R}_{\mathscr{C}}^{\,\text{I}}}'=\mathsf{R}_{\mathscr{C}}^{\,\text{I}}\setminus\{\mathscr{C}_{N}^{\text{I}},\mathscr{C}_{a_1}^{\text{I}},\dots,\mathscr{C}_{a_i}^{\text{I}}\}$,
$\mathscr{U}$ and $\mathscr{L}$ by the new obtained upper and lower blocks  $\mathscr{U}'$ and $\mathscr{L}'$  respectively.

{\bf Step-2} Pick out the highest-weight component $\mathscr{C}_{a'_{i'}}^{\text{I}}$ and arbitrary components $\mathscr{C}_{a'_1}^{\text{I}}$, ..., $\mathscr{C}_{a'_{i'-1}}^{\text{I}}$ from ${\mathsf{R}_{\mathscr{C}}^{\,\text{I}}}'$ defined in the previous step. Construct a chain, whose starting component is $\mathscr{C}_{a'_{i'}}^{\text{I}}$ and internal components are $\mathscr{C}_{a'_1}^{\text{I}}$, ..., $\mathscr{C}_{a'_{i'-1}}^{\text{I}}$, towards either the upper block $\mathscr{U}'$ or the lower one $\mathscr{L}'$:
\bea
(i).&&\mathscr{U}\to\mathscr{U}''=\bigl[\big(\mathscr{C}_{a_{i'}'}^{\text{I}}\bigr)_{t}-\big(\mathscr{C}_{a_{i'}'}^{\text{I}}\bigr)_{b}
,\big(\mathscr{C}_{a_{i'-1}'}^{\text{I}}\bigr)_{t\,(\text{or~}b)}-\big(\mathscr{C}_{a_{i'-1}'}^{\text{I}}\bigr)_{b\,(\text{or~}t)}
,\dots,\big(\mathscr{C}_{a_1'}^{\text{I}}\bigr)_{t\,(\text{or~}b)}-\big(\mathscr{C}_{a_1'}^{\text{I}}\bigr)_{b\,(\text{or~}t)}
,\mathscr{U}'\bigr],\nn
&&\mathscr{L}\to\mathscr{L}''=\mathscr{L}';\nn
(ii).&&\mathscr{U}\to\mathscr{U}''=\mathscr{U}',\Label{Eq:Rule-2Step-2}\\
&&\mathscr{L}\to\mathscr{L}''=\bigl[\big(\mathscr{C}_{a_{i'}'}^{\text{I}}\bigr)_{t}-\big(\mathscr{C}_{a_{i'}'}^{\text{I}}\bigr)_{b}
,\big(\mathscr{C}_{a_{i'-1}'}^{\text{I}}\bigr)_{t\,(\text{or~}b)}-\big(\mathscr{C}_{a_{i'-1'}}^{\text{I}}\bigr)_{b\,(\text{or~}t)}
,\dots,\big(\mathscr{C}_{a_1'}^{\text{I}}\bigr)_{t\,(\text{or~}b)}-\big(\mathscr{C}_{a_1'}^{\text{I}}\bigr)_{b\,(\text{or~}t)}
,\mathscr{L}'\bigr].\nonumber
\eea
Again, redefine $\mathsf{R}_{\mathscr{C}}$ by  $\mathsf{R}_{\mathscr{C}}''=\mathsf{R}_{\mathscr{C}}'\setminus\{\mathscr{C}_{a_1'}^{\text{I}},\dots,\mathscr{C}_{a_{i'}'}^{\text{I}}\}$.

{\bf Step-3} Repeat the above steps until the ordered set $\mathsf{R}_{\mathscr{C}}$ becomes empty. In each step, we construct a chain  which is led by the highest-weight component in the redefined $\mathsf{R}_{\mathscr{C}}$, towards either the upper block $\mathscr{U}$ or the lower block $\mathscr{L}$ obtained in the previous step. Then redefine the upper and lower blocks by the new constructed maximally connected subgraphs that containing the highest-weight node $h_{\rho(s)}$ and roots $\{1,\dots,r-1\}$, respectively.

{\bf Step-4} After the above steps, graphs with only two disjoint connected blocks (the final upper and lower blocks $\mathscr{U}$ and $\mathscr{L}$) are produced.
Pick out two nodes from the final blocks $\mathscr{U}$  and $\mathscr{L}$ respectively in each graph and connect them by a type-3 line. Then we get a connected graph containing a path from the highest-weight node $h_{\rho(s)}$ to a root. Considering the condition  that the structures \figref{Fig:GraphsNotAllowed} (a) and (b) must be avoided,
we only keep (physical) graphs where the highest-weight chain contains all kernels of the type-I components living on it (i.e. only the structures \figref{Fig:GraphsNotAllowed} (d) and (e) are allowed).

\begin{figure}
\centering
\includegraphics[width=1\textwidth]{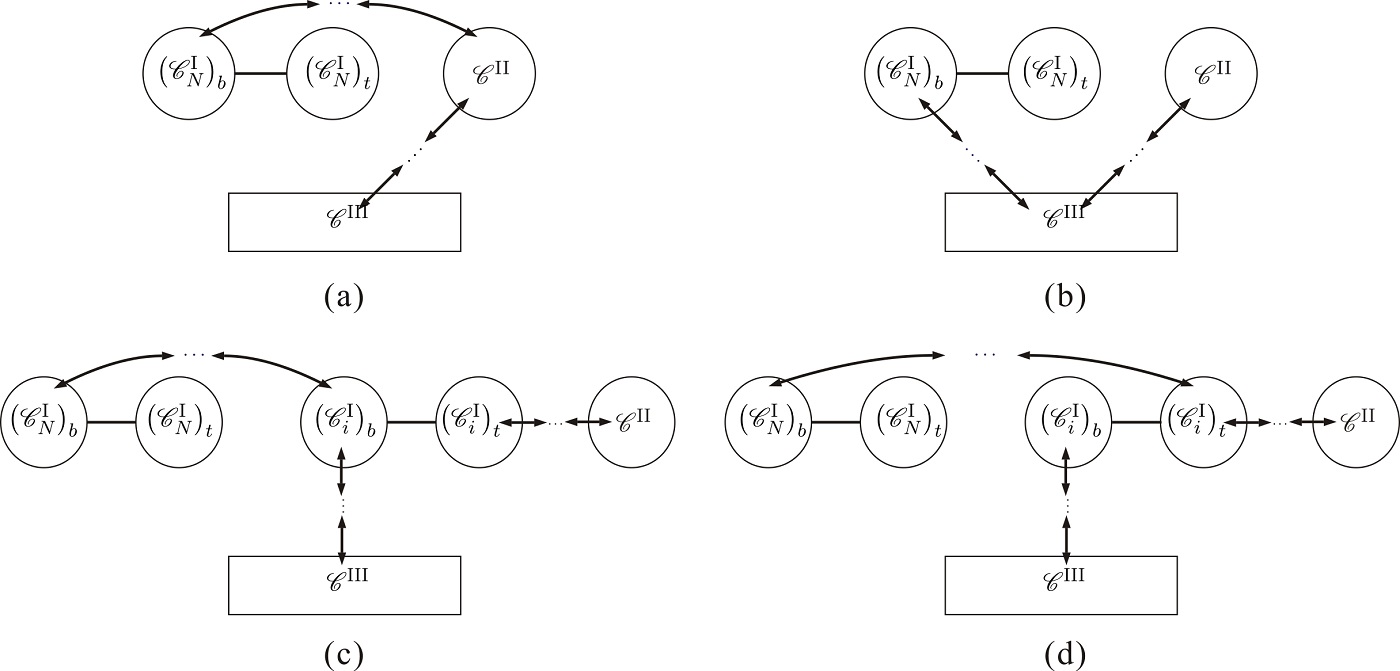}
\caption{According to rule-1, the chain starts at the component $\mathscr{C}_{N}^{\text{I}}$ must ends at (i) the starting component (graph (a)), (ii) the ending component (graph (b)) or (iii) an internal component (graph (c) and (d)) of the chain led by $\mathscr{C}^{\text{II}}$. According to rule-2, the chain led by $\mathscr{C}_{N}^{\text{I}}$ in graphs (a)-(d) must be of the form (i) $ \mathscr{U}'=\bigl[\bigl(\mathscr{C}_{N}^{\text{I}}\bigr)_t-\bigl(\mathscr{C}_{N}^{\text{I}}\bigr)_b,\dots,\mathscr{U}=\mathscr{C}^{\text{II}}\bigr]$ (graph (a)), (ii) $\mathscr{L}'=\bigl[\bigl(\mathscr{C}_{N}^{\text{I}}\bigr)_t-\bigl(\mathscr{C}_{N}^{\text{I}}\bigr)_b,\dots,\mathscr{L}=\mathscr{C}^{\text{III}}\bigr]$ (graph (b)), (iii) $\mathscr{U}'=\bigl[\bigl(\mathscr{C}_{N}^{\text{I}}\bigr)_t-\bigl(\mathscr{C}_{N}^{\text{I}}\bigr)_b,\dots,\mathscr{U}=\mathscr{C}^{\text{II}}\bigr]$ (graph (c)) and (iv) $ \mathscr{L}'=\bigl[\bigl(\mathscr{C}_{N}^{\text{I}}\bigr)_t-\bigl(\mathscr{C}_{N}^{\text{I}}\bigr)_b,\dots,\bigl(\mathscr{C}_{i}^{\text{I}}\bigr)_t-\bigl(\mathscr{C}_{i}^{\text{I}}\bigr)_b,\dots,\mathscr{L}=\mathscr{C}^{\text{III}}\bigr]$ (graph (d)).}\label{Fig:equivalence}
\end{figure}

All graphs $\mathcal{F}$ (all possible configurations of the final upper and lower blocks $\mathscr{U}$ and  $\mathscr{L}$ as well as all possible connections of these two blocks by a type-3 line for a given configuration of $\mathscr{U}$ and  $\mathscr{L}$) constructed in the above steps together form the set of physical graphs containing the skeleton $\mathcal{F}\,'$.

{\centering\subsubsection*{The equivalence between rule-1 and -2}}
Rule-1 naturally inherits the chain structures of the refined graphic rule. More specifically, the chains of components are constructed by keeping track of the chains led by the highest-weight nodes in them, as we have stated in the example. In the following, we prove that there exists a one-to-one correspondence between physical graphs constructed by rule-1 and rule-2.

Any physical graph constructed by rule-1 can be understood as follows. Define the upper block by $\mathscr{U}=\mathscr{C}^{\text{II}}$, the lower block by $\mathscr{L}=\mathscr{C}^{\text{III}}$ and the reference order $\mathsf{R}^{\,\text{I}}_{\mathscr{C}}\equiv\mathsf{R}_{\mathscr{C}}\setminus\mathscr{C}^{\text{II}}=\{\mathscr{C}_1^{\text{I}},\mathscr{C}_2^{\text{I}},\dots,\mathscr{C}_{N}^{\text{I}}\}$. For a given physical graph, we find out the chain led by the highest-weight type-I component $\mathscr{C}_{N}^{\text{I}}$ in the reference order $\mathsf{R}^{\,\text{I}}_{\mathscr{C}}$. According to rule-1, this chain can be ended at any component on the chain led by the type-II component $\mathscr{C}^{\text{II}}$ because the weight of $\mathscr{C}^{\text{II}}$ is higher than $\mathscr{C}_{N}^{\text{I}}$. Since the chain that is led by $\mathscr{C}^{\text{II}}$ starts from $\mathscr{C}^{\text{II}}$, ends at $\mathscr{C}^{\text{III}}$ and can also have possible internal type-I  components, the ending component of the chain led by $\mathscr{C}_{N}^{\text{I}}$ could be (i) $\mathscr{U}=\mathscr{C}^{\text{II}}$ (\figref{Fig:equivalence} (a)), (ii) $\mathscr{L}=\mathscr{C}^{\text{III}}$ (\figref{Fig:equivalence} (b)) or (iii) any internal type-I component of the chain led by $\mathscr{C}^{\text{II}}$ (\figref{Fig:equivalence} (c) and (d)). For the case (i), we redefine the upper block by the chain $\mathscr{U}\to \mathscr{U}'=\bigl[\bigl(\mathscr{C}_{N}^{\text{I}}\bigr)_t-\bigl(\mathscr{C}_{N}^{\text{I}}\bigr)_b,\dots,\mathscr{U}=\mathscr{C}^{\text{II}}\bigr]$ that starts at $\mathscr{C}_{N}^{\text{I}}$  and ends at the upper block $\mathscr{U}=\mathscr{C}^{\text{II}}$. For the case (ii), we redefine the lower block by the chain $\mathscr{L}\to \mathscr{L}'=\bigl[\bigl(\mathscr{C}_{N}^{\text{I}}\bigr)_t-\bigl(\mathscr{C}_{N}^{\text{I}}\bigr)_b,\dots,\mathscr{L}=\mathscr{C}^{\text{III}}\bigr]$ that starts from $\mathscr{C}_{N}^{\text{I}}$  and ends at the lower block $\mathscr{L}=\mathscr{C}^{\text{III}}$. For the case (iii), if the chain led by $\mathscr{C}_{N}^{\text{I}}$ (defined by rule-1) has the form $\bigl[\bigl(\mathscr{C}_{N}^{\text{I}}\bigr)_t-\bigl(\mathscr{C}_{N}^{\text{I}}\bigr)_b,\dots,\bigl(\mathscr{C}_{i}^{\text{I}}\bigr)_b-\bigl(\mathscr{C}_{i}^{\text{I}}\bigr)_t\bigr]$ (see \figref{Fig:equivalence} (c))  i.e. this chain ends at the bottom side $\bigl(\mathscr{C}_{i}^{\text{I}}\bigr)_b$ of a component belonging to the chain led by $\mathscr{C}^{\text{II}}$, we redefine the upper block by the chain (defined by rule-2) $\mathscr{U}\to \mathscr{U}'=\bigl[\bigl(\mathscr{C}_{N}^{\text{I}}\bigr)_t-\bigl(\mathscr{C}_{N}^{\text{I}}\bigr)_b,\dots,\bigl(\mathscr{C}_{i}^{\text{I}}\bigr)_b-\bigl(\mathscr{C}_{i}^{\text{I}}\bigr)_t,\dots,\mathscr{U}=\mathscr{C}^{\text{II}}\bigr]$. Else, if the chain led by $\mathscr{C}_{N}^{\text{I}}$ has the form $\bigl[\bigl(\mathscr{C}_{N}^{\text{I}}\bigr)_t-\bigl(\mathscr{C}_{N}^{\text{I}}\bigr)_b,\dots,\bigl(\mathscr{C}_{i}^{\text{I}}\bigr)_t-\bigl(\mathscr{C}_{i}^{\text{I}}\bigr)_b\bigr]$,
we redefine the lower block by the chain (defined by rule-2)
$\mathscr{L}\to \mathscr{L}'=\bigl[\bigl(\mathscr{C}_{N}^{\text{I}}\bigr)_t-\bigl(\mathscr{C}_{N}^{\text{I}}\bigr)_b,\dots,\bigl(\mathscr{C}_{i}^{\text{I}}\bigr)_t-\bigl(\mathscr{C}_{i}^{\text{I}}\bigr)_b,\dots,\mathscr{L}=\mathscr{C}^{\text{III}}\bigr]$.
After finding out the chain led by $\mathscr{C}_{N}^{\text{I}}$ defined by rule-2, we remove the starting and internal components of this chain from $\mathsf{R}^{\,\text{I}}_{\mathscr{C}}\equiv\mathsf{R}_{\mathscr{C}}\setminus\mathscr{C}^{\text{II}}=\{\mathscr{C}_1^{\text{I}},\mathscr{C}_2^{\text{I}},\dots,\mathscr{C}_{N}^{\text{I}}\}$.

We find out the chain that is led by the highest-weight component in the redefined $\mathsf{R}^{\,\text{I}}_{\mathscr{C}}$ by following the previous step but using the redefined $\mathsf{R}^{\,\text{I}}_{\mathscr{C}}$, $\mathscr{U}$ and $\mathscr{L}$. Repeating these steps until the ordered set $\mathsf{R}^{\,\text{I}}_{\mathscr{C}}$ becomes empty, we get graphs with only two disjoint maximally connected subgraphs: the final upper and lower blocks. Then the graph is given by connecting the final upper and lower blocks $\mathscr{U}$ and $\mathscr{L}$ via a type-3 line. This description of the physical graphs obtained by rule-1 precisely agrees with the rule-2.

Conversely, any graph constructed according to rule-2 can also be obtained by rule-1. We define reference order $\mathsf{R}_{\mathscr{C}}\equiv\{\mathscr{C}_1^{\text{I}},\mathscr{C}_2^{\text{I}},\dots,\mathscr{C}_{N}^{\text{I}},\mathscr{C}^{\text{II}}\}$.
One can always find out a path from the highest-weight component $\mathscr{C}^{\text{II}}$ in $\mathsf{R}_{\mathscr{C}}$ to $\mathscr{C}^{\text{III}}$. This path can be considered as the chain (defined by rule-1) $\bigl[\mathscr{C}^{\text{II}},\dots,\mathscr{C}^{\text{III}}\bigr]$ which starts at $\mathscr{C}^{\text{II}}$ and ends at $\mathscr{C}^{\text{III}}$. Redefine the ordered set $\mathsf{R}_{\mathscr{C}}$ by deleting the starting and internal components of this chain.
%, i.e. $\mathsf{R}_{\mathscr{C}}\equiv\{\mathscr{C}_1^{\text{I}},\mathscr{C}_2^{\text{I}},\dots,\mathscr{C}_{N}^{\text{I}},\mathscr{C}^{\text{II}}\}$.

In the same way, we find out the chain (defined by rule-1) led by the highest-weight component of the reference order $\mathsf{R}_{\mathscr{C}}$ that is defined in the previous step and redefine $\mathsf{R}_{\mathscr{C}}$ again. Repeat these steps until  $\mathsf{R}_{\mathscr{C}}$ becomes an empty set. Then we find that a graph constructed by rule-2 can also be obtained by rule-1.

\subsection{The construction of spurious graphs}\label{section:SpuriousGraphs}
In the last example in \secref{sec:|H|=3Example}, we have introduced spurious graphs (see \figref{Fig:Figure8} (c), (f)) which are helpful when studying the relationship between the gauge invariance induced identity and BCJ relation. In general, spurious graphs are introduced for skeletons with at least three components. To see this point, we recall rule-2 for the construction of physical graphs. In the last step of rule-2, we connect the final upper and lower blocks $\mathscr{U}$ and $\mathscr{L}$  via a type-3 line by avoiding structures \figref{Fig:GraphsNotAllowed} (a) and (b) because the highest-weight chain that starts from $\mathscr{C}^{\text{II}}$ and ends at $\mathscr{C}^{\text{III}}$ must pass through kernels of all type-I components on it. Thus the chain of components led by $\mathscr{C}^{\text{II}}$ in a physical graph only contains structures of the form \figref{Fig:GraphsNotAllowed} (c) and (d). \emph{Spurious graphs are introduced as graphs where the path starting from the highest-weight node $h_{\rho(s)}$ towards a root passes through singles-sides of some components (i.e. structures \figref{Fig:GraphsNotAllowed} (a) and/or (b) existing on this path) when connecting the final upper and lower blocks via a type-3 line.} A spurious graph must be associated with a proper sign. We will see soon,  the spurious graphs for any given configuration of the final upper and lower blocks do not vanish. Nevertheless, we find that spurious graphs for distinct configurations of the final upper and lower blocks precisely cancel out in pairs (for example \figref{Fig:Figure8} (c), (f)). Therefore, the sum over all physical graphs corresponding to a given skeleton $\mathcal{F}'$  equals to the sum over (i) all possible configurations of the final upper and lower blocks $\mathscr{U}$ and $\mathscr{L}$ for $\mathcal{F}'$, (ii) all possible (spurious and physical) graphs which are obtained by connecting arbitrary two nodes in the final upper and lower blocks respectively by a type-3 line for a given configuration of $\mathscr{U}$ and $\mathscr{L}$. In the following, we first take the spurious graphs for the skeleton  \figref{Fig:Figure9} as an example, then provide a general discussion on spurious graphs.

\subsubsection{An example: spurious graphs for the skeleton \figref{Fig:Figure9}}
All spurious graphs for  \figref{Fig:Figure9} are presented by \figref{Fig:Figure15-1} (a2)-(d2), \figref{Fig:Figure15} (a2)-(d2) and \figref{Fig:Figure16} (a2)-(d2). In each graph, the final upper and lower blocks are given by the two maximally connected subgraphs that contains only solid lines.
We take \figref{Fig:Figure15-1} (a2) for instance. The final upper and lower blocks in in \figref{Fig:Figure15-1} (a2) are correspondingly given by  $\mathscr{C}$ and the connected subgraph that is constructed by connecting both $\mathscr{A}$ and $\mathscr{B}$ with $\mathscr{R}$ by type-3 lines. When we further connect two nodes respectively belonging to  $\mathscr{C}$ (the final upper block) and $\mathscr{A}_{b}$ or $\mathscr{B}_{b}$ (in the final lower block) via a type-3 line (the dashed arrow lines), spurious graph, where the path starts from $h_{\rho(s)}$  passes through a single side $\mathscr{A}_{b}$ (\figref{Fig:Figure15-1} (a2) with the line $1+$) or  $\mathscr{B}_{b}$ (\figref{Fig:Figure15-1} (a2) with the line $2+$) and ends at a root, is obtained. The spurious graphs  \figref{Fig:Figure15-1} (a2) together with the physical graphs  \figref{Fig:Figure15-1} (a1) provide all possible graphs produced by connecting the final upper and lower blocks via a type-3 line.

An interesting observation is that each structure of spurious graph in \figref{Fig:Figure15-1} (a2)-(d2), \figref{Fig:Figure15} (a2)-(d2) and \figref{Fig:Figure16} (a2)-(d2) always appears twice. Thus we can dress the two graphs with the same structure opposite signs so that all spurious graphs cancel out in pairs after summation.
As an example, the graph \figref{Fig:Figure15-1} (a2) with `$1+$' line (the sign is $+$) and the graph \figref{Fig:Figure15-1}.(d2) with `$1-$' line (the sign is $-$) have the same structure but opposite signs.

General construction for spurious graphs with strictly defined signs is provided in the coming discussion.

%When all spurious graphs (with proper signs) \figref{Fig:Figure15-1} (a2)-(d2), \figref{Fig:Figure15} (a2)-(d2) and \figref{Fig:Figure16} (a2)-(d2) for all possible configurations of the final upper and lower blocks are constructed, we find that the sum of physical graphs and spurious graphs for each configuration is just given by the sum of all possible graphs by connecting arbitrary two nodes from each part via a type-3 line. Since all spurious graphs cancel out, the sum of all physical graphs containing the skeleton \figref{Fig:Figure9} is then given by  i) summing over all configurations of the two disjoint parts obtained by the step-3 of rule-2 and ii) summing over all possible (physical and spurious) graphs by connecting the two disjoint parts together for a given configuration.

\subsubsection{General rule for the construction of spurious graphs}
\begin{figure}
\centering
\includegraphics[width=1\textwidth]{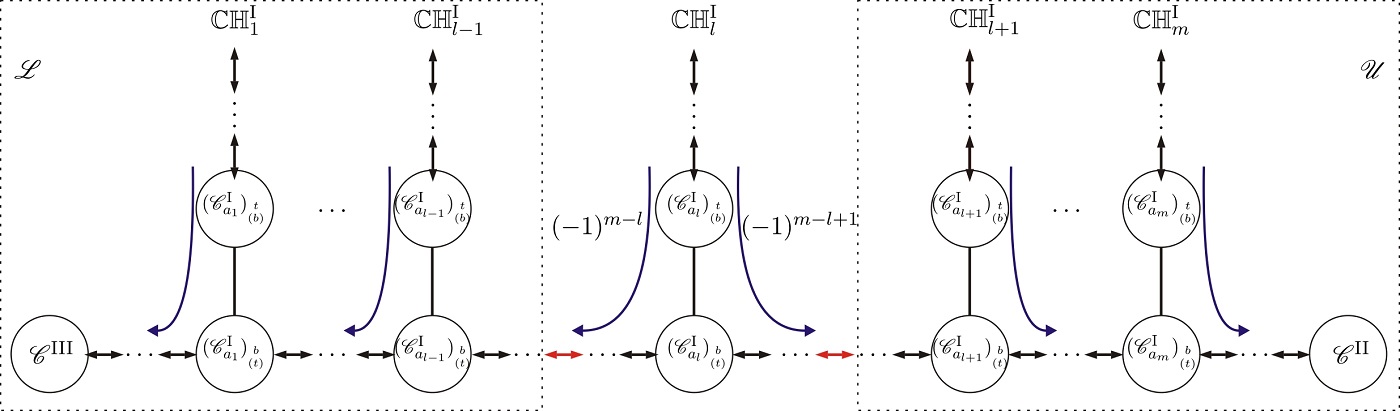}
\caption{A typical spurious graph.
There are $m$ spurious structures $\mathscr{C}_{a_2}^{\text{I}}$, ..., $\mathscr{C}_{a_m}^{\text{I}}$ on the path from the highest weight node $h_{\rho(s)}$ towards a root.
The chains  $\mathbb{CH}_{1}^{\text{I}}$, $\mathbb{CH}_{2}^{\text{I}}$, \dots, $\mathbb{CH}_{l-1}^{\text{I}}$ ($\mathbb{CH}_{l+1}^{\text{I}}$, $\mathbb{CH}_{l+2}^{\text{I}}$, \dots, $\mathbb{CH}_{m}^{\text{I}}$) containing  $\mathscr{C}_{a_1}^{\text{I}}$, $\mathscr{C}_{a_2}^{\text{I}}$, ..., $\mathscr{C}_{a_{l-1}}^{\text{I}}$ ($\mathscr{C}_{a_{l+1}}^{\text{I}}$, $\mathscr{C}_{a_{l+2}}^{\text{I}}$, ..., $\mathscr{C}_{a_{m}}^{\text{I}}$) must belong to the final lower (upper) block. The lowest-weight chain  $\mathbb{CH}_{l}^{\text{I}}$ among  $\mathbb{CH}_{i}^{\text{I}}$ $(i=1,\dots,m)$ can belong to either the final upper block (with sign $(-1)^{m-l+1}$) or the final lower one (with a sign $(-1)^{m-l}$). Thus two spurious graphs with the same configuration but opposite signs are obtained from two distinct configurations of the final upper and lower blocks, respectively.
 }\label{Fig:Spurious}
\end{figure}
In general, when we connect the final upper and lower blocks into a connected graph by a type-3 line, we can find the unique path starting from the highest-weight node $h_{\rho(s)}\in\mathscr{C}^{\text{II}}$ and ending at a root in $\{1,\dots,r-1\}$ (thus in $\mathscr{C}^{\text{III}}$). If there exist $m$ ($m\geq 1$) type-I components $\mathscr{C}_{a_1}^{\text{I}}$, $\mathscr{C}_{a_2}^{\text{I}}$, ..., $\mathscr{C}_{a_m}^{\text{I}}$  such that only a single side of each is on the path, as shown by  \figref{Fig:GraphsNotAllowed} (a), (b), the graph must be a spurious one.  Each component $\mathscr{C}_{a_i}^{\text{I}}$  in a spurious graph is called \emph{a spurious component}. The sign associated to the spurious graph is defined by $(-1)^{\mathcal{S}(\mathscr{U})}$, where $\mathcal{S}(\mathscr{U})$ is the number of spurious components in the final upper block. For example, in \figref{Fig:Figure15} (c2), $\mathcal{S}(\mathscr{U})=1,2,1$ for the graphs with lines $5-$, $12+$, $15-$ correspondingly.

To show that spurious graphs for distinct configurations always cancel in pairs, we assume that the path from $h_{\rho(s)}\in\mathscr{C}^{\text{II}}$  towards a root passes through spurious components $\mathscr{C}_{a_i}^{\text{I}}$ ($i=1,\dots,m$)  in the order $\mathscr{C}_{a_m}^{\text{I}}$, ..., $\mathscr{C}_{a_2}^{\text{I}}$, $\mathscr{C}_{a_1}^{\text{I}}$ for convenience. The chains (defined by rule-2) containing components\footnote{Note that the starting components of these chains defined by rule-2 can only be type-I components.} $\mathscr{C}_{a_1}^{\text{I}}$, $\mathscr{C}_{a_2}^{\text{I}}$, ..., $\mathscr{C}_{a_m}^{\text{I}}$ are correspondingly denoted by $\mathbb{CH}_{1}^{\text{I}}$, $\mathbb{CH}_{2}^{\text{I}}$, ..., $\mathbb{CH}_m^{\text{I}}$ and their weights are denoted by $\mathcal{W}_{1}^{\text{I}}$, $\mathcal{W}_{2}^{\text{I}}$, ..., $\mathcal{W}_{m}^{\text{I}}$. The lowest-weight chain  among  $\mathbb{CH}_{1}^{\text{I}}$, $\mathbb{CH}_{2}^{\text{I}}$, ..., $\mathbb{CH}_m^{\text{I}}$ is assumed to be $\mathbb{CH}_{l}^{\text{I}}$. We find that the following properties of chains  $\mathbb{CH}_{1}^{\text{I}}$, $\mathbb{CH}_{2}^{\text{I}}$, ..., $\mathbb{CH}_m^{\text{I}}$ in a given spurious graph must be satisfied:
\begin{itemize}
\item [(i)] The weights of chains $\mathcal{W}_{1}^{\text{I}}$, $\mathcal{W}_{2}^{\text{I}}$, ..., $\mathcal{W}_{m}^{\text{I}}$ must satisfy
\bea
\mathcal{W}_{1}^{\text{I}}>\mathcal{W}_{2}^{\text{I}}>\dots>\mathcal{W}_{l-1}^{\text{I}}>\mathcal{W}_{l}^{\text{I}},~~\mathcal{W}_{l}^{\text{I}}<\mathcal{W}_{l+1}^{\text{I}}<\dots<\mathcal{W}_{m-1}^{\text{I}}<\mathcal{W}_{m}^{\text{I}}.
\eea
This is a consequent result of the rule-2: we always connect chains, in the descending order of their weights, to either the upper or the lower block.

\item [(ii)] The chains $\mathbb{CH}_{1}^{\text{I}}$, $\mathbb{CH}_{2}^{\text{I}}$, \dots, $\mathbb{CH}_{l-1}^{\text{I}}$ ($\mathbb{CH}_{l+1}^{\text{I}}$, $\mathbb{CH}_{l+2}^{\text{I}}$, \dots, $\mathbb{CH}_{m}^{\text{I}}$) and structures attached to them belong to the final lower (upper) block (see \figref{Fig:Spurious}). If not, for example
the chain $\mathbb{CH}_{i}^{\text{I}}$ for given $i<l$  belongs to the final upper block, there must be at least one component (i.e. $\mathscr{C}_{a_l}^{\text{I}}$) with only a single side on $\mathbb{CH}_{i}^{\text{I}}$ before connecting the upper and lower blocks together. This conflicts with the definition of chains in rule-2.
\end{itemize}
According to rule-2, the lowest-weight chain $\mathbb{CH}_{l}^{\text{I}}$ (and structure attached to it) among $\mathbb{CH}_{1}^{\text{I}}$, $\mathbb{CH}_{2}^{\text{I}}$, ..., $\mathbb{CH}_m^{\text{I}}$ can be considered to be connected with either the upper or the lower block, which have been defined previously, via a type-3 line (see \figref{Fig:Spurious}). Thus the chain $\mathbb{CH}_{l}^{\text{I}}$ (and structure attached to it) in a given spurious graph can belong to either the final upper block or the final lower block. Correspondingly, the extra signs for these two spurious graphs, which have the same structure and distinct configurations of the final upper and lower blocks, are $(-1)^{m-l+1}$ and $(-1)^{m-l}$. As a result,  all spurious graphs must cancel out in pairs.

\subsection{The sum over all physical and spurious graphs}\label{section:SumPhySpu}

Now we are ready for rearranging the expression in the square brackets of \eqref{Eq:GaugeInv5} in an appropriate form for studying the relationship between the gauge invariance induced identity \eqref{Eq:GaugeInv2} and BCJ relations. For any skeleton $\mathcal{F}\,'$, the expression in the brackets is given by
\bea
I=\Sl_{\substack{\mathcal{F}\text{~s.t.}\\\mathcal{F}\supset\mathcal{F}'}}\mathcal{K}^{[{\mathcal{F}\setminus\mathcal{F}'}]}\Sl_{\pmb{\sigma}^{\mathcal{F}}}\,A(1,\pmb{\sigma}^{\mathcal{F}},r).
\eea
The sum over all (physical) graphs $\mathcal{F}$ containing the skeleton $\mathcal{F}'$ can be achieved by the following two summations.
\begin{itemize}
\item [(i)] Sum over all possible configurations $\mathscr{U}\oplus\mathscr{L}$\footnote{Here, $\oplus$ denotes the disjoint union.} of the final upper and lower blocks $\mathscr{U}$ and $\mathscr{L}$ constructed by rule-2 , including (1) all possible configurations of $\mathscr{U}$ and $\mathscr{L}$  constructed by chains of components when neglecting the inner structures of all type-II, type-III components and both sides of type-I components; (2) all possible choices of end nodes of the type-3 lines for each configuration in (1). The kinematic factor $\mathcal{K}^{\left[\mathscr{U}\oplus\mathscr{L}\setminus\mathcal{F}'\right]}$ with respect to a given configuration  $\mathscr{U}\oplus\mathscr{L}$ is the product of $k\cdot k$ factors corresponding to the type-3 lines in the final upper and lower blocks $\mathscr{U}$, $\mathscr{L}$.

\item[(ii)] For a given configuration $\mathscr{U}\oplus\mathscr{L}$ in the previous summation, sum over all \emph{physical graphs} $\mathcal{F}\in \text{Phy}\left(\mathscr{U},\mathscr{L}\right)$ which are obtained by connecting two nodes in distinct blocks via a type-3 line such that the chain led by $h_{\rho(s)}$ does not invlove structures \figref{Fig:GraphsNotAllowed} (a) and (b). The kinematic factor provided in this step is the $k\cdot k$ factor for the type-3 line between the two blocks and is denoted by $\mathcal{K}^{\left[\mathcal{F}\setminus\mathscr{U}\oplus\mathscr{L}\right]}$.
\end{itemize}
Thus $I$ is given by
\bea
I=\Sl_{\mathscr{U}\oplus\mathscr{L}}\mathcal{K}^{\left[\mathscr{U}\oplus\mathscr{L}\setminus\mathcal{F}'\right]}\Biggl[\Sl_{\mathcal{F}\in \text{Phy} \left(\mathscr{U},\mathscr{L}\right)}\mathcal{K}^{\left[\mathcal{F}\setminus\mathscr{U}\oplus\mathscr{L}\right]}\Sl_{\pmb{\sigma}^{\mathcal{F}}}\,A(1,\pmb{\sigma}^{\mathcal{F}},r)\Biggr].
\eea
Since all spurious graphs for distinct configurations of $\mathscr{U}\oplus\mathscr{L}$  corresponding to the same skeleton $\mathcal{F}'$ must cancel out in pairs,
 the summation over all physical graphs for a given  $\mathscr{U}\oplus\mathscr{L}$  can be extended to a summation over all physical and spurious graphs $\mathcal{F}\in\text{Phy} \left(\mathscr{U},\mathscr{L}\right)\cup\text{Spu}\left(\mathscr{U},\mathscr{L}\right)$ for a given $\mathscr{U}\oplus\mathscr{L}$. Hence we arrive
\bea
I=\Sl_{\mathscr{U}\oplus\mathscr{L}}\mathcal{K}^{\left[\mathscr{U}\oplus\mathscr{L}\setminus\mathcal{F}'\right]}\Biggl[\Sl_{{\substack{\mathcal{F}\in\text{Phy} \left(\mathscr{U},\mathscr{L}\right)\\
~~\cup\text{Spu}\left(\mathscr{U},\mathscr{L}\right)}}}\mathcal{K}^{\left[\mathcal{F}\setminus\mathscr{U}\oplus\mathscr{L}\right]}\Sl_{\pmb{\sigma}^{\mathcal{F}}}\,A(1,\pmb{\sigma}^{\mathcal{F}},r)\Biggr].
\eea
According to the definition, the sum over all physical and spurious graphs is nothing but the sum over all possible graphs, each of which is constructed by connecting two arbitrary nodes belonging to the final upper and lower blocks respectively via a type-3 line:
\bea
I=\Sl_{\mathscr{U}\oplus\mathscr{L}}\mathcal{K}^{\left[\mathscr{U}\oplus\mathscr{L}\setminus\mathcal{F}'\right]}I\left[\mathscr{U},\mathscr{L}\right].\Label{Eq:GaugeInv6}
\eea
The $I\left[\mathscr{U},\mathscr{L}\right]$ for a given configuration $\mathscr{U}\oplus\mathscr{L}$ in the above equation is defined by
 \bea
I\left[\mathscr{U},\mathscr{L}\right]\equiv\Sl_{ \mathcal{F}\in\left(\mathscr{U},\mathscr{L}\right)
}(-)^{\mathcal{F}}\mathcal{K}^{\left[\mathcal{F}\setminus\mathscr{U}\oplus\mathscr{L}\right]}\Sl_{\pmb{\sigma}^{\mathcal{F}}}\,A(1,\pmb{\sigma}^{\mathcal{F}},r),\Label{Eq:GaugeInv7}
\eea
where
\bea
(-)^{\mathcal{F}}=(-1)^{\mathcal{N}(\mathscr{U})+\mathcal{N}(\mathscr{L})+\mathcal{S}(\mathscr{U})+1}
\eea
is a proper sign depending on the total number of arrows  pointing deviate from the direction of root ($\mathcal{N}(\mathscr{U})$ for the final upper block, $\mathscr{N}(\mathscr{L})$ for the final lower block and $1$ for the type-3 line between the two blocks) and the number $\mathcal{S}(\mathscr{U})$ of spurious components in the final upper block of a graph.
\begin{figure}
\centering
\includegraphics[width=0.9\textwidth]{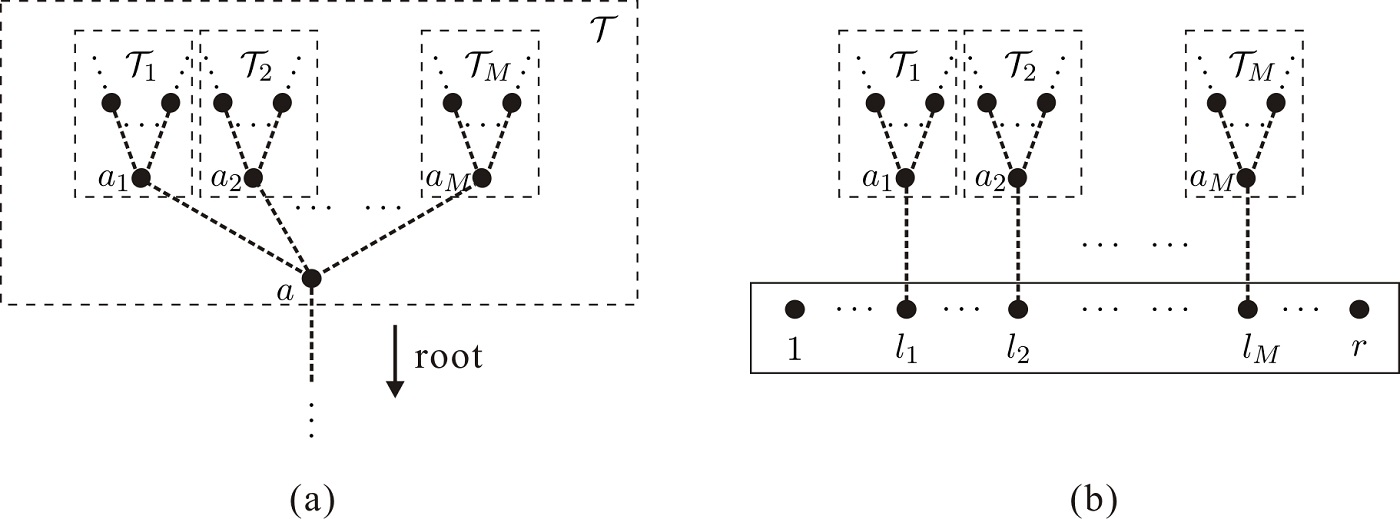}
\caption{(a): A typical tree structure with sub-trees planted at the node $a$, (b): A typical graph with trees planted at roots $l_1,\dots,l_M\in \{1,\dots,r-1\}$.}
 \label{Fig:Permutations1}
\end{figure}
\begin{figure}
\centering
\includegraphics[width=0.7\textwidth]{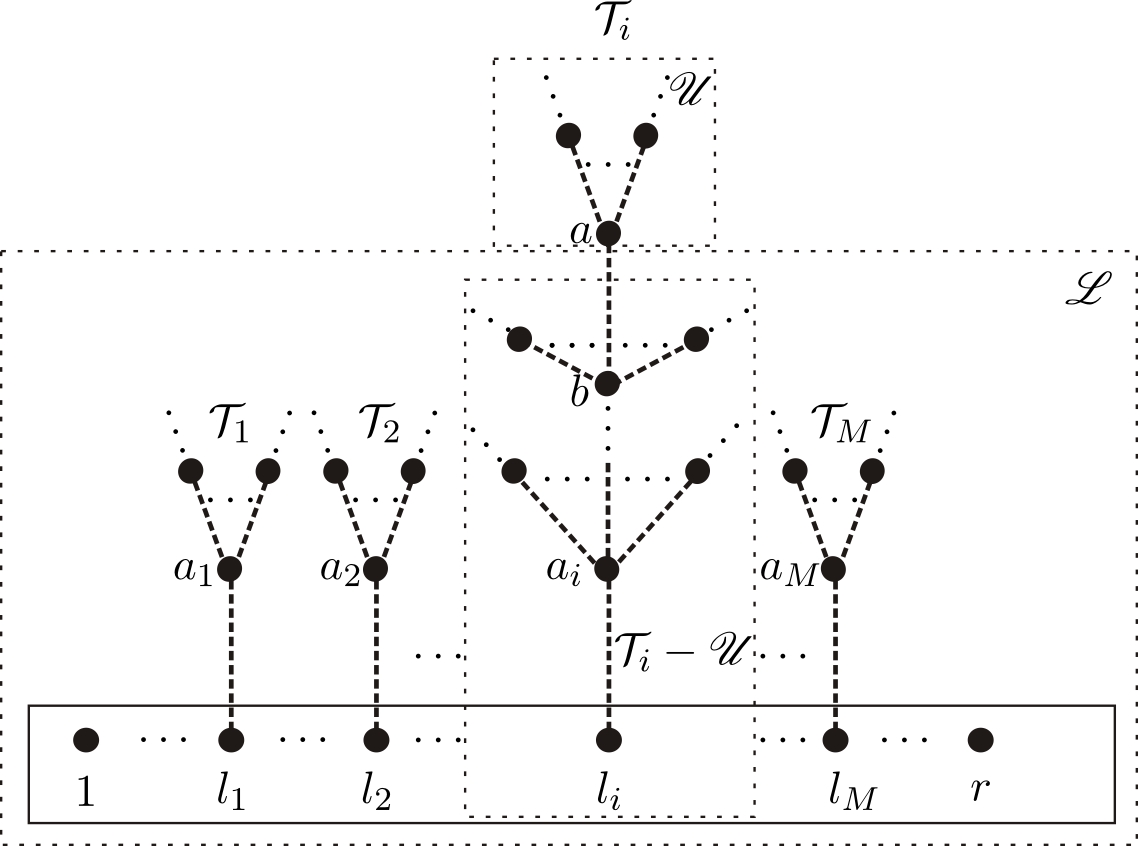}
\caption{Permutations established by a graph with trees $\mathcal{T}_1,\mathcal{T}_2,\dots,\mathcal{T}_M$ planted at roots $l_1, l_2, \dots, l_M\in\{1,\dots,r-1\}$ are given by \eqref{Eq:Permutation1}. The nearest-to-root nodes in these trees are correspondingly $a_1, a_2,\dots,a_M$. The final upper block belongs to the tree $\mathcal{T}_i$ which is rooted at $l_i\in\{1,\dots,r-1\}$. Nodes $b\in \mathscr{L}$ and  $a\in \mathscr{U}$ are the two nodes that belong to distinct blocks and connected by a type-3 line. According to whether the nodes belong to the final upper or lower block, the tree $\mathcal{T}_i$ is divided into two parts $\mathscr{U}$ and $\mathcal{T}_i-\mathscr{U}$.
 }\label{Fig:Permutations2}
\end{figure}

{\centering\subsubsection*{Comments on the expression \eqref{Eq:GaugeInv7}}}

{\bf Permutations $\bf \pmb{\sigma}^{\mathcal{F}}$}~~ From the graphic rule, we can see permutations $\bf \pmb{\sigma}^{\mathcal{F}}$ for a given graph $\mathcal{F}\in\left(\mathscr{U},\mathscr{L}\right)$ do not rely on types of lines. They only depend on the relative positions between nodes in a graph.  Therefore, in the following we replace all types of lines in $\mathcal{F}$ by dashed lines with no arrow when considering permutations $\pmb{\sigma}^{\mathcal{F}}$ established by a graph ${\mathcal{F}}$. For each graph $\mathcal{F}$, the possible permutations $\pmb{\sigma}^{\mathcal{F}}$ are collected as follows:
\begin{itemize}
%\item The elements $1$ and $r$ are respectively the first and the last elements. The relative order of elements in the set $\{1,2,\dots,r-1,r\}$ can be expressed by a simple chain on which if $i<j$, the node $i$ is nearer to $1$ than $j$. Since this chain does not involve Lorentz inner products, it is just introduced for keeping track with the relative orders between nodes, we use another type of line -dashed line (with no arrow) to connect nodes.

\item The relative order of roots is always $(\sigma^{\mathcal{F}})^{-1}(1)<(\sigma^{\mathcal{F}})^{-1}(2)<\dots<(\sigma^{\mathcal{F}})^{-1}(r-1)$. We set $\sigma^{-1}(r)=\infty$ to require that the element $r$ is always the last one in the permutation $\pmb{\sigma}^{\mathcal{F}}$.

\item For tree structures planted at roots $\{1,2,\dots,r-1\}$, if two elements $a, b\in \mathsf{H}$ are on a same path which starts from an element of $\mathsf{H}$ and end at an arbitrary root $l\in \{1,2,\dots,r-1\}$ and satisfy the condition that $a$ is nearer to the root $l$ than $b$, we have $(\sigma^{\mathcal{F}})^{-1}(l)<(\sigma^{\mathcal{F}})^{-1}(a)<(\sigma^{\mathcal{F}})^{-1}(b)$.
\item  If a node $a\in \mathsf{H}$ in a connected tree structure  $\mathcal{T}$ (shown by \figref{Fig:Permutations1} (a)) is nearer to root than all other nodes in  $\mathcal{T}$ and $a$ is attached by $M$ connected sub-tree structures (branches) $\mathcal{T}_{1}$, $\mathcal{T}_{2}$,...,$\mathcal{T}_{M}$ ($M\geq 2$) (where the nodes nearest to $a$ in these branches are $a_1$, $a_2$,...,$a_M$  correspondingly), the collection of all possible relative orders for nodes in $\mathcal{T}$ are recursively expressed by
\bea
{\mathcal{T}}\big|_{a}=\bigl\{a,{\mathcal{T}_1}\big|_{a_1}\shuffle{\mathcal{T}_2}\big|_{a_2}\shuffle\dots\shuffle{\mathcal{T}_M}\big|_{a_M}\bigr\},
\eea
where $\mathcal{T}\big|_{a}$ denotes all possible permutations established by $\mathcal{T}$ with the leftmost element $a$.

\item Suppose there are $M$ connected tree structures  $\mathcal{T}_{1}$, $\mathcal{T}_{2}$,...,$\mathcal{T}_{M}$ planted at roots $l_1,l_2,\dots,l_M\in\{1,\dots,r-1\}$ in a graph $\mathcal{F}$ correspondingly while the nearest-to-roots elements with respect to these tree structures are $a_{1}$, $a_{2}$,...,$a_{M}$ (see \figref{Fig:Permutations1} (b)). The permutations $\pmb{\sigma}^{\mathcal{F}}$ of elements in $\mathsf{H}\cup\{1,\dots,r\}$ satisfy
\bea
\pmb{\sigma}^{\mathcal{F}}\in\bigl\{1,\bigl\{2,\dots,r-1\bigr\}\shuffle{\mathcal{T}_1}\big|_{a_1}\shuffle{\mathcal{T}_2}\big|_{a_2}\shuffle\dots\shuffle{\mathcal{T}_M}\big|_{a_M},r\bigr\}\Big|_{\substack{l_i\prec a_i\\(i=1,\dots,M)}},
\eea
where $a\prec b$ in a permutation $\pmb{\sigma}$ means $\sigma^{-1}(a)<\sigma^{-1}(b)$.
\end{itemize}

{\bf Rearrangement of \eqref{Eq:GaugeInv7}}~~ Since the graph $\mathcal{F}$ is given by connecting any node $a$ in the final upper block $\mathscr{U}$ and any node $b$ in the final lower block $\mathscr{L}$ via a type-3 line, the factor $\mathcal{K}^{\left[\mathcal{F}\setminus\mathscr{U}\oplus\mathscr{L}\right]}$ is just $k_a\cdot k_b$ while the sum over all $\mathcal{F}\in\left(\mathscr{U},\mathscr{L}\right)$ is given by summing over all choices of nodes $a\in \mathscr{U}$ and $b\in \mathscr{L}$.
Permutations $\pmb{\sigma}^{\mathcal{F}}$ can be understood  as follows:
\begin{itemize}
\item [(i)] Suppose that all tree structures planted at roots $l_1,l_2,\dots,l_M\in\{1,\dots,r-1\}$ in a given (physical or spurious) graph $\mathcal{F}$ are $\mathcal{T}_{1}$, $\mathcal{T}_{2}$, ..., $\mathcal{T}_{M}$ and the nearest-to-root elements with respect to these tree structures are $a_{1}$, $a_{2}$, ..., $a_{M}$ correspondingly (see \figref{Fig:Permutations2}). The final upper block $\mathscr{U}$ (connected to the node $b\in \mathscr{L}$), when all lines therein are replaced by dashed lines, must belong to some connected tree structure $\mathcal{T}_i$ which is attached to a root $l_i$. Permutations $\pmb{\sigma}^{\mathcal{F}}$ then satisfy
\bea
&&\pmb{\sigma}^{\mathcal{F}}\in\Bigl\{1,\bigl\{2,\dots,r-1\bigr\}\shuffle{\mathcal{T}_1}\big|_{a_1}
\shuffle\dots\shuffle\bigl((\mathcal{T}_i-\mathscr{U})\big|_{a_i}\shuffle\mathscr{U}\big|_a\bigr)\big|_{b\prec a}\shuffle\dots\shuffle{\mathcal{T}_M}\big|_{a_M},r\Bigr\}\Big|_{\substack{l_i\prec a_i\\(i=1,\dots,M)}}.\Label{Eq:Permutation1}
\eea

\item [(ii)] Any permutation $\pmb{\sigma}^{\mathcal{F}}$ satisfying \eqref{Eq:Permutation1} can be obtained by shuffling $\pmb{\zeta}\in{\mathscr{U}}\big|_{a}$ (a relative order of elements in $\mathscr{U}$) with $\pmb{\gamma}$ (a relative order of elements in  $\mathscr{L}$)
\bea
&&\pmb{\gamma}\in\Bigl\{\bigl\{2,\dots,r-1\bigr\}\shuffle{\mathcal{T}_1}\big|_{a_1}
\shuffle\dots\shuffle(\mathcal{T}_i-\mathscr{U})\big|_{a_i}\shuffle\dots\shuffle{\mathcal{T}_M}\big|_{a_M}\Big\}\Big|_{\substack{l_i\prec a_i\\(i=1,\dots,M)}}\Label{Eq:Permutation2}
\eea
such that $({\sigma}^{\mathcal{F}})^{-1}(b)<({\sigma}^{\mathcal{F}})^{-1}(a)$. Thus all permutations $\pmb{\sigma}^{\mathcal{F}}$ in \eqref{Eq:GaugeInv7} satisfy  $\pmb{\sigma}^{\mathcal{F}}\in\big\{1,\pmb{\gamma}\shuffle\pmb{\zeta},r\big\}\big|_{({\sigma}^{\mathcal{F}})^{-1}(b)<({\sigma}^{\mathcal{F}})^{-1}(a)}$
for a given $\pmb{\gamma}$ satisfying \eqref{Eq:Permutation2} and a given $\pmb{\zeta}\in{\mathscr{U}}\big|_{a}$.
\end{itemize}
\begin{figure}
\centering
\includegraphics[width=0.8\textwidth]{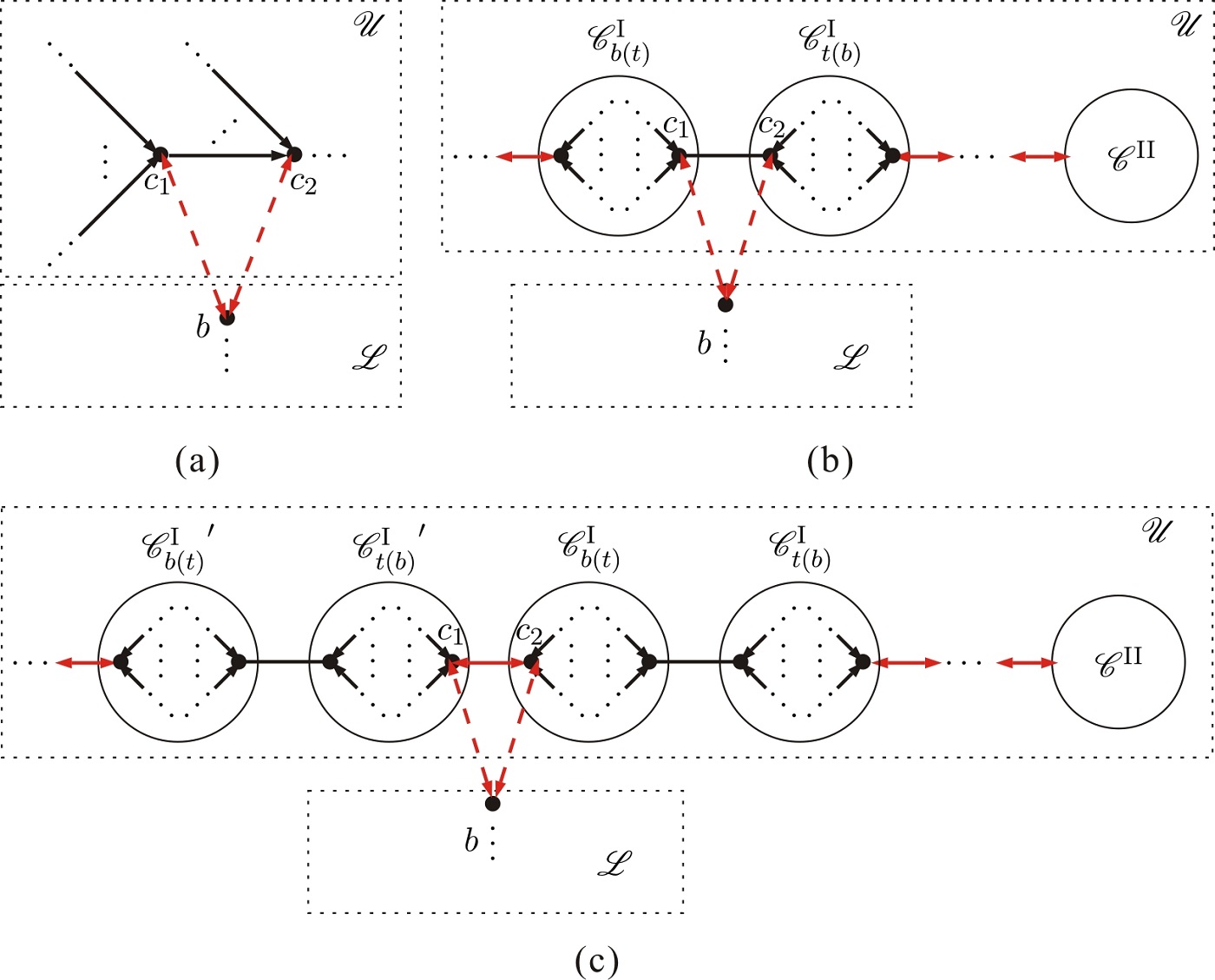}
\caption{For any given configuration of the final upper and lower blocks $\mathscr{U}$ and $\mathscr{L}$, the factors $(-1)^{\tiny\mathcal{N}(\mathscr{L})+\mathcal{N}(\mathscr{U}_c)+\mathcal{S}(\mathscr{U}_c)+1}$ for graphs with $a=c_1$ and $a=c_2$ ($c_1$, $c_2$ are two adjacent nodes in the final upper block) always have opposite signs. Particularly, the number $\mathcal{N}(\mathscr{L})$ is always independent of the choice of $a$. When we consider the number $\mathcal{N}(\mathscr{U}_c)+\mathcal{S}(\mathscr{U}_c)$, there are three possible cases:
(i) In (a), $c_1$ and $c_2$ are two end nodes of a type-3 line and they both belong to a same top or bottom side of some type-I component. The two graphs corresponding to $a=c_1$ and $a=c_2$ must have the same $\mathcal{S}(\mathscr{U}_c)$, but $\mathcal{N}(\mathscr{U}_{c_1})=\mathcal{N}(\mathscr{U}_{c_2})+1$. (ii) In (b), $c_1$ and $c_2$ are two end nodes of a type-1 line. Since a type-1 line does node carry arrow, we have $\mathcal{N}(\mathscr{U}_{c_1})=\mathcal{N}(\mathscr{U}_{c_2})$. The component $\mathscr{C}^{\text{I}}$ is a spurious one for the graph with $a=c_2$ but not spurious for the graph with $a=c_1$, thus $\mathcal{S}(\mathscr{U}_{c_1})=\mathcal{S}(\mathscr{U}_{c_2})-1$. (iii) Similarly, in (c), where $c_1$ and $c_2$ are two end nodes of a type-3 line, we have $\mathcal{N}(\mathscr{U}_{c_1})=\mathcal{N}(\mathscr{U}_{c_2})$ and $\mathcal{S}(\mathscr{U}_{c_1})=\mathcal{S}(\mathscr{U}_{c_2})+1$.
 }\label{Fig:Sign}
\end{figure}
Having the above observations in hand, we reexpress \eqref{Eq:GaugeInv7} by
 \bea
&&I\left[\mathscr{U},\mathscr{L}\right]~~~~\Label{Eq:GaugeInv8}\\
&=&\Sl_{a\in \mathscr{U}
}\Sl_{\substack{b\in \mathscr{L}\\b\neq r}
}(-1)^{\tiny\mathcal{N}(\mathscr{L})+\mathcal{N}(\mathscr{U}_a)+\mathcal{S}(\mathscr{U}_a)+1}(k_a\cdot k_b)\Sl_{\pmb{\zeta}\in{\mathscr{U}}|_{a}}\Sl_{\pmb{\gamma}}\Sl_{\pmb{\sigma}}A\bigl(1,\pmb{\sigma}\in(\pmb{\zeta}\shuffle\pmb{\gamma})\big|_{\tiny({\sigma})^{-1}(b)<({\sigma})^{-1}(a)},r\bigr),\nonumber
\eea
where all $\pmb{\gamma}$ satisfy \eqref{Eq:Permutation2} and $\pmb{\zeta}\in{\mathscr{U}}|_{a}$ are the permutation for the final upper block when considering $a$ is the nearest to root node in $\mathscr{U}$. For a given $\mathscr{U}$, $\mathcal{N}(\mathscr{U}_a)$ and $\mathcal{S}(\mathscr{U}_a)$ are the  $\mathcal{N}(\mathscr{U})$ and $\mathcal{S}(\mathscr{U})$ for choosing a given node $a\in\mathscr{U}$. Note that $\mathcal{N}(\mathscr{L})$ is independent of the choice of $b$ in the final lower block $\mathscr{L}$. Since the choices of nodes $a$ and $b$ are independent of each other, the above expression can be further arranged as
 \bea
&&I\left[\mathscr{U},\mathscr{L}\right]\Label{Eq:GaugeInv9}\\
&=&(-1)^{\tiny\mathcal{N}(\mathscr{L})}\Sl_{\pmb{\gamma}}\biggl\{\Sl_{a\in \mathscr{U}
}\Sl_{\pmb{\zeta}\in{\mathscr{U}}|_{a}}(-1)^{\tiny\mathcal{N}(\mathscr{U}_a)+\mathcal{S}(\mathscr{U}_a)+1}\biggl[\Sl_{\substack{b\in \mathscr{L}\\b\neq r}
}(k_a\cdot k_b)\Sl_{\pmb{\sigma}}A\bigl(1,\pmb{\sigma}\in(\pmb{\zeta}\shuffle\pmb{\gamma})\big|_{\tiny({\sigma})^{-1}(b)<({\sigma})^{-1}(a)},r\bigr)\biggr]\biggr\}.\nonumber
\eea
As we have shown in examples in \secref{sec:DirectEvalutations}, when we collect coefficients for any given permutation $\pmb{\sigma}\in\pmb{\gamma}\shuffle\pmb{\zeta}$ together,
the expression in the square brackets in \eqref{Eq:GaugeInv9} is given by
\bea
(k_a\cdot Y_a(\pmb{\sigma}))A\bigl(1,\pmb{\sigma}\in(\pmb{\zeta}\shuffle\pmb{\gamma}),r\bigr).~~\Label{Eq:GaugeInv10}
\eea
Here $Y_a(\pmb{\sigma})$ is the sum of all momenta of elements in $\pmb{\gamma}$ (the element $1$ is also included) appearing on the LHS of the node $a$ in $\pmb{\sigma}$.
%The phase factor $(-1)^{\tiny\mathcal{N}(\mathscr{L})}$ for a given configuration of the final lower block $\mathscr{L}$ is independent of the choice of node $b$, thus it can be %extracted as a total factor in \eqref{Eq:GaugeInv9}.
The factor $(-1)^{\tiny\mathcal{N}(\mathscr{U}_a)+\mathcal{S}(\mathscr{U}_a)+1}$ in \eqref{Eq:GaugeInv9} depends on the choice of $a$. Once the node $a\in\mathscr{U}$ (i.e. the node nearest to root in $\mathscr{U}$) under the first summation in the braces of \eqref{Eq:GaugeInv9} is fixed, both the number $\mathcal{N}(\mathscr{U}_a)$ of arrows pointing deviate from the direction of root and the number $\mathcal{S}(\mathscr{U}_a)$ of spurious components in $\mathscr{U}$ are fixed. To analyze the relation between signs corresponding to distinct choices of $a$, we consider two adjacent nodes $c_1,c_2\in\mathscr{U}$ which can be connected via all three types of lines:
\begin{itemize}
\item [(i)] If the line between $c_1$ and $c_2$ is a type-2 line, as shown by \figref{Fig:Sign} (a), the factors for choosing  $a=c_1$ and  $a=c_2$ are related by flipping the sign, i.e. $(-1)^{\mathcal{N}(\mathscr{U}_{c_1})}=(-1)^{\mathcal{N}(\mathscr{U}_{c_2})\pm 1}$ because the difference of the numbers of all arrows deviating from root for these two choices is $1$.
\item [(ii)] If $c_1$ and $c_2$ are two ends of a type-1 line or a type-3 line, as shown by \figref{Fig:Sign} (b) and (c), the number of arrows that deviate from root are same for $a=c_1$ and  $a=c_2$. Thus  $(-1)^{\mathcal{N}(\mathscr{U}_{c_1})}=(-1)^{\mathcal{N}(\mathscr{U}_{c_2})}$. Nevertheless, we should also count the number $\mathcal{S}(\mathscr{U})$ of spurious chains in a (spurious) graph. Assume that the distance between $c_1$ and the (highest weight) node $h_{\rho(s)}$ is always larger than that between  $c_2$ and  $h_{\rho(s)}$. If $c_1$, $c_2$ are two end nodes of a type-1 (or type-3) line,  the numbers of spurious components $\mathcal{S}(\mathscr{U})$ for these two cases always differ by $1$. Thus the factors $(-1)^{\mathcal{S}(\mathscr{U}_a)}$ for choosing $a=c_1$ and $a=c_2$ always have the opposite sign.
\end{itemize}
To sum up, the factors   $(-1)^{\tiny\mathcal{N}(\mathscr{U}_a)+\mathcal{S}(\mathscr{U}_a)+1}$ associated with graphs corresponding to $a=c_1$ and $a=c_2$, where $c_1,c_2\in\mathcal{U}$ are two adjacent nodes connected by an arbitrary type of line, always differ by a factor $(-1)$. Replacing the expression in the square brackets of \eqref{Eq:GaugeInv9} by  \eqref{Eq:GaugeInv10} and considering the relative signs, we rewrite \eqref{Eq:GaugeInv10} as
\bea
I\left[\mathscr{U},\mathscr{L}\right]
=(-1)^{\tiny\mathcal{N}(\mathscr{L})+\mathcal{N}(\mathscr{U}_c)+\mathcal{S}(\mathscr{U}_c)+1}\Sl_{\pmb{\gamma}}\biggl[\Sl_{a\in \mathscr{U}
}f^a\Sl_{\pmb{\zeta}\in{\mathscr{U}}|_{a}}\Sl_{\pmb{\sigma}}(k_a\cdot Y_a(\pmb{\sigma}))A\bigl(1,\pmb{\sigma}\in(\pmb{\zeta}\shuffle\pmb{\gamma}),r\bigr)\biggr],\Label{Eq:GaugeInv11}
\eea
where the factor $(-1)^{\tiny\mathcal{N}(\mathscr{L})+\mathcal{N}(\mathscr{U}_c)+\mathcal{S}(\mathscr{U}_c)+1}$ for a fixed $c\in\mathscr{U}$  \eqref{Eq:GaugeInv9} has been exacted out. The relative signs $f^a$ for choosing $a$ as other nodes can be fully fixed because the factor for any two adjacent choices of $a$ must be differ by a factor $(-1)$. In the next section, we will prove that for any given $\pmb{\gamma}$, the expression in the brackets of \eqref{Eq:GaugeInv11} must vanish because it can always be written as a combination of BCJ relations.

%%%%%%%%%%%%%%%%%%%%%%%%%%%%%%%%%%%%%%%%%%%%%%%%%%%%%%%%%%%%%%%%%%%%%%%%%%%%%
\section{Graph-based BCJ relation as a combination of traditional BCJ relations}\label{sec:GraphBasedBCJ}
%%%%%%%%%%%%%%%%%%%%%%%%%%%%%%%%%%%%%%%%%%%%%%%%%%%%%%%%%%%%%%%%%%%%%%%%%%%%%%

In this section, we introduce the following \emph{graph-based BCJ relation}\footnote{Similar relations have also been discussed by Chen, Johanssion, Teng and Wang \cite{ChenToAppear} via kinematic algebra. }
\bea
\Sl_{a\in \mathcal{T}}f^{a}\Sl_{\pmb{\zeta}\in{\mathcal{T}}|_a}\Sl_{\pmb{\sigma}}\left[k_{a}\cdot Y_{a}(\pmb{\sigma})\right]\,A\left(1,\pmb{\sigma}\in(\pmb{\zeta}\shuffle\,\pmb{\gamma}),r\right)=0.~~\Label{Eq:Graph-based-BCJ}
\eea
Here, $\pmb{\gamma}$ is an arbitrary permutation of elements in $\{2,\dots,r-1\}$ and $\mathcal{T}$ is an arbitrary connected tree graph. We use $\mathcal{T}|_a$ to denote the relative orders between nodes of $\mathcal{T}$ when choosing the node $a$ as the leftmost one (see \secref{section:SumPhySpu}).
For a given graph $\mathcal{T}$, the factor $f^a$ is a relative sign depending on the choice of $a$. This factor is fixed as follows: (i) Choose an arbitrary node $c$ and require $f^c=1$, (ii) For arbitrary two adjacent nodes $c_1$ and $c_2$, we have $f^{c_1}=-f^{c_2}$. In the following, we use $\mathcal{B}^{\,c}_G(1\,|\,\mathcal{T},\pmb{\gamma}\,|\,r)$ to stand for the LHS of \eqref{Eq:Graph-based-BCJ} with choosing $f^{c}=1$ ($c\in \mathcal{T}$).
We will prove that the $\mathcal{B}^{\,c}_G(1\,|\,\mathcal{T},\pmb{\gamma}\,|\,r)$ (thus \eqref{Eq:GaugeInv11} and \eqref{Eq:GaugeInv6}) is a combination of the LHS of traditional BCJ relations \eqref{Eq:BCJRelation} (see \eqref{Eq:LHSBCJ}). As a result, the gauge invariance induced identity \eqref{Eq:GaugeInv2} can always be expanded in terms of traditional BCJ relations.

%%%%%%%%%%%%%%%%%%%%%%
\subsection{Examples}
%%%%%%%%%%%%%%%%%%%%%%
Now we present several examples for the graph-based BCJ relation \eqref{Eq:Graph-based-BCJ}.
\newline
{\bf Example-1}
\newline
The simplest example is that the tree graph $\mathcal{T}$ consists of only a single node $h_1$. The LHS of \eqref{Eq:Graph-based-BCJ} is nothing but the LHS of a fundamental BCJ relation
\bea
\mathcal{B}^{\,h_1}_G\big(1\big|\includegraphics[width=0.2cm]{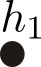},\pmb{\gamma}\big|r\big)=\mathcal{B}(1|\{h_1\},\pmb{\gamma}|r).\Label{Eq:GraphBaseEG1}
\eea
%\newline
{\bf Example-2}
\newline
The next simplest example is that $\mathcal{T}$ consists of two nodes $h_1$ and $h_2$ with one (dashed) line between them. The LHS of  \eqref{Eq:Graph-based-BCJ} in this case reads
\bea
\mathcal{B}^{h_1}_G\big(1\big|
\includegraphics[width=0.4in]{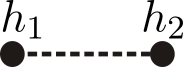},\pmb{\gamma}\big|r\big)=\Sl_{\shuffle}(k_{h_1}\cdot Y_{h_1}(\shuffle))A(1,\{h_1,h_2\}\shuffle \pmb{\gamma},r)-\Sl_{\shuffle}(k_{h_2}\cdot Y_{h_2}(\shuffle))A(1,\{h_2,h_1\}\shuffle \pmb{\gamma},r).\Label{Eq:GraphBaseEG2}
\eea
Notice that the last term can be replaced by
\bea
\mathcal{B}^{h_2}_G\big(1\big|\includegraphics[width=0.1915cm]{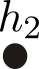},\{h_1\}\shuffle\pmb{\gamma}\big|r\big)-\Sl_{\shuffle}(k_{h_2}\cdot X_{h_2}(\shuffle))A(1,\{h_1,h_2\}\shuffle \pmb{\gamma},r).
\eea
The second term of the above equation together with the first term of \eqref{Eq:GraphBaseEG2} produces the LHS of the traditional BCJ relation \eqref{Eq:BCJRelation} with $\pmb{\beta}=\{h_1,h_2\}$, i.e.  $\mathcal{B}(1\,|\,\{h_1,h_2\},\pmb{\gamma}\,|\,r)$. Thus \eqref{Eq:GraphBaseEG2} is finally given by
\bea
\mathcal{B}^{h_1}_G\big(1\big|
\includegraphics[width=0.4in]{GraphBasedEG2.jpg},\pmb{\gamma}\big|r\big)&=&\mathcal{B}(1|\{h_1,h_2\},\pmb{\gamma}|r)-\mathcal{B}^{h_2}_G\big(1\big|\includegraphics[width=0.1915cm]{GraphBasedEG3.jpg},\{h_1\}\shuffle\pmb{\gamma}\big|r\big)=\mathcal{B}(1|\{h_1,h_2\},\pmb{\gamma}|r)-\mathcal{B}(1|\{h_2\},\{h_1\}\shuffle\pmb{\gamma}|r),\Label{Eq:GraphBaseEG2-1}\nn
\eea
which is a combination of the LHS of traditional BCJ relations. It is worth pointing out that the above expression is not unique.
When exchanging the roles of $h_1$ and $h_2$ in \eqref{Eq:GraphBaseEG2-1}, we get another equivalent expansion
\bea
\mathcal{B}^{h_2}_G\big(1\big|
\includegraphics[width=0.4in]{GraphBasedEG2.jpg},\pmb{\gamma}\big|r\big)=\mathcal{B}(1|\{h_2,h_1\},\pmb{\gamma}|r)+\mathcal{B}(1|\{h_1\},\{h_2\}\shuffle\pmb{\gamma}|r)=-\mathcal{B}^{h_1}_G\big(1\big|
\includegraphics[width=0.4in]{GraphBasedEG2.jpg},\pmb{\gamma}\big|r\big).\Label{Eq:GraphBaseEG2-2}
\eea
%\newline
{\bf Example-3}
\newline
Now we consider a graph with three nodes
%$\{h_1,h_2,h_3\}$, the only possible structure is a simple chain with three nodes on it. Since our discussion does not rely on the labeling of nodes, we now consider the tree graph %
$\mathcal{T}=\includegraphics[width=1.6cm]{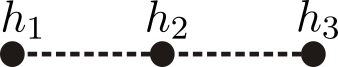}$. The LHS of \eqref{Eq:Graph-based-BCJ} for this graph reads
\bea
&&\mathcal{B}^{h_1}_G\big(1\big|
\includegraphics[width=1.6cm]{GraphBasedEG4.jpg},\pmb{\gamma}\big|r\big)\Label{Eq:GraphBaseEG3}\\
&=&\Sl_{\shuffle}(k_{h_1}\cdot Y_{h_1}(\shuffle))A(1,\{h_1,h_2,h_3\}\shuffle \pmb{\gamma},r)-\Bigl[\Sl_{\shuffle_1,\shuffle_2}(k_{h_2}\cdot Y_{h_2}(\shuffle_1,\shuffle_2))A(1,\{h_2,\{h_1\}\shuffle_1\{h_3\}\}\shuffle_2 \pmb{\gamma},r)\nn
&&-\Sl_{\shuffle}(k_{h_3}\cdot Y_{h_3}(\shuffle))A(1,\{h_3,h_2,h_1\}\shuffle \pmb{\gamma},r)\Bigr].\nonumber
\eea
The expression in the square brackets can be considered as $\Sl_{\shuffle}\mathcal{B}^{h_2}_G\big(1\big|
\includegraphics[width=0.4in]{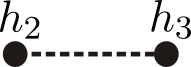},\{h_1\}\shuffle\pmb{\gamma}\big|r\big)\big|_{h_2\prec h_1}$, where $h_2\prec h_1$ for a given permutation $\pmb{\xi}$ means that $\xi^{-1}(h_2)<\xi^{-1}(h_1)$. When the first equality of \eqref{Eq:GraphBaseEG2-1} in example-2 is applied, $\Sl_{\shuffle}\mathcal{B}^{h_2}_G\big(1\big|
\includegraphics[width=0.4in]{GraphBasedEG5.jpg},\{h_1\}\shuffle\pmb{\gamma}\big|r\big)\big|_{h_2\prec h_1}$ turns to
\bea
&&\Sl_{\shuffle}\mathcal{B}^{h_2}_G\big(1\big|
\includegraphics[width=0.4in]{GraphBasedEG5.jpg},\{h_1\}\shuffle\pmb{\gamma}\big|r\big)\big|_{h_2\prec h_1}\\
&=&\Sl_{\shuffle}\mathcal{B}(1|\{h_2,h_3\},\{h_1\}\shuffle\pmb{\gamma}|r)|_{h_2\prec h_1}-\Sl_{\shuffle}\mathcal{B}^{h_3}_G\big(1\big|\includegraphics[width=0.2cm]{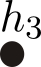},\{h_2,h_1\}\shuffle\pmb{\gamma}\big|r\big)\nn
&=&\Sl_{\shuffle}\mathcal{B}(1|\{h_2,h_3\},\{h_1\}\shuffle\pmb{\gamma}|r)-\mathcal{B}(1|\{h_2,h_3\},\{h_1\}\shuffle\pmb{\gamma}|r)|_{h_1\prec h_2}-\Sl_{\shuffle}\mathcal{B}^{h_3}_G\big(1\big|\includegraphics[width=0.2cm]{GraphBasedEG9.jpg},\{h_2,h_1\}\shuffle\pmb{\gamma}\big|r\big).\nonumber
\eea
The first term and the third term on the last line are just a combination of LHS of BCJ relations with $\pmb{\beta}=\{h_2,h_3\}$ and  $\pmb{\beta}=\{h_3\}$ respectively. The sum of the second term on the last line of the above equation and the first term in \eqref{Eq:GraphBaseEG3} is nothing but the LHS of traditional BCJ relation \eqref{Eq:BCJRelation} with $\pmb{\beta}=\{h_1,h_2,h_3\}$. Hence \eqref{Eq:GraphBaseEG3} is finally expanded into the following two parts
\bea
&&\mathcal{B}^{h_1}_G\big(1\big|
\includegraphics[width=1.6cm]{GraphBasedEG4.jpg},\pmb{\gamma}\big|r\big)=\mathcal{B}^{{(1)}}_G\big(1\big|
\includegraphics[width=1.6cm]{GraphBasedEG4.jpg}\big|_{h_1},\pmb{\gamma}\big|r\big)+\Sl_{\shuffle}\mathcal{B}^{{(2)}}_G\big(1\big|
\includegraphics[width=0.4in]{GraphBasedEG5.jpg},\{h_1\}\shuffle\pmb{\gamma}\big|r\big)\Label{Eq:GraphBaseEG3-1},
\eea
where $\mathcal{B}^{{(1)}}_G\big(1\big|
\includegraphics[width=1.6cm]{GraphBasedEG4.jpg}\big|_{h_1},\pmb{\gamma}\big|r\big)$ and $\Sl_{\shuffle}\mathcal{B}^{{(2)}}_G\big(1\big|
\includegraphics[width=0.4in]{GraphBasedEG5.jpg},\{h_1\}\shuffle\pmb{\gamma}\big|r\big)$ are defined by
\bea
\mathcal{B}^{{(1)}}_G\big(1\big|
\includegraphics[width=1.6cm]{GraphBasedEG4.jpg}\big|_{h_1},\pmb{\gamma}\big|r\big)&\equiv&\mathcal{B}(1|\{h_1,h_2,h_3\},\pmb{\gamma}|r),\nn
\Sl_{\shuffle}\mathcal{B}^{{(2)}}_G\big(1\big|
\includegraphics[width=0.4in]{GraphBasedEG5.jpg},\{h_1\}\shuffle\pmb{\gamma}\big|r\big)&\equiv&\Sl_{\shuffle}\big[-\mathcal{B}(1|\{h_2,h_3\},\{h_1\}\shuffle\pmb{\gamma}|r)+\mathcal{B}(1|\{h_3\},\{h_2,h_1\}\shuffle\pmb{\gamma}|r)\big].
\eea
 The expression of $\mathcal{B}^{{(1)}}_G\big(1\big|
\includegraphics[width=1.6cm]{GraphBasedEG4.jpg}\big|_{h_1},\pmb{\gamma}\big|r\big)$ is the LHS of \eqref{Eq:BCJRelation} where all three nodes $h_1$, $h_2$ and $h_3$ are in the $\pmb{\beta}$ set, particularly $\pmb{\beta}=\{h_1,h_2,h_3\}=\includegraphics[width=1.6cm]{GraphBasedEG4.jpg}|_{h_1}$. The expression of $\Sl_{\shuffle}\mathcal{B}^{{(2)}}_G\big(1\big|
\includegraphics[width=0.4in]{GraphBasedEG5.jpg},\{h_1\}\shuffle\pmb{\gamma}\big|r\big)$  is a combination of BCJ relations with fewer ${\beta}$'s. The node $h_1$ is always considered as an element of the $\pmb{\alpha}$ set in the expression of $\Sl_{\shuffle}\mathcal{B}^{{(2)}}_G\big(1\big|
\includegraphics[width=0.4in]{GraphBasedEG5.jpg},\{h_1\}\shuffle\pmb{\gamma}\big|r\big)$.

Similar with example-2, $\mathcal{B}^{h_1}_G\big(1\big|
\includegraphics[width=1.6cm]{GraphBasedEG4.jpg},\pmb{\gamma}\big|r\big)$ can also be expanded in other equivalent forms. For example, we consider $\mathcal{B}^{h_2}_G\big(1\big|
\includegraphics[width=1.6cm]{GraphBasedEG4.jpg},\pmb{\gamma}\big|r\big)$ which is expressed by \eqref{Eq:GraphBaseEG3} associated with a minus sign. We keep the first term (the term where $h_2$ is considered as the leftmost node in the graph $\mathcal{T}$) in the square brackets  of \eqref{Eq:GraphBaseEG3} and rewrite other two terms according to \eqref{Eq:GraphBaseEG1} as follows
\bea
&&\Sl_{\shuffle}(k_{h_1}\cdot Y_{h_1}(\shuffle))A(1,\{h_1,h_2,h_3\}\shuffle \pmb{\gamma},r)\nn
&=&\Sl_{\shuffle}\mathcal{B}^{\,h_1}_G\big(1\big|\includegraphics[width=0.2cm]{GraphBasedEG1.jpg},\{h_2,h_3\}\shuffle\pmb{\gamma}\big|r\big)-\Sl_{\shuffle_1,\shuffle_2}(k_{h_1}\cdot X_{h_1})A(1,\{h_2,\{h_1\}\shuffle_1\{h_3\}\}\shuffle_2 \pmb{\gamma},r),\nn
&&\Sl_{\shuffle}(k_{h_3}\cdot Y_{h_3}(\shuffle))A(1,\{h_3,h_2,h_1\}\shuffle \pmb{\gamma},r)\nn
&=&\Sl_{\shuffle}\mathcal{B}^{\,h_3}_G\big(1\big|\includegraphics[width=0.2cm]{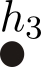},\{h_2,h_1\}\shuffle\pmb{\gamma}\big|r\big)-\Sl_{\shuffle_1,\shuffle_2}(k_{h_3}\cdot X_{h_3})A(1,\{h_2,\{h_1\}\shuffle_1\{h_3\}\}\shuffle_2 \pmb{\gamma},r).
\eea
Thus $\mathcal{B}^{h_2}_G\big(1\big|
\includegraphics[width=1.6cm]{GraphBasedEG4.jpg},\pmb{\gamma}\big|r\big)=-\mathcal{B}^{h_1}_G\big(1\big|
\includegraphics[width=1.6cm]{GraphBasedEG4.jpg},\pmb{\gamma}\big|r\big)$ is written as
\bea
\mathcal{B}^{h_2}_G\big(1\big|\includegraphics[width=1.6cm]{GraphBasedEG4.jpg},\pmb{\gamma}\big|r\big)&=&\mathcal{B}^{{{(1)}}}_G\big(1\big|
\includegraphics[width=1.6cm]{GraphBasedEG4.jpg}\big|_{h_2},\pmb{\gamma}\big|r\big)+\Sl_{\shuffle}\mathcal{B}^{{{(2)}}}_G\big(1\big|
\includegraphics[width=0.2cm]{GraphBasedEG6.jpg}~~~\includegraphics[width=0.2cm]{GraphBasedEG1.jpg},\{h_2\}\shuffle\pmb{\gamma}\big|r\big),\Label{Eq:GraphBaseEG3-2}
\eea
where
\bea
\mathcal{B}^{{{(1)}}}_G\big(1\big|
\includegraphics[width=1.6cm]{GraphBasedEG4.jpg}\big|_{h_2},\pmb{\gamma}\big|r\big)&\equiv&\Sl_{\shuffle}\mathcal{B}(1|\{h_2,\{h_1\}\shuffle\{h_3\}\},\pmb{\gamma}|r),\nn
\Sl_{\shuffle}\mathcal{B}^{{{(2)}}}_G\big(1\big|
\includegraphics[width=0.2cm]{GraphBasedEG6.jpg}~~~\includegraphics[width=0.2cm]{GraphBasedEG1.jpg},\{h_2\}\shuffle\pmb{\gamma}\big|r\big)&\equiv&-\left[\Sl_{\shuffle}\mathcal{B}^{\,h_2}_G\big(1\big|\includegraphics[width=0.2cm]{GraphBasedEG1.jpg},\{h_2,h_3\}\shuffle\pmb{\gamma}\big|r\big)+\Sl_{\shuffle}\mathcal{B}^{\,h_3}_G\big(1\big|\includegraphics[width=0.2cm]{GraphBasedEG6.jpg},\{h_2,h_1\}\shuffle\pmb{\gamma}\big|r\big)\right].\Label{Eq:GraphBaseEG3-3}
\eea
In \eqref{Eq:GraphBaseEG3-2}, $\mathcal{B}^{{{(1)}}}_G\big(1\big|
\includegraphics[width=1.6cm]{GraphBasedEG4.jpg}\big|_{h_2},\pmb{\gamma}\big|r\big)$ is just the sum of the LHS of BCJ relations \eqref{Eq:BCJRelation} with $\pmb{\beta}\in\{h_2,h_1\shuffle h_3\}$.  The term $\Sl_{\shuffle}\mathcal{B}^{{{(2)}}}_G\big(1\big|
\includegraphics[width=0.2cm]{GraphBasedEG6.jpg}~~~\includegraphics[width=0.2cm]{GraphBasedEG1.jpg},\{h_2\}\shuffle\pmb{\gamma}\big|r\big)$ in \eqref{Eq:GraphBaseEG3-2} is a combination of the LHS of graph-based BCJ relations with one-node subtrees (here, $h_2$ is in the $\pmb{\alpha}$ set), thus it is also a combination of traditional BCJ relations (according to \eqref{Eq:GraphBaseEG1}).
\newline
\newline
{\bf Example-4}
\newline
The first example of graphs with complex chains is the star graph with four nodes. The LHS of the graph-based-BCJ  relation \eqref{Eq:Graph-based-BCJ} for this example reads:
\bea
&&\mathcal{B}^{h_1}_G\left(1\left|{\begin{minipage}{1.6cm}
\includegraphics[width=1.6cm]{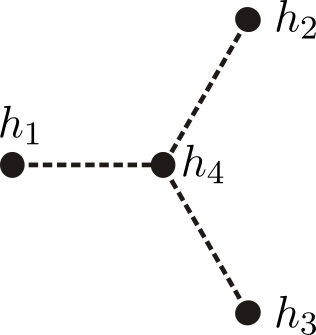} \end{minipage} }\,,\,\pmb{\gamma}\right|r\right)\Label{Eq:GraphBaseEG4}\\
&=&\Sl_{\shuffle_1,\shuffle_2}(k_{h_1}\cdot Y_{h_1})A(1,\{h_1,h_4,\{h_2\}\shuffle_1\{h_3\}\}\shuffle_2\pmb{\gamma},r)-\Sl_{\shuffle}\mathcal{B}^{h_4}_G\big(1\big|
\includegraphics[width=1.6cm]{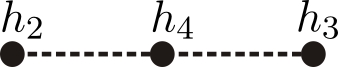},\{h_1\}\shuffle\pmb{\gamma}\big|r\big)\big|_{h_4\prec h_1}.\nonumber
\eea
According to \eqref{Eq:GraphBaseEG3-2}, the last term is given by
\bea
&&\Sl_{\shuffle}\mathcal{B}^{h_4}_G\big(1\big|
\includegraphics[width=1.6cm]{GraphBasedEG8.jpg},\{h_1\}\shuffle\pmb{\gamma}\big|r\big)\big|_{h_4\prec h_1}\Label{Eq:GraphBaseEG4-1}\\
&=&\Sl_{\shuffle}\mathcal{B}^{(1)}_G\big(1\big|
\includegraphics[width=1.6cm]{GraphBasedEG8.jpg}\big|_{h_4},\{h_1\}\shuffle\pmb{\gamma}\big|r\big)\big|_{h_4\prec h_1}
+\Sl_{\shuffle_1,\shuffle_2}\mathcal{B}^{(2)}_G\big(1\big|
\includegraphics[width=0.1915cm]{GraphBasedEG3.jpg}~~~\includegraphics[width=0.2cm]{GraphBasedEG6.jpg},\{h_4\}\shuffle_1\{\{h_1\}\shuffle_2\pmb{\gamma}\}\big|r\big)\big|_{h_4\prec h_1}.\nonumber
\eea
From \eqref{Eq:GraphBaseEG3-3}, we know that the second term of the above equation is expressed in terms of the LHS of those BCJ relations where  $h_4$ belongs to the $\pmb{\alpha}$ set. Since $h_1$ is also in the $\pmb{\alpha}$ set (of the traditonal BCJ expansion) of the second term, $\Sl_{\shuffle_1,\shuffle_2}\mathcal{B}^{(2)}_G\big(1\big|
\includegraphics[width=0.1915cm]{GraphBasedEG3.jpg}~~~\includegraphics[width=0.2cm]{GraphBasedEG6.jpg},\{h_4\}\shuffle_1\{\{h_1\}\shuffle_2\pmb{\gamma}\}\big|r\big)\big|_{h_4\prec h_1}$ is already a combination of the LHS of BCJ relations where both $h_4$ and $h_1$ belong to  $\pmb{\alpha}$ set and satisfy $h_4\prec h_1$. The first term in \eqref{Eq:GraphBaseEG4-1} reads
\bea
%&&\Sl_{\shuffle}\left[\mathcal{B}^{h_4}_G\big(1\big|
%\includegraphics[width=1.6cm]{GraphBasedEG8.jpg},\{h_1\}\shuffle\pmb{\rho}\big|r\big)\right]_{\text{I}}\Big|_{h_4\prec h_1}\Label{Eq:GraphBaseEG4-2}\\
%
%&=&
\Sl_{\shuffle_1,\shuffle_2}\mathcal{B}(1|\{h_4,\{h_2\}\shuffle_1\{h_3\}\},\{h_1\}\shuffle_2\pmb{\gamma}|r)-\Sl_{\shuffle_1,\shuffle_2}\mathcal{B}(1|\{h_4,\{h_2\}\shuffle_1\{h_3\}\},\{h_1\}\shuffle_2\pmb{\gamma}|r)|_{h_1\prec h_4}.\Label{Eq:GraphBaseEG4-2}
\eea
The sum of the last term in \eqref{Eq:GraphBaseEG4-2} and  the first term in \eqref{Eq:GraphBaseEG4} is
\bea
\mathcal{B}^{(1)}_G\left(1\left|{\begin{minipage}{1.6cm}
\includegraphics[width=1.6cm]{GraphBasedEG7.jpg} \end{minipage} }\,\Bigg|_{h_1},\,\pmb{\gamma}\right|r\right)
\equiv \Sl_{\pmb{\zeta}}\mathcal{B}\left(1\left|\pmb{\zeta}\in{\begin{minipage}{1.6cm}
\includegraphics[width=1.6cm]{GraphBasedEG7.jpg} \end{minipage} }\,\Bigg|_{h_1},\pmb{\gamma}\right|r\right)=\Sl_{\shuffle}\mathcal{B}(1|\{h_1,h_4,\{h_2\}\shuffle\{h_3\}\},\pmb{\gamma}|r),\nn
\eea
which is a combination of the LHS of traditional BCJ relations with all the four nodes belonging to the $\pmb{\beta}$ set. The sum of the first term in \eqref{Eq:GraphBaseEG4-2} and the last term in \eqref{Eq:GraphBaseEG4-1} defines
\bea
\Sl_{\shuffle}\mathcal{B}^{(2)}_G\left(1\left|\includegraphics[width=1.6cm]{GraphBasedEG8.jpg},\{h_1\}\shuffle\pmb{\gamma}\right|r\right)
&\equiv&-\Sl_{\shuffle_1,\shuffle_2}\mathcal{B}^{(2)}_G\big(1\big|
\includegraphics[width=0.1915cm]{GraphBasedEG3.jpg}~~~\includegraphics[width=0.2cm]{GraphBasedEG6.jpg},\{h_4\}\shuffle_1\{\{h_1\}\shuffle_2\pmb{\gamma}\}\big|r\big)\big|_{h_4\prec h_1}\nn
&&-\Sl_{\shuffle}\mathcal{B}(1|\{h_4,\{h_2\}\shuffle\{h_3\}\},\{h_1\}\shuffle\pmb{\gamma}|r),
\eea
which is a combination of the LHS of traditional BCJ relations where $h_1$ is an element of $\pmb{\alpha}$ set.

The combination of amplitudes $\mathcal{B}^{h_1}_G\left(1\left|{\begin{minipage}{1.6cm}
\includegraphics[width=1.6cm]{GraphBasedEG7.jpg} \end{minipage} }\,,\,\pmb{\gamma}\right|r\right)=-\mathcal{B}^{h_4}_G\left(1\left|{\begin{minipage}{1.6cm}
\includegraphics[width=1.6cm]{GraphBasedEG7.jpg} \end{minipage} }\,,\,\pmb{\gamma}\right|r\right)$ can also be expressed in another way when we consider
\bea
&&\mathcal{B}^{h_4}_G\left(1\left|{\begin{minipage}{1.6cm}
\includegraphics[width=1.6cm]{GraphBasedEG7.jpg} \end{minipage} }\,,\,\pmb{\gamma}\right|r\right)\Label{Eq:GraphBaseEG4-3}\\
&=&\Sl_{\shuffle_1,\shuffle_2,\shuffle_3}(k_{h_4}\cdot Y_{h_4})A(1,\{h_4,\{h_1\}\shuffle_1\{h_2\}\shuffle_2\{h_3\}\}\shuffle_3\pmb{\gamma},r)\nn
&&-\Bigl[\Sl_{\shuffle_1,\shuffle_2}\mathcal{B}^{h_4}_G\big(1\big|
\includegraphics[width=0.2cm]{GraphBasedEG1.jpg},\{h_4,\{h_2\}\shuffle_1\{h_3\}\}\shuffle_2\pmb{\gamma}\big|r\big)\big|_{h_1\prec h_4}+\Sl_{\shuffle_1,\shuffle_2}\mathcal{B}^{h_4}_G\big(1\big|
\includegraphics[width=0.1915cm]{GraphBasedEG3.jpg},\{h_4,\{h_1\}\shuffle_1\{h_3\}\}\shuffle_2\pmb{\gamma}\big|r\big)\big|_{h_2\prec h_4}\nn
&&+\,\,\Sl_{\shuffle_1,\shuffle_2}\mathcal{B}^{h_4}_G\big(1\big|
\includegraphics[width=0.2cm]{GraphBasedEG6.jpg},\{h_4,\{h_1\}\shuffle_1\{h_2\}\}\shuffle_2\pmb{\gamma}\big|r\big)\big|_{h_3\prec h_4}\Bigr].
\eea
Applying \eqref{Eq:GraphBaseEG1}, we rewrite the last three terms of the above equation as follows
\bea
&&\Sl_{\shuffle_1,\shuffle_2}\mathcal{B}^{h_4}_G\big(1\big|
\includegraphics[width=0.2cm]{GraphBasedEG1.jpg},\{h_4,\{h_2\}\shuffle_1\{h_3\}\}\shuffle_2\pmb{\gamma}\big|r\big)\big|_{h_1\prec h_4}\\
&=&\Sl_{\shuffle_1,\shuffle_2}\mathcal{B}(1|\{h_1\},\{h_4,\{h_2\}\shuffle_1\{h_3\}\}\shuffle_2\pmb{\gamma}|r)-\Sl_{\shuffle_1,\shuffle_2}\mathcal{B}(1|\{h_1\},\{h_4,\{h_2\}\shuffle_1\{h_3\}\}\shuffle_2\pmb{\gamma}|r)\big|_{h_1\succ h_4},\nn
&&\Sl_{\shuffle_1,\shuffle_2}\mathcal{B}^{h_4}_G\big(1\big|
\includegraphics[width=0.1915cm]{GraphBasedEG3.jpg},\{h_4,\{h_1\}\shuffle_1\{h_3\}\}\shuffle_2\pmb{\gamma}\big|r\big)\big|_{h_2\prec h_4}\\
&=&\Sl_{\shuffle_1,\shuffle_2}\mathcal{B}(1|\{h_2\},\{h_4,\{h_1\}\shuffle_1\{h_3\}\}\shuffle_2\pmb{\gamma}|r)-\Sl_{\shuffle_1,\shuffle_2}\mathcal{B}(1|\{h_2\},\{h_4,\{h_1\}\shuffle_1\{h_3\}\}\shuffle_2\pmb{\gamma}|r)\big|_{h_2\succ h_4},\nn
&&\Sl_{\shuffle_1,\shuffle_2}\mathcal{B}^{h_4}_G\big(1\big|
\includegraphics[width=0.2cm]{GraphBasedEG6.jpg},\{h_4,\{h_1\}\shuffle_1\{h_2\}\}\shuffle_2\pmb{\gamma}\big|r\big)\big|_{h_3\prec h_4}\\
&=&\Sl_{\shuffle_1,\shuffle_2}\mathcal{B}(1|\{h_3\},\{h_4,\{h_1\}\shuffle_1\{h_2\}\}\shuffle_2\pmb{\gamma}|r)-\Sl_{\shuffle_1,\shuffle_2}\mathcal{B}(1|\{h_3\},\{h_4,\{h_1\}\shuffle_1\{h_2\}\}\shuffle_2\pmb{\gamma}|r)\big|_{h_3\succ h_4}.\nonumber
\eea
The sum of the last terms of the above equations and the first term of \eqref{Eq:GraphBaseEG4-3} gives rise
\bea
\mathcal{B}^{(1)}_G\left(1\left|{\begin{minipage}{1.6cm}
\includegraphics[width=1.6cm]{GraphBasedEG7.jpg} \end{minipage} }\,\Bigg|_{h_4},\,\pmb{\gamma}\,\right|\,r\right)&\equiv&\Sl_{\pmb{\zeta}}\mathcal{B}\left(1\,\left|\pmb{\zeta}\in\,{\begin{minipage}{1.6cm}
\includegraphics[width=1.6cm]{GraphBasedEG7.jpg} \end{minipage} }\,\Bigg|_{h_4},\pmb{\gamma}\right|r\right)\nn
&=&
\Sl_{\shuffle_1,\shuffle_2}\mathcal{B}(1|\{h_4,\{h_1\}\shuffle_1\{h_2\}\shuffle_2\{h_3\}\},\pmb{\gamma}|r),
\eea
which is a combination of the LHS of BCJ relations \eqref{Eq:BCJRelation} with all nodes are in the $\pmb{\beta}$ set. The sum of other contributions are collected as
\bea
&&\Sl_{\shuffle}\mathcal{B}^{(2)}_G\left(1\left|\includegraphics[width=0.2cm]{GraphBasedEG1.jpg}~~\,\includegraphics[width=0.1915cm]{GraphBasedEG3.jpg}~~~\includegraphics[width=0.2cm]{GraphBasedEG6.jpg}\,,\,\{h_4\}\shuffle\pmb{\gamma}\right|r\right)\nn
&\equiv&\Sl_{\shuffle_1,\shuffle_2}\mathcal{B}(1|\{h_1\},\{h_4,\{h_2\}\shuffle_1\{h_3\}\}\shuffle_2\pmb{\gamma}|r)+\Sl_{\shuffle_1,\shuffle_2}\mathcal{B}(1|\{h_2\},\{h_4,\{h_1\}\shuffle_1\{h_3\}\}\shuffle_2\pmb{\gamma}|r)\nn
&&+\Sl_{\shuffle_1,\shuffle_2}\mathcal{B}(1|\{h_3\},\{h_4,\{h_1\}\shuffle_1\{h_2\}\}\shuffle_2\pmb{\gamma}|r),
\eea
which is a combination of the LHS of BCJ relations  \eqref{Eq:BCJRelation} where $h_4$ always belongs to the $\pmb{\alpha}$ set.
%%%%%%%%%%%%%%%%%%%%%%%%%%%%%%%%%
\subsection{The general proof}
%%%%%%%%%%%%%%%%%%%%%%%%%%%%%%%
\begin{figure}
\centering
\includegraphics[width=0.5\textwidth]{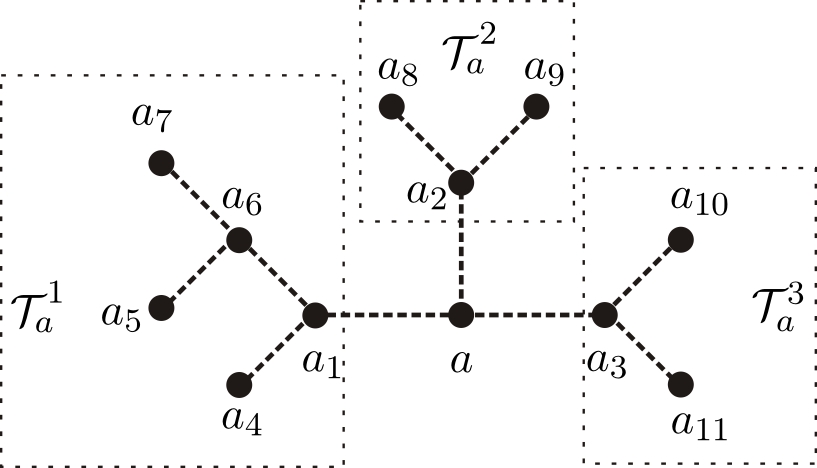}
\caption{A tree graph where $a$ is connected with three subtrees $\mathcal{T}_a^1$, $\mathcal{T}_a^2$ and $\mathcal{T}_a^3$.} \label{Fig:GraphBasedGEN1-1}
\end{figure}
In general, the LHS of graph-based BCJ relation \eqref{Eq:Graph-based-BCJ} for an arbitrary tree $\mathcal{T}$ can be expanded in terms of the LHS of BCJ relations \eqref{Eq:BCJRelation}.
Particularly, we pick out an arbitrary node $a\in \mathcal{T}$ and assume that there are $N$ subtrees attached to $a$, say $\mathcal{T}_a^{1}$, $\mathcal{T}_a^{2}$,...,$\mathcal{T}_a^{N}$. The corresponding nodes adjacent to $a$ in  $\mathcal{T}_a^{1}$, $\mathcal{T}_a^{2}$,...,$\mathcal{T}_a^{N}$ are
$a_1$, $a_2$, ..., $a_N$. We will prove that the LHS of  \eqref{Eq:Graph-based-BCJ} can be expanded as
\bea
&&~~~\mathcal{B}^{a}_G\left(1\left|{\begin{minipage}{1.8cm}
\includegraphics[width=2.6cm]{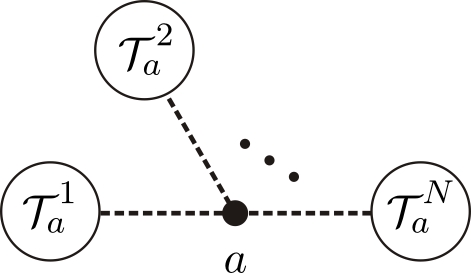} \end{minipage} }~~~~~~~,\pmb{\gamma}\right|r\right)\nn
&=&\mathcal{B}^{(1)}_G\left(1\left|{\begin{minipage}{1.8cm}
\includegraphics[width=2.6cm]{GraphBasedGEN1.jpg} \end{minipage} }~~~~~~~\Bigg|_a,\pmb{\gamma}\right|r\right)+\Sl_{\shuffle}\mathcal{B}^{(2)}_G\left(1\left|{\begin{minipage}{1.6cm}
\includegraphics[width=2.2cm]{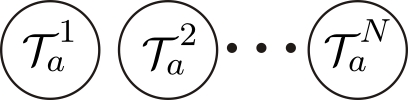} \end{minipage} }~~~~~,\{a\}\shuffle\pmb{\gamma}\right|r\right)\Label{Eq:GraphBaseGen}.
\eea
The first term of \eqref{Eq:GraphBaseGen} is the following combination
\bea
\mathcal{B}^{(1)}_G\left(1\left|{\begin{minipage}{1.8cm}
\includegraphics[width=2.6cm]{GraphBasedGEN1.jpg} \end{minipage} }~~~~~~~\Bigg|_a,\pmb{\gamma}\right|r\right)\equiv\Sl_{\pmb{\zeta}}\mathcal{B}\left(1\left|\pmb{\zeta}\in{\begin{minipage}{1.8cm}
\includegraphics[width=2.6cm]{GraphBasedGEN1.jpg} \end{minipage} }~~~~~~~\Bigg|_a,\pmb{\gamma}\right|r\right)
\eea
in which all nodes of $\mathcal{T}$ are in the $\pmb{\beta}$ set. The second term of \eqref{Eq:GraphBaseGen}
 is a proper combination of BCJ relations \eqref{Eq:Graph-based-BCJ} where the node $a$ is always in the $\pmb{\alpha}$ set. The explicit expression of this part can be determined recursively.  In the following, we will assume \eqref{Eq:GraphBaseGen} holds for all subtrees $\mathcal{T}_a^{1}$, $\mathcal{T}_a^{2}$, ...,$\mathcal{T}_a^{N}$ and prove \eqref{Eq:GraphBaseGen} by induction.

 To prove \eqref{Eq:GraphBaseGen}, we first show that $\mathcal{B}^{a}_G(1|\mathcal{T},\pmb{\gamma}|r)$ can be expressed by those $\mathcal{B}^{a}_G$'s with subtrees $\mathcal{T}_a^1,\dots,\mathcal{T}_a^N$ connecting to the node $a$. Let us explain this point explicitly by considering the example with  $\mathcal{T}=\text{\figref{Fig:GraphBasedGEN1-1}}$.
According to the definition, $\mathcal{B}^{a}_G\left(1\left|\mathcal{T}=\text{\figref{Fig:GraphBasedGEN1-1}},\pmb{\gamma}\right|r\right)$ can be written as
\bea
\mathcal{B}^{a}_G\left(1\left|\mathcal{T}=\text{\figref{Fig:GraphBasedGEN1-1}},\pmb{\gamma}\right|r\right)=\Big[\Sl_{\pmb{\sigma}}\Sl_{\pmb{\zeta}\in \mathcal{T}|_a}(k_a \cdot Y_a)+\Sl_{i=1}^3\Sl_{c\in \mathcal{T}_{a}^i}f^{c}\Sl_{\pmb{\sigma}}\Sl_{\pmb{\zeta}\in \mathcal{T}|_c}(k_c \cdot Y_c)\Big]A(1,\pmb{\sigma}\in(\pmb{\zeta}\shuffle \pmb{\gamma}),r).\Label{Eq:GraphBaseGenEG1}
 \eea
Here $a_1$, $a_2$ and $a_3$ are the nearest-to-$a$ nodes which belong to substructures $\mathcal{T}_a^1$, $\mathcal{T}_a^2$ and $\mathcal{T}_a^3$ correspondingly. The total contributions from nodes in $\mathcal{T}_{a}^1$ (i.e. contributions with $i=1$ in the second term of \eqref{Eq:GraphBaseGenEG1}) is given by
\bea
\Sl_{i\in \mathcal{T}_a^1}f^i\Sl_{\pmb{\zeta}^{(1)}\in \mathcal{T}_a^1|_{i}}\Bigl[\Sl_{\pmb{\zeta}^{(2)}\in\mathcal{T}_a^2|_{a_2}}\Sl_{\pmb{\zeta}^{(3)}\in\mathcal{T}_a^3|_{a_3}}\Sl_{\shuffle}\Sl_{\pmb{\sigma}}(k_i\cdot Y_i)A
\big(1,\pmb{\sigma}\in(\pmb{\zeta}^{(1)}\shuffle'\{a,\pmb{\zeta}^{(2)}\shuffle\pmb{\zeta}^{(3)}\shuffle\pmb{\gamma}\})\big|_{a_1\prec a},r\big)\Bigr],\Label{Eq:GraphBaseGenEG2}
\eea
where $\mathcal{T}_a^2|_{a_2}=\{a_2,\{a_8\}\shuffle\{a_9\}\}$,  $\mathcal{T}_a^3|_{a_3}=\{a_3,\{a_{10}\}\shuffle\{a_{11}\}\}$. Permutations $\mathcal{T}_a^1|_i$ in \eqref{Eq:GraphBaseGenEG2} are those established by the subgraph $\mathcal{T}_a^1$ when $i\in\mathcal{T}_a^1$ is considered as the leftmost element. Specifically,
\bea
\mathcal{T}_a^1|_{a_1}&=&\{a_1,\{a_4\}\shuffle\{a_6,\{a_5\}\shuffle\{a_7\}\}\},~\mathcal{T}_a^1|_{a_4}=\{a_4, a_1, a_6, \{a_5\}\shuffle\{a_7\}\}\\
\mathcal{T}_a^1|_{a_5}&=&\{a_5,a_6,\{a_7\}\shuffle\{a_1,a_4\}\},~\mathcal{T}_a^1|_{a_6}=\{a_6,\{a_5\}\shuffle\{a_7\}\shuffle\{a_1,a_4\}\},~\mathcal{T}_a^1|_{a_7}=\{a_7,a_6,\{a_5\}\shuffle\{a_1,a_4\}\}.\nonumber
\eea
For fixed permutations $\pmb{\zeta}^{(2)}$, $\pmb{\zeta}^{(3)}$ and a given permutation $\pmb{\xi}\in\{a,\pmb{\zeta}^{(2)}\shuffle\pmb{\zeta}^{(3)}\}\shuffle\pmb{\gamma}$, terms in \eqref{Eq:GraphBaseGenEG2} are collected as
\bea
\Sl_{i\in \mathcal{T}_a^1}f^i\Sl_{\pmb{\zeta}^{(1)}\in \mathcal{T}_a^1|_{i}}\Sl_{\pmb{\sigma}}(k_i\cdot Y_i)A\big(1,\pmb{\sigma}\in(\pmb{\zeta}^{(1)}\shuffle'\pmb{\xi})|_{a_1\prec a},r\big)=-\mathcal{B}^{a_1}_G\big(1\big|\mathcal{T}_a^1,\pmb{\xi}\big|r\big).
\eea
Thus the sum of all contributions from $\mathcal{T}_a^1$ in \eqref{Eq:GraphBaseGenEG2} reads (recall that $f^{a_1}=-f^a=-1$)
\bea
-\Sl_{\pmb{\zeta}^{(2)}\in\mathcal{T}_a^2|_{a_2}}\Sl_{\pmb{\zeta}^{(3)}\in\mathcal{T}_a^3|_{a_3}}\Sl_{\pmb{\xi}\in\{a,\pmb{\zeta}^{(2)}\shuffle\pmb{\zeta}^{(3)}\}\shuffle\pmb{\gamma}}\mathcal{B}^{a_1}_G\big(1\big|\mathcal{T}_a^1,\pmb{\xi}\big|r\big).\Label{Eq:GraphBaseGenEG3}
\eea
Following similar discussions, the sum of contributions from $\mathcal{T}_a^2$ and $\mathcal{T}_a^3$ are respectively given by the replacements of labels $1\leftrightarrow2$ and $1\leftrightarrow3$. The expression \eqref{Eq:GraphBaseGenEG1} is then expressed by terms with respect to subtree structures
\bea
&&\Sl_{\pmb{\sigma}}\Sl_{\pmb{\zeta}\in \mathcal{T}|_a}(k_a \cdot Y_a)A(1,\pmb{\sigma}\in(\pmb{\zeta}\shuffle \pmb{\gamma}),r)-\Bigl[\Sl_{\pmb{\zeta}^{(2)}\in\mathcal{T}_a^2|_{a_2}}\Sl_{\pmb{\zeta}^{(3)}\in\mathcal{T}_a^3|_{a_3}}\Sl_{\pmb{\xi}\in\{a,\pmb{\zeta}^{(2)}\shuffle\pmb{\zeta}^{(3)}\}\shuffle\pmb{\gamma}}\mathcal{B}^{a_1}_G\big(1\big|\mathcal{T}_a^1,\pmb{\xi}\big|r\big)\\
&+&\Sl_{\pmb{\zeta}^{(1)}\in\mathcal{T}_a^1|_{a_1}}\Sl_{\pmb{\zeta}^{(3)}\in\mathcal{T}_a^3|_{a_3}}\Sl_{\pmb{\xi}\in\{a,\pmb{\zeta}^{(1)}\shuffle\pmb{\zeta}^{(3)}\}\shuffle\pmb{\gamma}}\mathcal{B}^{a_2}_G\big(1\big|\mathcal{T}_a^2,\pmb{\xi}\big|r\big)
+\Sl_{\pmb{\zeta}^{(2)}\in\mathcal{T}_a^2|_{a_2}}\Sl_{\pmb{\zeta}^{(1)}\in\mathcal{T}_a^1|_{a_1}}\Sl_{\pmb{\xi}\in\{a,\pmb{\zeta}^{(2)}\shuffle\pmb{\zeta}^{(1)}\}\shuffle\pmb{\gamma}}\mathcal{B}^{a_3}_G\big(1\big|\mathcal{T}_a^3,\pmb{\xi}\big|r\big)\Bigr].\nonumber
\eea

 The above discussion is naturally generalized to arbitrary tree structure $\mathcal{T}$. For any $\mathcal{B}^{a}_G$, we have
\bea
&&~~~~\mathcal{B}^{a}_G\left(1\left|{\begin{minipage}{1.8cm}
\includegraphics[width=2.6cm]{GraphBasedGEN1.jpg} \end{minipage} }~~~~~~~,\pmb{\gamma}\right|r\right)\Label{Eq:GraphBaseGen1}\\
&=&\Sl_{\pmb{\sigma}}\Sl_{\pmb{\zeta}\in \mathcal{T}|_a}(k_a\cdot Y_a)A(1,\pmb{\sigma}\in(\pmb{\zeta}\shuffle\pmb{\gamma}),r)\nn
&-&\biggl[\Sl_{\substack{\pmb{\zeta}^{(j)}\in\mathcal{T}_a^{j}|_{a_j}\\\text{for}~j=2,\dots,N}}\Sl_{\shuffle}\mathcal{B}^{a_1}_G\bigl(1\big|
\mathcal{T}_a^{1},\bigl\{a, \bcancel{\pmb{\zeta}^{(1)}}\shuffle\pmb{\zeta}^{(2)}\shuffle\dots\shuffle\pmb{\zeta}^{(N)}\bigr\}\shuffle\pmb{\gamma}\big|r\bigr)\big|_{a_1\prec a}\nn
&&~~~~~~~+\Sl_{\substack{\pmb{\zeta}^{(j)}\in\mathcal{T}_a^{j}|_{a_j}\\\text{for}~j=1,3,\dots,N}}\Sl_{\shuffle}\mathcal{B}^{a_2}_G\bigl(1\big|
\mathcal{T}_a^{2},\bigl\{a, \pmb{\zeta}^{(1)}\shuffle\bcancel{\pmb{\zeta}^{(2)}}\shuffle\dots\shuffle\pmb{\zeta}^{(N)}\bigr\}\shuffle\pmb{\gamma}\big|r\bigr)\big|_{a_2\prec a}\nn
&+&~\dots~+\Sl_{\substack{\pmb{\zeta}^{(j)}\in\mathcal{T}_a^{j}|_{a_j}\\\text{for}~j=1,\dots,N-1}}\Sl_{\shuffle}\mathcal{B}^{a_N}_G\bigl(1\big|
\mathcal{T}_a^{N},\bigl\{a,\pmb{\zeta}^{(1)}\shuffle\pmb{\zeta}^{(2)}\shuffle\dots\shuffle\bcancel{\pmb{\zeta}^{(N)}}\bigr\}\shuffle\pmb{\gamma}\big|r\bigr)\big|_{a_N\prec a}\biggr],\nonumber
\eea
where we wrote the sum over multi-shuffle permutations by $\Sl_{\shuffle}$ for short. According to the inductive assumption, each term in the square brackets for a given $i$ can be expressed by the sum of the following two terms:
\bea
\Sl_{\substack{\pmb{\zeta}^{(j)}\in\mathcal{T}_a^{j}|_{a_j}\\\text{for}~j=1,\dots,N,j\neq i}}\Sl_{\pmb{\xi}^{(i)}}\mathcal{B}^{(1)}_G\bigl(1\big|
\mathcal{T}_a^{i}\big|_{a_i},\pmb{\xi}^{(i)}\big|r\bigr)\big|_{a_i\prec a}&=&\Sl_{\substack{\pmb{\zeta}^{(j)}\in\mathcal{T}_a^{j}|_{a_j}\\\text{for}~j=1,\dots,N}}\Sl_{\pmb{\xi}^{(i)}}\biggl[\mathcal{B}\bigl(1\big|\pmb{\zeta}^{(i)},\pmb{\xi}^{(i)}\big|r\bigr)-\mathcal{B}\bigl(1\big|\pmb{\zeta}^{(i)},\pmb{\xi}^{(i)}\big|r\bigr)\big|_{a_i\succ a}\biggr]\Label{Eq:GraphBaseGen2}\nn
\eea
 and
\bea
\Sl_{\substack{\pmb{\zeta}^{(j)}\in\mathcal{T}_a^{j}|_{a_j}\\\text{for}~j=1,\dots,N,j\neq i}}\Sl_{\pmb{\xi}^{(i)}}\biggl[\Sl_{\shuffle}\mathcal{B}^{(2)}_G\bigl(1\big|
\mathcal{T}_a^{i}-a_i,\{a_i\}\shuffle\pmb{\xi}^{(i)}\big|r\bigr)\big|_{a_i\prec a}\biggr].\Label{Eq:GraphBaseGen3}
\eea
In  \eqref{Eq:GraphBaseGen2} and \eqref{Eq:GraphBaseGen3}, the summations over $\pmb{\xi}^{(i)}$ are taken by summing over all
\bea
\pmb{\xi}^{(i)}\in\big\{\bigl\{{a,\pmb{\zeta}^{(1)}\shuffle\dots\shuffle\bcancel{\pmb{\zeta}^{(i)}}\shuffle\dots\shuffle\pmb{\zeta}^{(N)}}\bigr\}\shuffle\pmb{\gamma}\big\},\Label{Eq:GraphBaseGen4}
\eea
Eq. (\ref{Eq:GraphBaseGen1}) is then given by summing the following two contributions together
\begin{itemize}
\item \emph{The sum of all the last terms on the RHS of \eqref{Eq:GraphBaseGen2} for all $i=1,\dots,N$ and the first term of \eqref{Eq:GraphBaseGen1}}
~~ According to \eqref{Eq:LHSBCJ}, the last term in the brackets of \eqref{Eq:GraphBaseGen2} for given $\pmb{\zeta}^{(j)}\in\mathcal{T}_a^{j}|_{a_j}$ ($j=1,2,\dots,N$) when summing over $\pmb{\xi}^{(i)}$ can be explicitly written as
\bea
\Sl_{\pmb{\xi}^{(i)}}\mathcal{B}\bigl(1\big|\pmb{\zeta}^{(i)},\pmb{\xi}^{(i)}\big|r\bigr)\big|_{a_i\succ a}=\Sl_{\pmb{\xi}^{(i)}}\Sl_{\pmb{\eta}\in\pmb{\zeta}^{(i)}\shuffle\pmb{\xi}^{(i)}|_{a_i\succ a}}\Sl_{x\in\pmb{\zeta}^{(i)}}(k_x\cdot X_x(\pmb{\eta}))A(1,\pmb{\eta},r).
\eea
Here,  $\Sl_{\pmb{\xi}^{(i)}}\Sl_{\pmb{\eta}\in\pmb{\zeta}^{(i)}\shuffle\pmb{\xi}^{(i)}|_{a_i\succ a}}$ is given by summing over permutations
\bea
\bigl(\pmb{\zeta}^{(i)}\shuffle\big\{\bigl\{{a,\pmb{\zeta}^{(1)}\shuffle\dots\shuffle\bcancel{\pmb{\zeta}^{(i)}}\shuffle\dots\shuffle\pmb{\zeta}^{(N)}}\bigr\}\shuffle\pmb{\gamma}\big\}\bigr)\big|_{a_i\succ a}.
\eea
Recalling that $a_i$ is the leftmost element in $\pmb{\zeta}^{(i)}\in\mathcal{T}_a^{i}|_{a_i}$, the above expression turns to
\bea
\bigl\{{a,\pmb{\zeta}^{(1)}\shuffle\dots\shuffle{\pmb{\zeta}^{(i)}}\shuffle\dots\shuffle\pmb{\zeta}^{(N)}}\bigr\}\shuffle\pmb{\gamma}.
\eea
Therefore, the sum of all the last terms of in \eqref{Eq:GraphBaseGen2} over all $i=1,\dots,N$ and the first term of \eqref{Eq:GraphBaseGen1} gives
\bea
&&\Sl_{\substack{\pmb{\zeta}^{(j)}\in\mathcal{T}_a^{j}|_{a_j}\\\text{for}~j=1,\dots,N}}\Sl_{\substack{\pmb{\eta}\in\{a,\pmb{\zeta}^{(1)}\shuffle\dots\\\shuffle{\pmb{\zeta}^{(i)}}\shuffle\dots\shuffle\pmb{\zeta}^{(N)}\}}}\left[\Big(\Sl_{i=1}^N\Sl_{x\in\pmb{\zeta}^{(i)}}k_x\cdot X_x(\pmb{\eta})\Big)+k_a\cdot X_a(\pmb{\eta})\right]A(1,\pmb{\eta}\shuffle\pmb{\gamma},r)\nn
&=&\Sl_{\substack{\pmb{\zeta}^{(j)}\in\mathcal{T}_a^{j}|_{a_j}\\\text{for}~j=1,\dots,N}}\Sl_{\shuffle}\mathcal{B}\bigl(1\big|\bigl\{a,\pmb{\zeta}^{(1)}\shuffle\dots\shuffle\pmb{\zeta}^{(N)}\bigr\},\pmb{\gamma}\big|r\bigr)\nn
&=&\Sl_{\pmb{\zeta}}\mathcal{B}\left(1\left|\pmb{\zeta}\in{\begin{minipage}{1.8cm}
\includegraphics[width=2.6cm]{GraphBasedGEN1.jpg} \end{minipage} }~~~~~~~\Bigg|_a,\pmb{\gamma}\right|r\right),
\eea
which is just $\mathcal{B}^{(1)}_G$ in \eqref{Eq:GraphBaseGen}.

\item \emph{The sum of the first term in \eqref{Eq:GraphBaseGen2} and \eqref{Eq:GraphBaseGen3} over all $i=1,\dots,N$}~~ This part defines
\bea
&&\Sl_{\shuffle}\mathcal{B}^{(2)}_G\left(1\left|{\begin{minipage}{1.6cm}
\includegraphics[width=2.2cm]{GraphBasedGEN2.jpg} \end{minipage} }~~~~~,\{a\}\shuffle\pmb{\gamma}\right|r\right)\\
&=&\Sl_{i=1}^N\Sl_{\substack{\pmb{\zeta}^{(j)}\in\mathcal{T}_a^{j}|_{a_j}\\\text{for}~j=1,\dots,N,j\neq i}}\Sl_{\pmb{\xi}^{(i)}}\biggl[\Sl_{\pmb{\zeta}^{(i)}}\mathcal{B}\bigl(1\big|\pmb{\zeta}^{(i)}\in
\mathcal{T}_a^{i}\big|_{a_i},\pmb{\xi}^{(i)}\big|r\bigr)+\Sl_{\shuffle}\mathcal{B}^{(2)}_G\bigl(1\big|
\mathcal{T}_a^{i}-a_i,\{a_i\}\shuffle\pmb{\xi}^{(i)}\big|r\bigr)\big|_{a_i\prec a}\biggr],\Label{Eq:GraphBaseGen4}\nonumber
\eea
where $\pmb{\xi}^{(i)}$ is summed over all permutations satisfying \eqref{Eq:GraphBaseGen4}.
In the square brackets, the first term is precisely the LHS of BCJ relation \eqref{Eq:BCJRelation} where $a$ is in the $\pmb{\alpha}$ set.  According to the inductive assumption, the second term is a combination of BCJ relations where both $a_i$ is in the $\pmb{\alpha}$ set. Since $a\in \pmb{\xi}^{(i)}$, $a$ must also be in the $\pmb{\alpha}$ set. Thus the second term of \eqref{Eq:GraphBaseGen4} is a combination of the LHS of those BCJ relations with $a_i\prec a$.
\end{itemize}
All together, we have proven that the LHS of the graph-based BCJ relation is a combination of the LHS of traditional BCJ relations.
\section{Gauge invariance identities of tree-level YM and GR amplitudes}\label{sec:GRandYM}
%%%%%%%%%%%%%%%%%%%%%%%%%%%%%%%%%%%%%%%%%%%%%%%%%%%%%%%%%%%%%%%%%%%%%%%%%%%%%%%%%%%%%%%%%%%%%%
A tree level $n$ graviton amplitude $M(1,2,\dots,n)$ can be expanded as a combination of tree level single-trace EYM amplitudes. Particularly, we can pick out arbitrary two gravitons, for example, $1$ and $n$ and then write the amplitude $M(1,2,\dots,n)$ in terms of single-trace EYM amplitudes, where $1$ and $n$ are always considered as two gluons, as follows
\bea
M(1,2,\dots,n)&=&\Sl_{\text{Splits}}\,\Sl_{\pmb{\sigma}\in\text{perms}\,\mathsf{A}}(-1)^{|\mathsf{A}|}\big(\epsilon_1\cdot F_{\sigma(1)}\cdot\dots\cdot F_{\sigma(|\mathsf{A}|)}\cdot\epsilon_n\big)A(1,\pmb{\sigma},n\Vert\,{\mathsf{H}}).\Label{Eq:GREYM}
\eea
Here, $|\mathsf{A}|$ is the number of elements in $\mathsf{A}$. We have summed over all splits $\{2,\dots,n-1\}\to \mathsf{A}, \mathsf{H}$ in the first summation. For a given $\mathsf{A}$, we summed over all possible permutations $\pmb{\sigma}$ of elements in $\mathsf{A}$. When the EYM amplitudes in \eqref{Eq:GREYM} are expressed by \eqref{Eq:PureYMExpansion}, \eqref{Eq:GREYM} is then expanded in terms of pure YM amplitudes
\bea
M(1,2,\dots,n)&=&\Sl_{\text{Splits}}\,\Sl_{\pmb{\sigma}\in\text{perms}\,\mathsf{A}}
\Sl_{\pmb{\xi}\,\in\,\pmb{\sigma}\shuffle\,\text{perms}\,{\mathsf{H}}}W(1,\pmb{\sigma},n)\mathcal{C}(1,\pmb{\xi},n)A(1,\pmb{\xi},n),\Label{Eq:GRYM}
\eea
where
\bea
W(1,\pmb{\sigma},n)\equiv\epsilon_n\cdot F_{\sigma(|\mathsf{A}|)}\cdot\dots\cdot F_{\sigma(1)}\cdot\epsilon_1.\Label{Eq:WChain}
\eea
The factor $(-1)^{|\mathsf{A}|}$ in \eqref{Eq:GREYM} is absorbed when the antisymmetry of the strength tensors are considered. If we collect all coefficients $n_{1|\pmb{\zeta}|n}$ corresponding to a given permutation $\pmb{\zeta}$ of elements in $\{2,\dots,n-1\}$ together, a more brief form (dual DDM form) is obtained
\bea
M(1,2,\dots,n)=\Sl_{\pmb{\zeta}\in S_{n-2}}n_{1|\pmb{\zeta}|n}A(1,\pmb{\zeta},n),\Label{Eq:GRYM1}
\eea
where $n_{1|\pmb{\zeta}|n}$ are the \emph{BCJ numerators corresponding to half-ladder diagrams.}

The gauge invariance condition for any graviton $a$ states that the GR amplitude $M(1,2,\dots,n)$ must vanish under the replacement $\epsilon_a\to k_a$.
If the graviton $a$ belongs to $\{2,\dots,n-1\}$,  it can be an element of either $\mathsf{A}$ or $\mathsf{H}$ for a given split in \eqref{Eq:GREYM}. For the former case, the gauge invariance  is already included in the coefficients $W(1,\pmb{\sigma},n)$ due to the antisymmetry of the strength tensors. For the latter case, $a$ is considered as a graviton in an EYM amplitude whose gauge invariance induced identity has been studied in the previous sections. Therefore, only the identity induced by the gauge invariance condition for $a=1\,\text{or}\,n$ requires further study.

In the following, we review the old-version graphic rule for the construction of BCJ numerators $n_{1|\pmb{\zeta}|n}$ in \eqref{Eq:GRYM1}. Then we expand these graphs by refined graphs and prove that the gauge invariance induced identity
\bea
\Sl_{\pmb{\zeta}\in S_{n-2}}n_{1|\pmb{\zeta}|n}\big|_{\epsilon_1\to k_1}A(1,\pmb{\zeta},n)=0~~\Label{Eq:GRYM2}
\eea
 is a combination of BCJ relations. The identity induced by the condition $\epsilon_n\to k_n$ can be studied in a similar way.
Once all GR, EYM and YM amplitudes are replaced by YM, YMS (Yang-Mills scalar) and BS (biscalar) amplitudes correspondingly, the discussions are immediately extended to the identities (for BS amplitudes) induced by the gauge invariance of YM amplitudes.

\subsection{The old-version graphic rule for the BCJ numerators $n_{1|\pmb{\zeta}|n}$}
 BCJ numerators $n_{1|\pmb{\zeta}|n}$ obtained from \eqref{Eq:GRYM} can be constructed according to the following (old fasion) graphic rule \cite{Du:2017kpo}\footnote{Although the interpretation of the rule in this paper is different from the version in \cite{Du:2017kpo}, they essentially provide the same construction.}:

{\bf Step-1}~~Define a reference order of elements $2$, $3$, ..., $n-1$, $n$ by $\mathsf{R}=\{\rho(2),\rho(3),\dots,\rho(n-1),n\}$ where the graviton $n$ is considered the last element, $\pmb{\rho}$ is a permutation of  $2$, $3$, ..., $n-1$. Define \emph{weight} of elements $2$, $3$, ..., $n-1$, $n$ by their positions in  $\mathsf{R}$.

{\bf Step-2}~~For a given permutation $\pmb{\zeta}$, pick out  $i_1$, $i_2$, ..., $i_j$ and $n$ from $\mathsf{R}=\{\rho(2),\rho(3),\dots,\rho(n-1),n\}$ such that the following condition is satisfied
\bea
\zeta^{-1}(1)<\zeta^{-1}(i_1)<\dots<\zeta^{-1}(i_j)<\zeta^{-1}(n).
\eea
Then we construct a chain $\mathbb{CH}=[n,i_j,i_{j-1},\dots,i_1,1]$ which starts from the node $n$, passes through internal nodes $i_j$, ..., $i_1$ and ended at $1$. The node $1$ is considered as the root. The factor corresponding to the chain  $\mathbb{CH}=[n,i_j,i_{j-1},\dots,i_1,1]$ is
\bea
W(1,i_1,\dots,i_j,n)=\epsilon_n\cdot F_{i_j}\cdot\dots\cdot F_{i_1}\cdot \epsilon_1.
\eea
Redefine the reference order by $\mathsf{R}\to \mathsf{R}'=\mathsf{R}\setminus\{i_1,\dots,i_j,n\}\equiv\{\rho'(1),\dots,\rho'({n'})\}$.

{\bf Step-3} Pick out $\rho'(n')$, $i'_1$, ..., $i'_{j'}$ from  $\mathsf{R}'$. Construct a chain $\mathbb{CH}=[\rho'(n'),i'_{j'},i'_{j'-1},\dots,i'_1,a]$ towards $a\in\{1\}\cup\{i_1, i_2, ...,i_j\}$ (noting that $a$ cannot be $n$ because $n$ is always considered as the last element) such that
\bea
\zeta^{-1}(a)<\zeta^{-1}(i'_1)<\dots<\zeta^{-1}(i'_{j'})<\zeta^{-1}(\rho'(n'))
\eea
is satisfied. This chain corresponds to a factor
\bea
\epsilon_{\rho'(n')}\cdot F_{i'_{j'}}\cdot\dots F_{i'_1}\cdot k_a.
\eea
Redefine the ordered set by $\mathsf{R}\to \mathsf{R}''=\mathsf{R}'\setminus\{i'_1,\dots,i'_{j'},\rho'(n')\}\equiv\{\rho''(1),\dots,\rho''({n''})\}$.

{\bf Step-4}~~Repeat the above step. In each step, we define a chain which starts from the highest-weight element in the new defined $\mathsf{R}$ and ends at any node on chains constructed previously except the node $n$ (the root $1$ should be included). Redefine $\mathsf{R}$ by removing the starting and internal nodes which have been used in the current step. This process is terminated when $\mathsf{R}$  is empty. All chains together form a connected tree graph with the root $1$. Multiplying all factors corresponding to the chains and summing over all possible graphs for the permutation $\pmb{\zeta}$, we finally get the BCJ numerator  $n_{1|\pmb{\zeta}|n}$.

\subsection{Understanding the gauge invariance induced identity \eqref{Eq:GRYM2}}

When the condition $\epsilon_1\to k_1$ is imposed and all strength tensors $F_i^{\mu\nu}$ are expanded explicitly according to the definition \eqref{Eq:StrengthTensor},
the three types of chains (see \figref{Lines}) are distinguished. Hence the old-version graphic rule becomes a refined rule.
The coefficients $n_{1|\pmb{\zeta}|n}\big|_{\epsilon_1\to k_1}$ in the identity \eqref{Eq:GRYM2} is then provided by
\bea
n_{1|\pmb{\zeta}|n}\big|_{\epsilon_1\to k_1}=\Sl_{\mathcal{F}\in \W{\mathbb{G}}(1,\pmb{\zeta},n)}{\W{\mathcal{D}}}^{[\mathcal{F}]},
\eea
where $\W{\mathbb{G}}(1,\pmb{\zeta},n)$ is the set of physical graphs defined by the refined graphic rule and we summed over all possible graphs $\mathcal{F}\in \W{\mathbb{G}}(1,\pmb{\zeta},n)$. The factor ${\W{\mathcal{D}}}^{[\mathcal{F}]}$ denotes the coefficient for a given graph constructed by the refined rule. The identity \eqref{Eq:GRYM2} then turns to
\bea
\Sl_{\pmb{\zeta}\in S_{n-2}}\left[\Sl_{\mathcal{F}\in \W{\mathbb{G}}(1,\pmb{\zeta},n)}{\W{\mathcal{D}}}^{[\mathcal{F}]}\right]A(1,\pmb{\zeta},n)=0~~\Label{Eq:GRYM3}.
\eea
Before showing the relationship between the identity \eqref{Eq:GRYM3} and BCJ relation \eqref{Eq:BCJRelation}, we point out the following features of graphs $\mathcal{F}$ in \eqref{Eq:GRYM3}:
\begin{itemize}
\item [(i)] For any permutation $1,\pmb{\zeta},n$, the nodes $1$ and $n$ are always the leftmost and the rightmost nodes respectively. The node $1$ always plays as the root of a graph $\mathcal{F}$. The node $n$ cannot be the end node of other chains.

\item [(ii)] Since $\epsilon_1$ is replaced by $k_1$, the node $1$ must be connected by lines with arrows pointing to it. Hence only type-2 lines (with the arrows pointing to $1$) and type-3 lines can be connected to $1$.
\item [(iii)] Any starting node of a chain is associated with a polarization, thus each of them starts a type-1 line or a type-2 line pointing towards the direction of the root.
\item [(iv)] Being different from the graphs for the identity induced from EYM amplitudes, graphs that do not involve any type-3 line are allowed. In such a graph, all lines are type-2 lines pointing to the direction of the root $1$.
\end{itemize}

\begin{figure}
\centering
\includegraphics[width=0.9\textwidth]{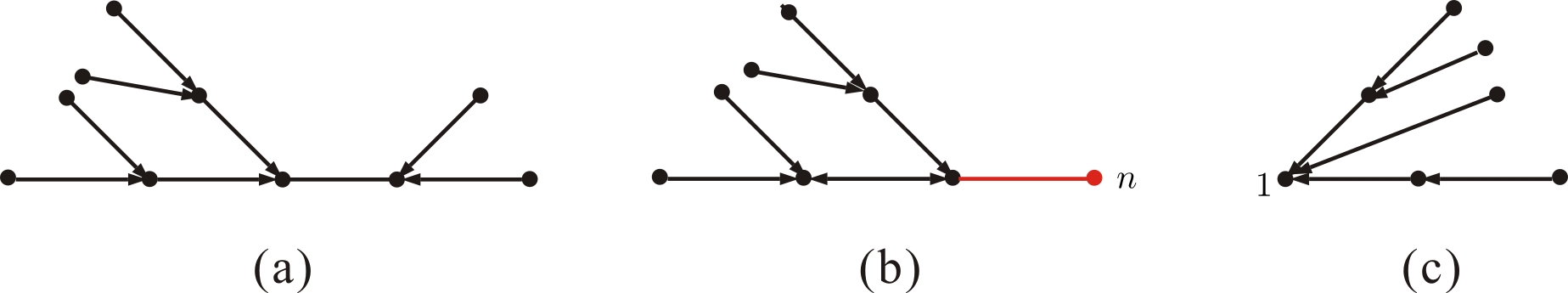}
\caption{Graphs (a), (b) and (c) are correspondingly typical graphs of type-I, -II,-III components in the skeleton of the $T_1$ part in \eqref{Eq:GRYMT0}. }\label{Fig:GRYMT2}
\end{figure}
\begin{figure}
\centering
\includegraphics[width=0.9\textwidth]{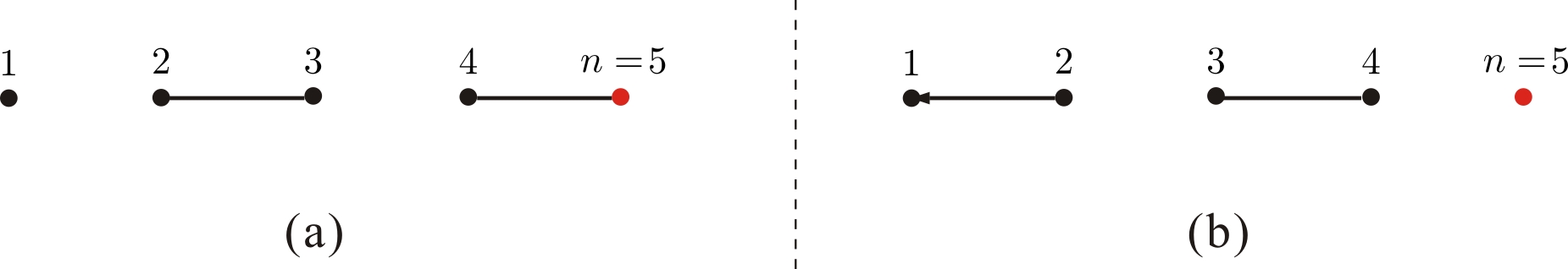}
\caption{Graphs (a) and (b) are typical skeletons (with three components) for graphs contributing to $T_1$ and $T_2$ (see \eqref{Eq:GRYMT0}) with $n=5$, respectively. Here the reference order is chosen as $\mathsf{R}=\{1,2,3,4,5\}$. In graph (a), the type-1 line with its two ends $2$ and $3$, the type-1 line with its two ends $4$ and $5$, the single node $a$ are respectively type-I, -II and -III components. In graph (b), the type-2 line with its two ends $1$ and $2$ (the arrow points to $1$), the type-1 line with its ends $3$ and $4$, the single node $5$ are respectively type-I, -II and -III components. }\label{Fig:GRYMEG}
\end{figure}
\begin{figure}
\centering
\includegraphics[width=\textwidth]{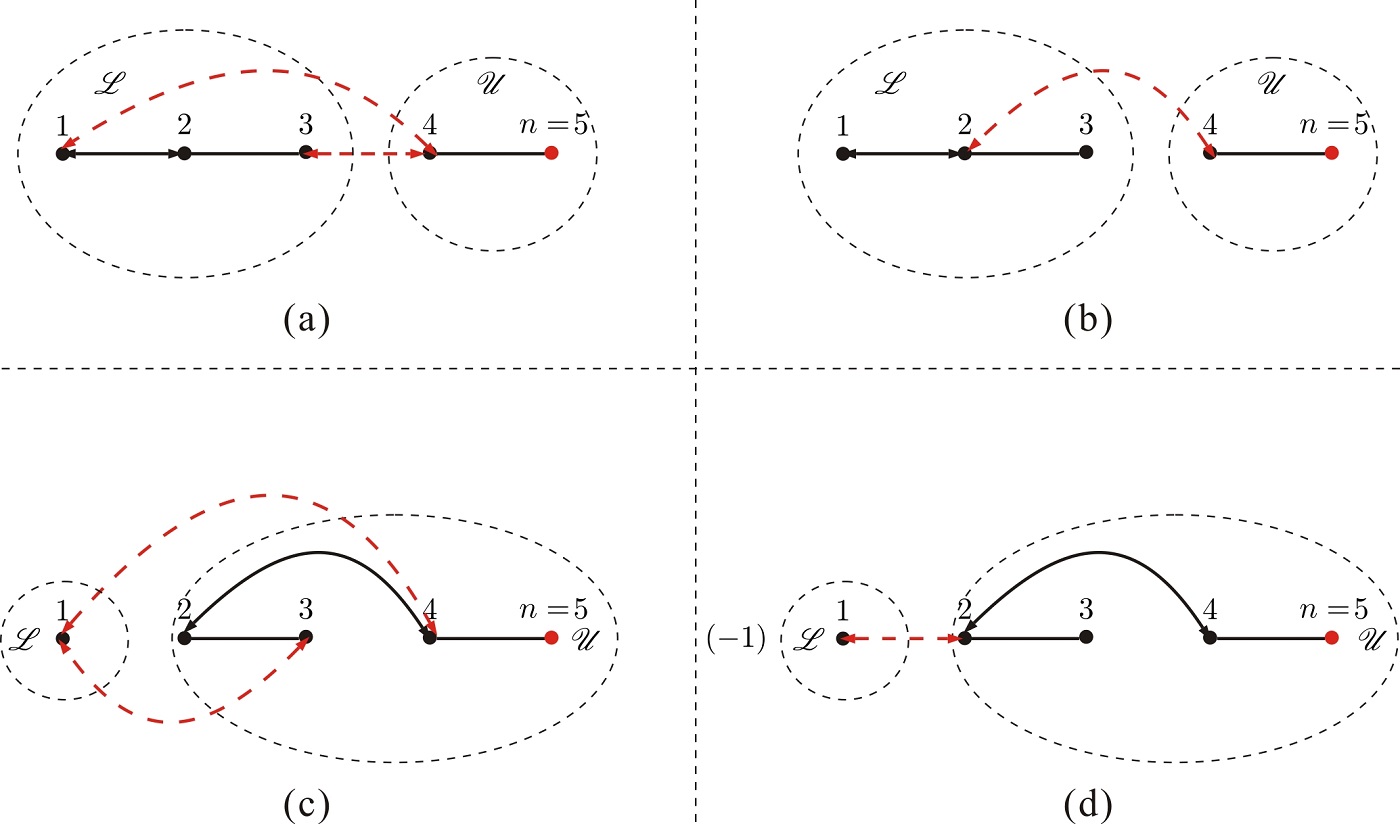}
\caption{Graphs (a) and (c) are physical graphs corresponding to the skeleton \figref{Fig:GRYMEG} (a). They are obtained by constructing a chain of component, which starts from the type-I component, towards the lower and upper blocks respectively according to rule-2 (noting that the node $n$ cannot be the ending node of any chain). Thus there are two distinct configurations of the final upper and lower blocks $\mathscr{U}$ and $\mathscr{L}$. Spurious graphs are given by (b) and (d). Since the spurious component, whose only a single side (i.e. the node $2$) on the part from the highest weight node $5$ to the root $1$, belongs to the final upper block in (d), we should have an extra minus. Apparently graphs (b) and (d) cancel with each other.
%The sum of graphs (a) and (b) provides $(-1)^2(\epsilon_2\cdot\epsilon_3)(\epsilon_4\cdot\epsilon_5)(k_1\cdot k_2)\mathcal{B}^{4}_{G}\big(1\big|{\begin{minipage}{0.15cm}
%\includegraphics[width=0.2cm]{GRYMEG4.jpg}\end{minipage}},\{2,3\}\big|5\big)$. The sum of graphs (c) and (d) contributes $\mathcal{B}^{3}_G\big(1\big|{\begin{minipage}{1.6cm}\includegraphics[width=1.4cm]{GRYMEG3.jpg}\end{minipage}},\emptyset\big|5\big)$.
}\label{Fig:GRYMEG1}
\end{figure}
To study \eqref{Eq:GRYM3}, we classify all graphs into two categories according to wether the node $n$ is connected by a type-1 line or a type-2 line.
Then the LHS of $\eqref{Eq:GRYM3}$ is written as the sum of $T_1$ and $T_2$:
\bea
T_1\equiv\Sl_{\pmb{\zeta}\in S_{n-2}}\left[\Sl_{\mathcal{F}\in \W{\mathbb{G}}_1(1,\pmb{\zeta},n)}{\W{\mathcal{D}}}^{[\mathcal{F}]}\right]A(1,\pmb{\zeta},n),~~~T_2&\equiv&\Sl_{\pmb{\zeta}\in S_{n-2}}\left[\Sl_{\mathcal{F}\in \W{\mathbb{G}}_2(1,\pmb{\zeta},n)}{\W{\mathcal{D}}}^{[\mathcal{F}]}\right]A(1,\pmb{\zeta},n),\Label{Eq:GRYMT0}
\eea
where $\W{\mathbb{G}}_1(1,\pmb{\zeta},n)$ and $\W{\mathbb{G}}_2(1,\pmb{\zeta},n)$ denote the set of graphs where $n$ is connected by a type-1 line and a type-2 line (pointing to the direction of root), respectively. We study these two parts separately.

{\bf The $T_1$ part}~~If the node $n$ is connected by a type-1 line, the chain $\mathbb{CH}=[n,i_j,i_{j-1},\dots,i_1,1]$ contributing to $T_1$ (see \eqref{Eq:GRYMT0}) can be described by \figref{Fig:SectorsOfChains} (b) when replacing the nodes $b$ and $h_{\rho(s)}$ therin respectively by the node $1$ and the node $i_j$ that is further connected by a type-1 line starting at the node $n$. According to discussions in \secref{section:GeneralStructure}, such a chain must involve at least one type-3 line. As a result, the skeleton (the graph obtained by removing all type-3 lines) for a graph contributing to $T_1$ must be a disconnected graph. Following a similar discussion with  \secref{section:GeneralStructure}, we find that there are three types of components corresponding to a given skeleton in general (as in the EYM case):
\begin{itemize}
\item\emph{Type-I component: A component which contains neither $1$ nor $n$ and involves a type-1 line (see \figref{Fig:GRYMT2} (a))}~~~In general, such a component can have type-2 lines whose arrows point to the direction of the ends of the type-1 line (i.e. the kernel).

\item\emph{Type-II component: A component containing the node $n$ (see \figref{Fig:GRYMT2} (b))}~~~Components of this type also have a kernel: the type-1 line with two ends $n$ and $a$ ($a\neq n,1$). The difference from type-I components is that all possible type-2 lines must be only on the bottom side (i.e. the side containing node $a$) of this component. The node $n$ is the starting node of the kernel (the type-1 line) of this component, while it cannot be the ending node of any chain. Thus there cannot be lines ended at $n$.

\item \emph{Type-III component: A component containing the root $1$ (see \figref{Fig:GRYMT2} (c))}~~~Such a component only involve type-2 lines whose arrows point to the direction of the root $1$.
\end{itemize}
A typical skeleton for graphs contributing to $T_1$ with $n=5$ is presented by \figref{Fig:GRYMEG} (a).

Having defined these three types of components, we can follow all the discussions in \secref{section:GraphsForASkeleton} including the rule-1 and -2 for physical graphs as well as the construction and cancellation of spurious graphs (see \figref{Fig:GRYMEG1} for the skeleton \figref{Fig:GRYMEG} (a)). The only thing we need to pay more attention to is that the node $n$ in the type-II component cannot be connected with nodes in other components via type-3 lines and it must be the rightmost element in any permutation. For any configuration of final upper and lower blocks, the sum of all physical graphs and spurious graphs is proportional to the LHS of the graph-based BCJ relation. For example, the sum of graphs \figref{Fig:GRYMEG1} (a) and (b) provides $(-1)^2(\epsilon_2\cdot\epsilon_3)(\epsilon_4\cdot\epsilon_5)(k_1\cdot k_2)\mathcal{B}^{4}_{G}\,\big(1\big|\includegraphics[width=0.15cm]{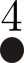},\{2,3\}\big|5\big)$ while the sum of graphs (c) and (d) contributes $(\epsilon_2\cdot\epsilon_3)(\epsilon_4\cdot\epsilon_5)(k_2\cdot k_4)\mathcal{B}^{3}_G\,\big(1\big|\includegraphics[width=1.6cm]{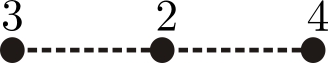},\emptyset\big|5\big)$.
Therefore, we conclude that the $T_1$ part can always be written as a combination of the LHS of graph-based BCJ relations \eqref{Eq:Graph-based-BCJ} which have been proven to be a combination of traditional BCJ relations \eqref{Eq:BCJRelation}.

%\begin{figure}
%\centering
%\includegraphics[width=0.7\textwidth]{GRYMT1.jpg}
%\caption{In this graph, $\mathcal{F}''$ is given by the subgraph with all black lines and all nodes $1,\dots,9$. When connecting the node $9$ to a node $a\in\{1,\dots,8\}$ via a type-2 line pointing towards the direction of the root (a read dashed arrow lines), we arrive a graph $\mathcal{F}\supset\mathcal{F}''$. In this example, ${\W{\mathcal{D}}}^{[\mathcal{F}'']}=$   }\label{Fig:GRYMT1}
%\end{figure}
%
%
\begin{figure}
\centering
\includegraphics[width=\textwidth]{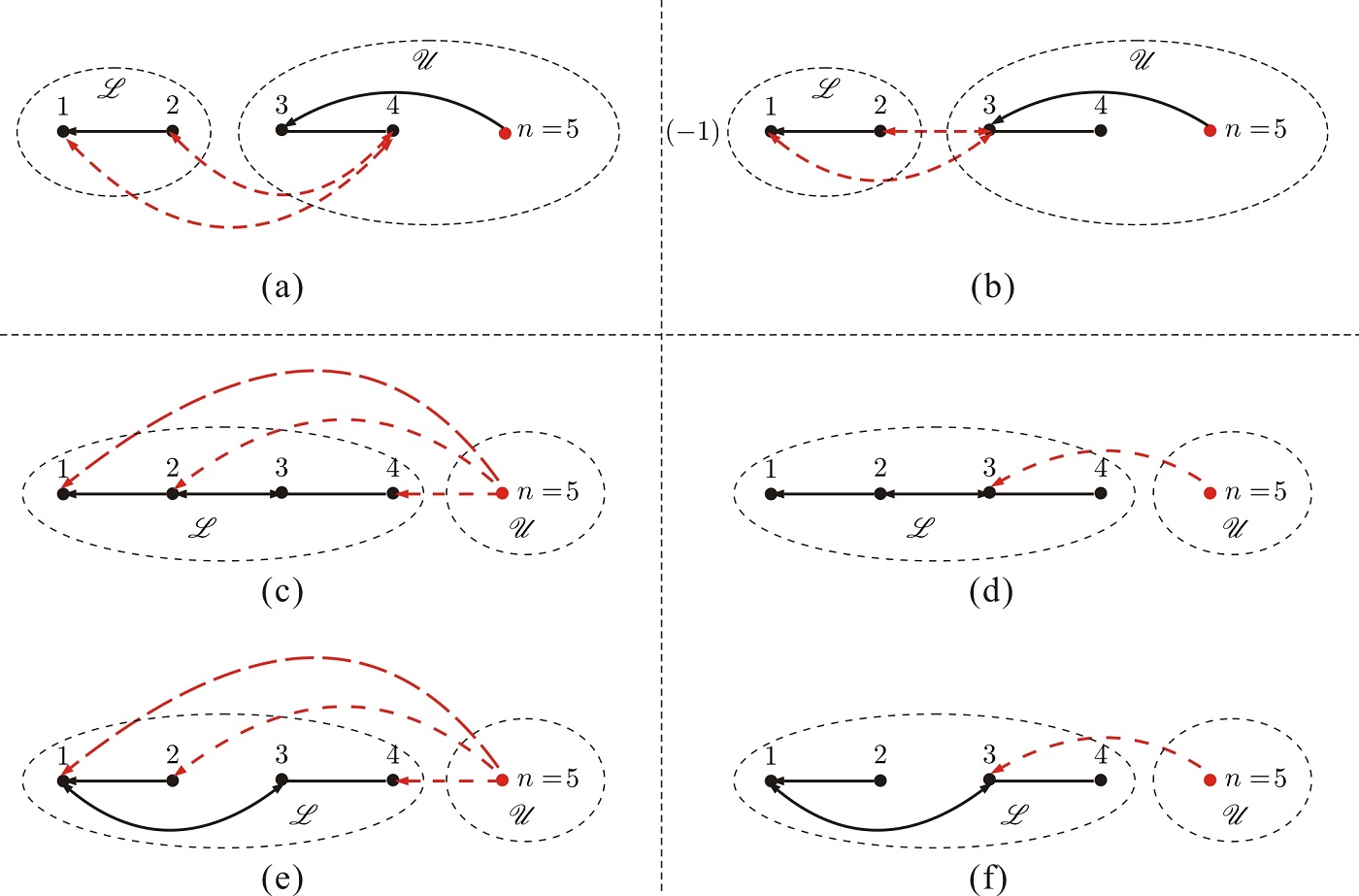}
\caption{Graphs (a), (c) and (e) are physical graphs corresponding to the skeleton \figref{Fig:GRYMEG} (b). The type-I chain belongs to the final upper (lower) block in (a)((c)and (e)). Corresponding spurious graphs are given by (b), (d) and (f). The final upper block in (b) contains a spurious component (i.e. the type-I component), thus there should be an extra minus. In graphs (d) and (f), the final upper block does not contain any spurious component, thus the sign is plus. All spurious graphs (b), (d) and (f) cancel out (noting that (b) stands for two graphs).}\label{Fig:GRYMEG2}
\end{figure}
{\bf The $T_2$ part}~~Graphs contributing to $T_2$ involve the node $n$ which is connected by a type-2 line (pointing towards the direction of root).
To expand $T_2$ by BCJ relations, \emph{we should adjust the definition of skeleton by those graphs which are obtained by removing both type-3 lines and the type-2 line attached to the node $n$.} Any skeleton under this definition is a disconnected graph. The definition of type-I and type-III components are same with those in the $T_1$ case. \emph{The type-II component is defined by the single node $n$}. Along the line pointed in rule-2 in \secref{section:GraphsForASkeleton}, we define the upper and lower blocks by the maximally connected graphs containing the node $n$ and $1$ respectively. At the beginning, the upper and lower blocks are correspondingly the type-II and the type-III components. According to rule-2 in \secref{section:PhysicalGraphs}, we attach a chain of components to either the upper or the lower bock and redefine these blocks in each step. \emph{In this case, we should require that the node $n$ can only be connected by the first chain that is attached to the upper block, via a type-2 line (with the arrows pointing towards root).} When all components are used, we arrive the final upper and lower blocks. As done in \secref{section:GraphsForASkeleton}, we connect these two blocks into a fully connected graph:
\begin{itemize}
\item [(i)] If the final upper block consists of more than one node, such as  \figref{Fig:GRYMEG2} (a), (b), we should connect a node $b$ in the final lower block with a node $a$ ($b\neq n$) in the final upper block via a type-3 line.
\item [(ii)] If the final upper block only contains the node $n$, such as \figref{Fig:GRYMEG2} (c), (d), (e) and (f), we should connect a node $b$ in the final lower block with $n$ via a type-2 line whose arrow points to $a$.
\end{itemize}
After this step, a graph is a spurious one (for example \figref{Fig:GRYMEG2} (b), (d) and (f)) if it contains spurious components (whose only a single side is passed through by the path from $n$ to $1$), otherwise, it is a physical graph (for example \figref{Fig:GRYMEG2} (a), (c) and (e)). The extra sign associated with a spurious graph is given by $(-1)^{\mathcal{S}(\mathscr{U})}$, where $\mathcal{S}(\mathscr{U})$ is the number of spurious components in the final upper block. Spurious graphs belonging to distinct final upper and lower blocks cancel in pairs as shown in \secref{section:SpuriousGraphs}. For example the two graphs included by \figref{Fig:GRYMEG2} (b) cancel with \figref{Fig:GRYMEG2} (d) and (e), respectively. Hence, the $T_2$ can be given by  summing over (1) all possible configurations of the final upper and lower blocks, (2) all possible graphs constructed by connecting these two blocks according to (i) and (ii) pointed above for a given configuration. Particularly,
\begin{itemize}
\item [(i)] If the final upper block consists of more than one node, the sum of all graphs given by (i) is expanded in terms of the LHS of the graph-based BCJ relations \eqref{Eq:Graph-based-BCJ} (hence the LHS of traditional BCJ relations \eqref{Eq:BCJRelation}). For example the sum of graphs included in \figref{Fig:GRYMEG2} (a) and (b) is proportional to the LHS of a graph-based BCJ relation $(-1)(\epsilon_1\cdot\epsilon_2)(\epsilon_3\cdot\epsilon_4)(\epsilon_5\cdot k_3)B^{4}_G\big(1\big|\includegraphics[width=0.4in]{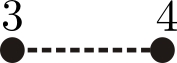},\{2\}\big|5\big)$.

\item [(ii)] If the final upper block consists of only the node $n$, $n$ can be connected to any node $b$ in the final lower block via a type-2 line $\epsilon_n\cdot k_a$ (see \figref{Fig:GRYMEG2} (c), (d) and (e), (f) for distinct configurations of the final lower block). The permutations established by graphs with all choices of $b$ are same. Thus we can extract a total factor
\bea
\Sl_{a\in\{1,2,\dots,n-1\}}\epsilon_n\cdot k_a=\epsilon_n\cdot(-k_n)=0,
\eea
where the transversality of polarization and momentum conservation have been applied. Hence this part must vanish.
\end{itemize}

To sum up, the identity \eqref{Eq:GRYM2} which is induced by the gauge invariance of gravity can always been expressed as a combination of BCJ relations.

%%%%%%%%%%%%%%%%%%
\section{Conclusions}\label{sec:Conclusions}
%%%%%%%%%%%%%%%%%%
In this paper, a graphic approach to the relationship between gauge invariance induced identity (where coefficients are polynomials of Lorentz inner products $\epsilon\cdot\epsilon$, $\epsilon\cdot k$ and $k\cdot k$) and BCJ relations (where coefficients are functions of only Mandelstam variables $k\cdot k$) is provided. By establishing a refined graphic rule, the three types of Lorentz inner products $\epsilon\cdot\epsilon$, $\epsilon\cdot k$ and $k\cdot k$ are represented by type-1, -2 and -3 lines correspondingly. To find out the relationship between the identity induced from the gauge invariance of tree level single-trace EYM amplitude and BCJ relations, we collected terms containing the same skeleton (the subgraph obtained by deleting all type-3 lines). We further proved that the sum of physical graphs for a given skeleton can be given by (a) summing over all possible configurations of final upper and lower blocks (defined by rule-2); (b) for a given configuration, summing over all possible (physical and spurious) graphs constructed by connecting a type-3 line between the final upper and lower blocks. Each summation in (b) is in fact proportional to a graph-based BCJ relation which was further proved to be a combination of traditional BCJ relations. Following a similar discussion, we proved that the identity induced by the gauge invariance of tree level GR (YM) amplitudes could also be expanded in terms of traditional BCJ relations.

We close this paper by presenting the following two related topics that deserve further study:
\begin{itemize}
\item First, how to extend the discussions to the identities induced from multi-trace EYM amplitudes? Two types of recursive expansion of tree level multi-trace EYM amplitudes were established in \cite{Du:2017gnh}. Correspondingly, there are two identities (mentioned as type-I and II in  \cite{Du:2017gnh}) induced by the gauge invariance condition of the fiducial graviton and the cyclic symmetry of a gluon trace. The type-II identity for amplitudes with two gluon traces and no graviton can be understood as the graph-based BCJ relation \eqref{Eq:Graph-based-BCJ} in which the tree graph $\mathcal{T}$ is a simple chain. However, this observation cannot be straightforwardly generalized to amplitudes with an arbitrary number of traces and gravitons. Thus the relationship between identities induced from arbitrary multi-trace EYM amplitudes still deserves further consideration.
\item Second, how to calculate helicity amplitudes in four dimensions by the refined graphic rule? In four dimensions, the tree level single-trace EYM and GR amplitudes with maximally-helicity-violating (MHV) configurations were shown to be proportional to (Hodges) determinants (see \cite{Hodges:2012ym} for GR and \cite{Du:2016wkt} for EYM) which can be further expanded by graphs \cite{Nguyen:2009jk,Feng:2012sy,Du:2016wkt}. Such brief formulas cannot be trivially extended to arbitrary helicity configurations. It is worth studying the relationship between the graphic expansion of the Hodges determinants \cite{Hodges:2012ym} and the refined graphic expansion of amplitudes in the current paper. This may provide hints for finding simple formulas of EYM and GR amplitudes beyond MHV.

\end{itemize}

%%%%%%%%%%%%%%%%%%%%%%%
\section*{Acknowledgments}
%%%%%%%%%%%%%%%%%%%%%%
The authors are grateful to Gang Chen, Bo Feng, Chih-Hao Fu, Hui Luo, Fei Teng and Yong Zhang for helpful discussions or/and valuable comments. This work is supported by NSFC under Grant No. 11875206, Jiangsu Ministry of Science and Technology under contract
BK20170410 as well as the ``Fundamental Research Funds for the Central Universities".

%%%%%%%%%%%%%%%%%%%%%
\appendix
%%%%%%%%%%%%%%%%%%%%%

\section{Conventions and definitions}\label{sec:convention}
%%%%%%%%%%%%%%%%%%%%%%%%%%%%%%%%%%%%%%%%%%%%%%%%%%%%%%%%%%

\noindent
$\mathsf{H}$: the graviton set~~~~~~~~~~~~~~~~~~~~~~~~~~~~~$F_i^{\mu\nu}$: the strength tensor $k_i^{\mu}\epsilon_i^{\nu}-\epsilon_i^{\mu}k_i^{\nu}$
%%%%%%%%%%%%%%%%%%%%%%%%%%%%%%%%%%%%%%%%%%%%%%%%%%%%%%%%%%
\newline

\noindent
Permutations: $\pmb{\alpha},\pmb{\beta},\pmb{\rho},\pmb{\sigma},\pmb{\xi},\pmb{\zeta}$~~~$\sigma(i)$: the $i$-th element in $\pmb{\sigma}$~~~$\sigma^{-1}(l)$: the position of $l$ in permutation $\pmb{\sigma}$
%\noindent
%$\pmb{h}$ denotes an arbitrary subset of $\mathsf{H}$.
%%%%%%%%%%%%%%%%%%%%%%%%%%%%%%%%%%%%%%%%%%%%%%%%%%%%%%%%%%
\newline

\noindent
$\mathsf{A}\shuffle \mathsf{B}$ for ordered sets $\mathsf{A}$ and $\mathsf{B}$: permutations by merging  $\mathsf{A}$ and $\mathsf{B}$ s.t. the relative order of elements in each set is kept, e.g. $\{1,2\}\shuffle\{a,b\}=\{\{1,2,a,b\},\{1,a,2,b\},\{1,a,b,2\} ,\{a,1,2,b\} ,\{a,1,b,2\},\{a,b,1,2\}\}$.
%%%%%%%%%%%%%%%%%%%%%%%%%%%%%%%%%%%%%%%%%%%%%%%%%%%%%%%%%%
\newline

%\noindent
%$C_{h_a}(\pmb{h})$: the {coefficients} of the recursive expansion \eqref{Eq:RecursiveExpansion} of single trace EYM amplitude.% the subscript labels the leaded graviton of this chain.
%%%%%%%%%%%%%%%%%%%%%%%%%%%%%%%%%%%%%%%%%%%%%%%%%%%%%%%%%%
%\newline

%\noindent
%$\mathcal{C}(1,\pmb{\sigma},r)$: the {coefficient} in the pure YM expansion \eqref{Eq:PureYMExpansion} of single trace EYM amplitudes.
%%%%%%%%%%%%%%%%%%%%%%%%%%%%%%%%%%%%%%%%%%%%%%%%%%%%%%%%%%
%\newline

\noindent
$Y_{i}(\pmb{\sigma})$ for $\pmb{\sigma}\in\pmb{\alpha}\shuffle\pmb{\beta}$ and $i\in \pmb{\beta}$: the sum of all momenta of elements in $\pmb{\sigma}$ (including the element $1$) on the LHS of $i$ in permutation $\pmb{\sigma}$.
%%%%%%%%%%%%%%%%%%%%%%%%%%%%%%%%%%%%%%%%%%%%%%%%%%%%%%%%%%
\newline

\noindent
$X_{i}(\pmb{\sigma})$ for $\pmb{\sigma}\in\pmb{\alpha}\shuffle\pmb{\beta}$ and $i\in \pmb{\beta}$: the sum of all momenta of elements in both $\pmb{\beta}$ and $\pmb{\alpha}$ (including the element $1$) that appear on the LHS of $i$ in permutation $\pmb{\sigma}$.
%%%%%%%%%%%%%%%%%%%%%%%%%%%%%%%%%%%%%%%%%%%%%%%%%%%%%%%%%%
\newline

\noindent
$\mathcal{F}$: Graphs~~~~~~~~~~~~~~~~~~~~~~~~~~$\mathsf{R}=\{h_{\rho(1)},\dots,h_{\rho(s)}\}$: a reference order of gravitons $h_1,\dots,h_s$
\newline

\noindent
Weight of a graviton $h_i$: the position of $h_i$ in the reference order $\mathsf{R}$
\newline

\noindent
$\mathbb{CH}=\left[a,i_j,\dots,i_1,b\right]$: a chain of nodes with starting node $a$, ending node $b$ and internal nodes $i_1,\dots,i_j$
%%%%%%%%%%%%%%%%%%%%%%%%%%%%%%%%%%%%%%%%%%%%%%%%%%%%%%%%%%
\newline

\noindent
$\mathcal{B}(1\,|\,\pmb{\beta},\pmb{\alpha}\,|\,r)$: the LHS of BCJ relation \eqref{Eq:BCJRelation}
\newline

\noindent
$\mathcal{B}^{\,c}_G(1\,|\,\mathcal{T},\pmb{\gamma}\,|\,r)$: the LHS of the graph based BCJ relation \eqref{Eq:Graph-based-BCJ} with choosing $f^c=1$
\newline

\noindent
Type-1, 2 and 3 lines: lines corresponding to factors of forms $\epsilon_a\cdot\epsilon_b$, $\epsilon_a\cdot k_b$ and $k_a\cdot k_b$ in the refined graphic rule
\newline

\noindent
$\mathcal{F}'\subset\mathcal{F}$: skeleton which is defined by removing all type-3 lines from a graph $\mathcal{F}$ for the relation
%%%%%%%%%%%%%%%%%%%%%%%%%%%%%%%%%%%%%%%%%%%%%%%%%%%%%%%%%%
\newline

\noindent
${\mathcal{F}\setminus\mathcal{F}'}$ for \eqref{Eq:GaugeInv3}\footnote{As pointed in \secref{sec:GRandYM}, the definition of skeleton for the identity \eqref{Eq:GRYM3} is adjusted.}: the subgraph of $\mathcal{F}$ which is obtained by removing all lines of $\mathcal{F}'$ from $\mathcal{F}$
\newline

%\noindent
%$\mathcal{D}^{[\mathcal{F}]}$ denotes the part of coefficient exchanging the highest weight node's polarization $\epsilon_{h_{\rho(s)}}$ to it's momentum $k_{h_{\rho(s)}}$, after summing about all possible $\mathcal{F}$ we got the whole coefficient.
%%%%%%%%%%%%%%%%%%%%%%%%%%%%%%%%%%%%%%%%%%%%%%%%%%%%%%%%%%

\noindent
$\mathcal{P}^{[\mathcal{F}']}$: coefficient associated to the skeleton $\mathcal{F}'$~~~~~$\mathcal{K}^{[{\mathcal{F}\setminus\mathcal{F}'}]}$:  coefficient associated to the graph ${\mathcal{F}\setminus\mathcal{F}'}$
%%%%%%%%%%%%%%%%%%%%%%%%%%%%%%%%%%%%%%%%%%%%%%%%%%%%%%%%%%
\newline

\noindent
$\mathcal{N}(\mathcal{F})$:  the number of arrows (excluding the one which is attached to and points to the highest-weight node $h_{\rho(s)}$ in the gauge invariance induced identity \eqref{Eq:GaugeInv3}) pointing deviate from the direction of roots.
%%%%%%%%%%%%%%%%%%%%%%%%%%%%%%%%%%%%%%%%%%%%%%%%%%%%%%%%%%
\newline

\noindent
$\pmb{\sigma}^{\mathcal{F}}$: permutations allowed by graph $\mathcal{F}$
%%%%%%%%%%%%%%%%%%%%%%%%%%%%%%%%%%%%%%%%%%%%%%%%%%%%%%%%%%
\newline

\noindent
Type-I sector of a chain: A sector containing a type-1 line and possible type-2 lines whose arrows point to the two ends of the type-1 line
\newline

\noindent
Type-II sector of a chain (the chain led by the highest weight node $h_{\rho(s)}$): A sector only containing type-2 lines whose arrows point to the direction of the starting node $h_{\rho(s)}$ of the chain.
\newline

\noindent
Type-III sector of a chain: A sector only containing type-2 lines whose arrows point to the direction of the
ending node of the chain
%%%%%%%%%%%%%%%%%%%%%%%%%%%%%%%%%%%%%%%%%%%%%%%%%%%%%%%%%%
\newline

\noindent
Type-I component $\mathscr{C}^{\text{I}}$: component containing a type-1 line and possible type-2 lines whose arrows pointing to the direction of the two ends of the type-1 line.
\newline

Kernel of a type-I component: the type-1 line in the type-I component
\newline

The top and bottom sides $\mathscr{C}^{\text{I}}_t$ and $\mathscr{C}^{\text{I}}_b$ of a type-I component $\mathscr{C}^{\text{I}}$: the two sides connected by the type-1 line. The side containing the highest-weight node in $\mathscr{C}^{\text{I}}$ is the top side $\mathscr{C}^{\text{I}}_t$. The side which does not contain the highest-weight node in $\mathscr{C}^{\text{I}}$ is the bottom side $\mathscr{C}^{\text{I}}_b$.
\newline

\noindent
Type-II component $\mathscr{C}^{\text{II}}$: the component containing the highest-weight graviton $h_{\rho(s)}$.
\newline

\noindent
Type-III component $\mathscr{C}^{\text{III}}$ \footnote{In the example of \figref{Fig:Figure9} we use $\mathscr{A}$ and $\mathscr{B}$ to denote two Type-I components,use $\mathscr{C}$ to denote Type-II component and use $\mathscr{R}$ denote Type-III component.}: the component containing roots.
\newline

%%%%%%%%%%%%%%%%%%%%%%%%%%%%%%%%%%%%%%%%%%%%%%%%%%%%%%%%%%

\noindent
$\mathsf{R}_{\mathscr{C}}$: the reference order of components ~~~The weight of a component: the position of a component in $\mathsf{R}_{\mathscr{C}}$
\newline

%%%%%%%%%%%%%%%%%%%%%%%%%%%%%%%%%%%%%%%%%%%%%%%%%%%%%%%%%%
\noindent
The upper and lower blocks $\mathscr{U}$ and $\mathscr{L}$ in rule-2: the maximally connected graphs involving the type-II and the type-III components $\mathscr{C}^{\text{II}}$ and $\mathscr{C}^{\text{III}}$ respectively.
%%%%%%%%%%%%%%%%%%%%%%%%%%%%%%%%%%%%%%%%%%%%%%%%%%%%%%%%%%
\newline

\noindent
$\mathcal{S}(\mathscr{U})$: the number of spurious components in the final upper block
\newline
%\noindent
%$\mathcal{W}_{1}^{\text{I}}$ represent the weights of a component chain $\mathbb{CH}_m^{\text{I}}$.
%%%%%%%%%%%%%%%%%%%%%%%%%%%%%%%%%%%%%%%%%%%%%%%%%%%%%%%%%%

\noindent
$\mathcal{T}$:  a tree graph with dashed lines (lines only imply the relative positions of nodes)
\newline

\noindent
$\mathcal{T}|_a$: permutations established by $\mathcal{T}$ when considering $a$ as the leftmost node
%%%%%%%%%%%%%%%%%%%%%%%%%%%%%%%%%%%%%%%%%%%%%%%%%%%%%%%%%%
\newline

\noindent
$\oplus$: union of two disjoint graphs
\newline

\noindent
$\mathcal{T}-\mathcal{T}'$ for a graph $\mathcal{T}$ its subgraph $\mathcal{T}'$: the graph obtained by removing all nodes in $\mathcal{T}'$ and the lines attached to these nodes from $\mathcal{T}$
%\noindent
%For a given graph $\mathcal{T}$, the factor $f^a$ is a relative sign depending on the choice of $a$.

%%%%%%%%%%%%%%%%%%%%%%%%%%%%%%%%%%%%%%%%%%%%%%%%%%%%%%%%%%%%%%%%%%%%%%%%%%%%%%%%
\section{Graphs for the $\mathsf{H}=11$ example}\label{sec:ComplicatedGraphs}
%%%%%%%%%%%%%%%%%%%%%%%%%%%%%%%%%%%%%%%%%%%%%%%%%%%%%%%%%%%%%%%%%%%%%%%%%%%%%%%%%

\begin{figure}
\centering
\includegraphics[width=0.9\textwidth]{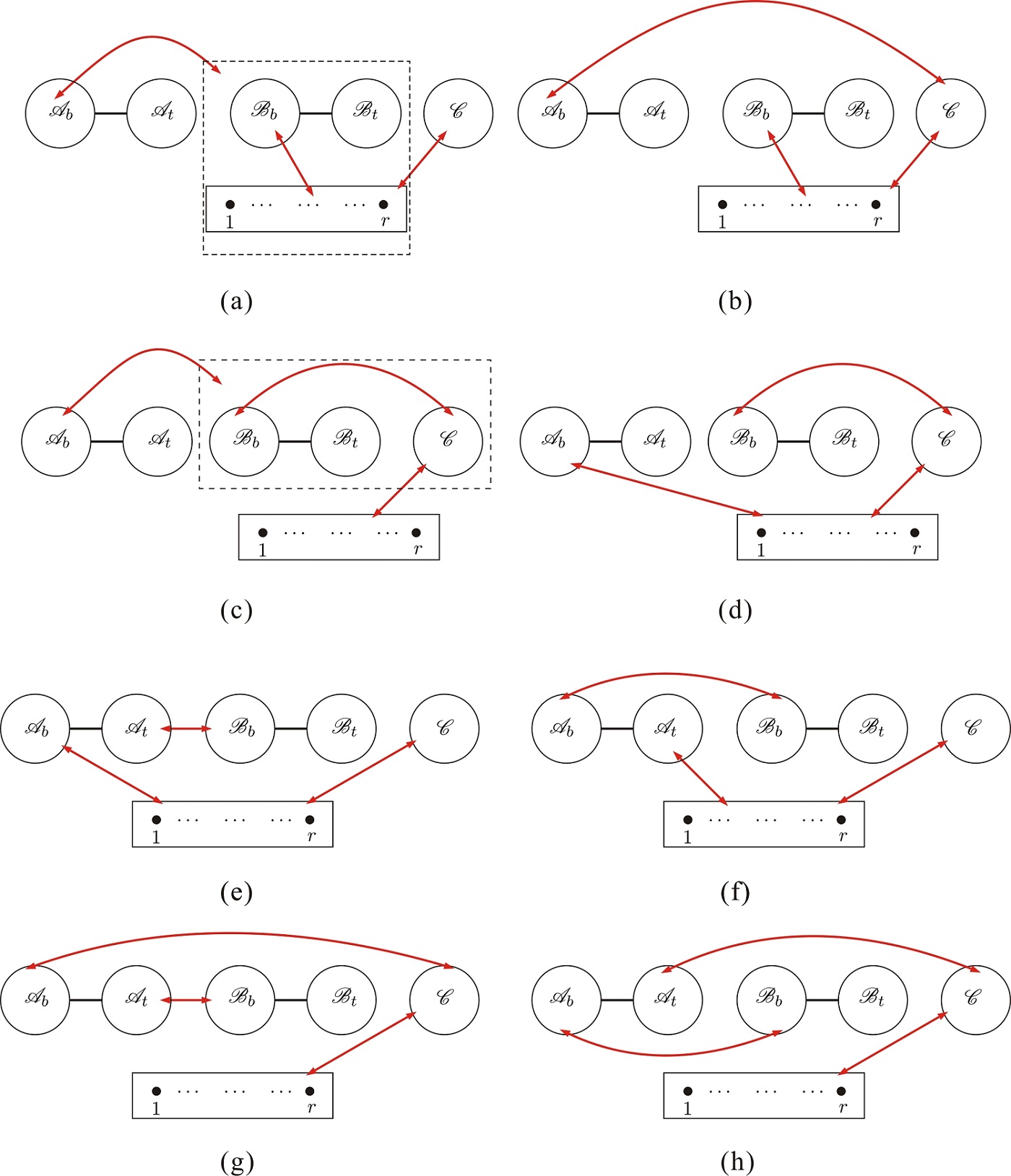}
\caption{Physical graphs for the skeleton \figref{Fig:Figure9} by { construction-1}: part 1. }\label{Fig:Figure12}
\end{figure}
\begin{figure}
\centering
\includegraphics[width=0.8\textwidth]{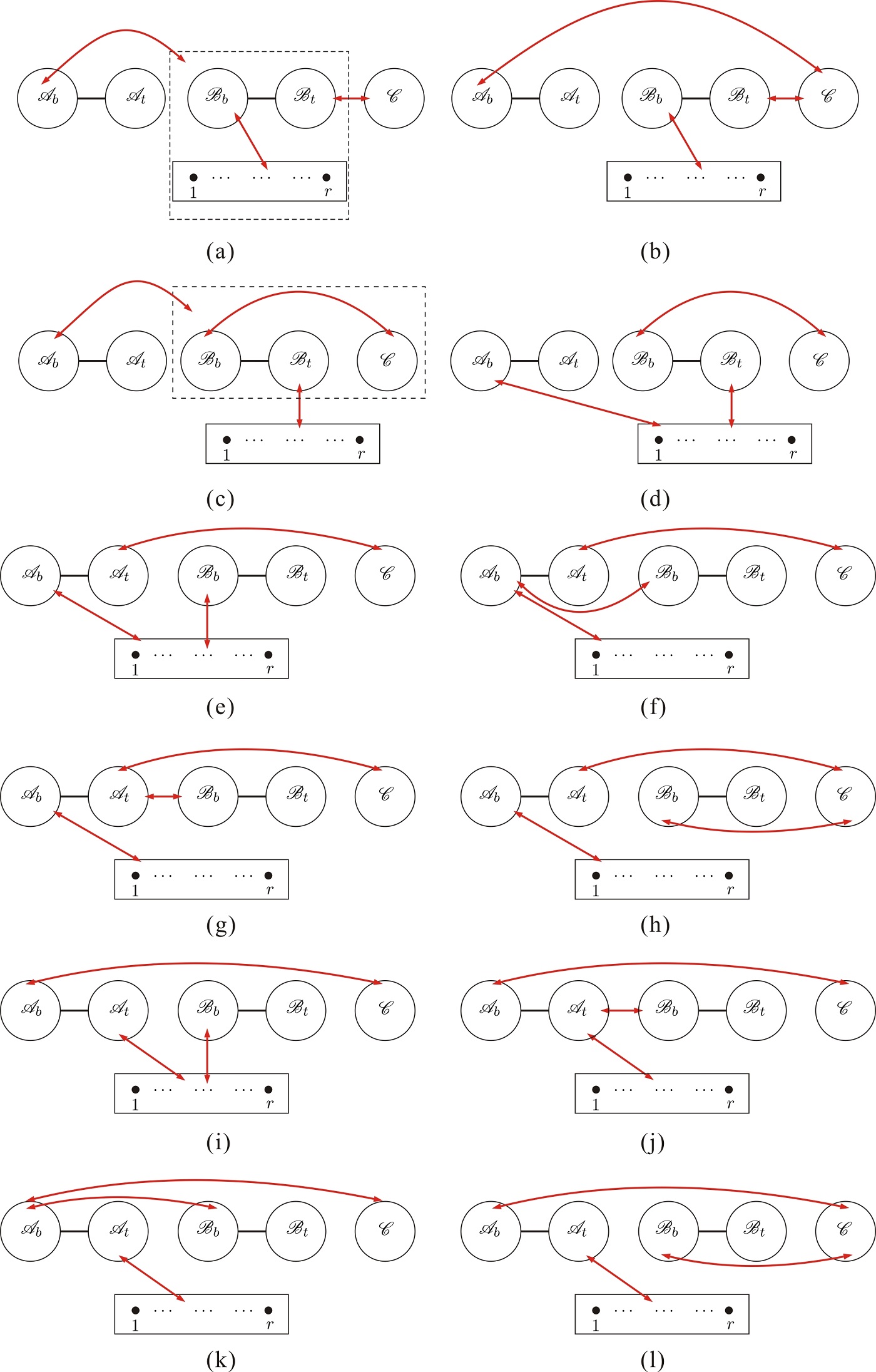}
\caption{Physical graphs for the skeleton \figref{Fig:Figure9} by { construction-1}: part 2.}\label{Fig:Figure13}
\end{figure}
\begin{figure}
\centering
\includegraphics[width=0.9\textwidth]{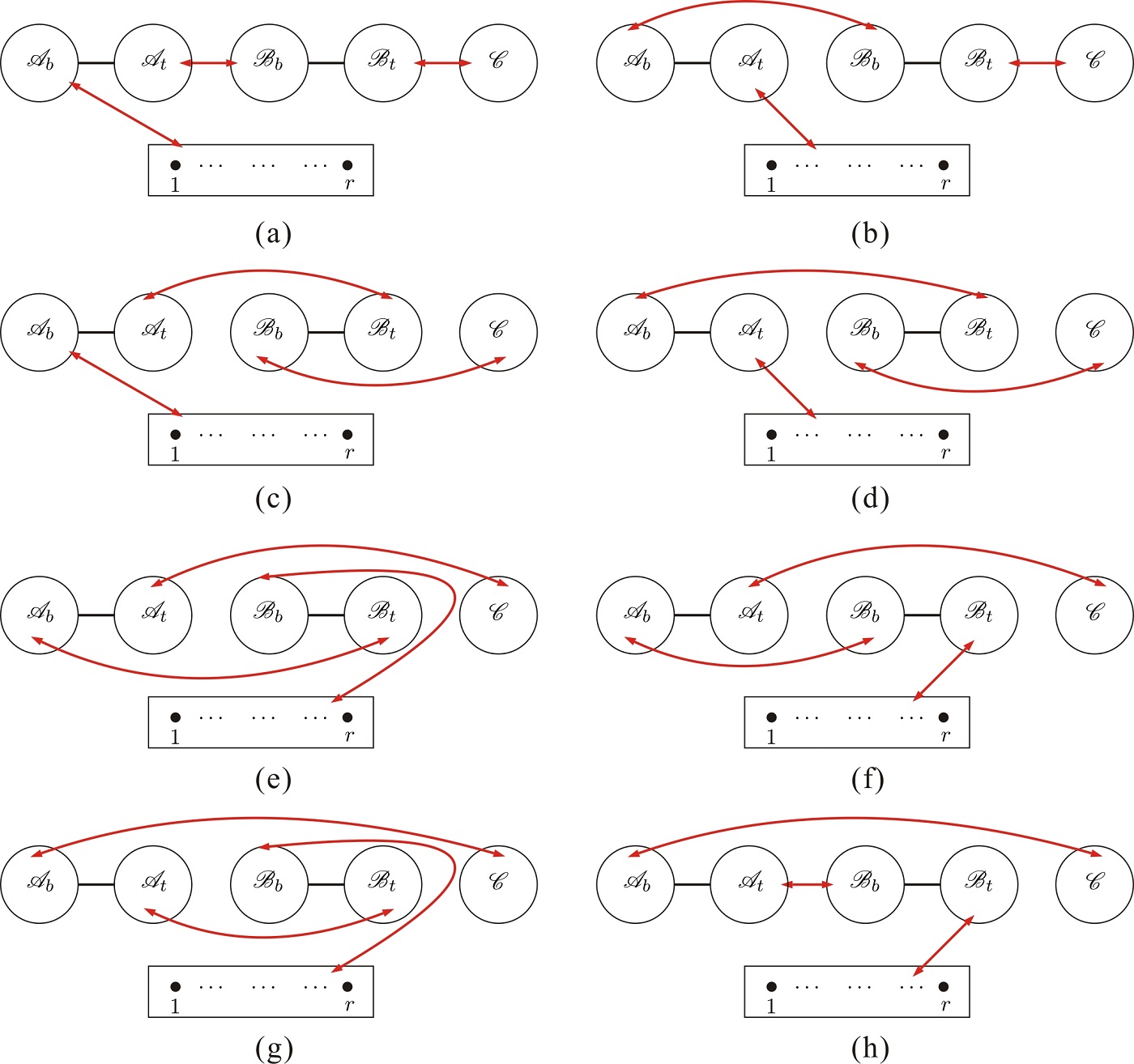}
\caption{Physical graphs for the skeleton \figref{Fig:Figure9} by {construction-1}: part 3.}\label{Fig:Figure14}
\end{figure}
 \begin{figure}
 \centering
 \includegraphics[width=0.9\textwidth]{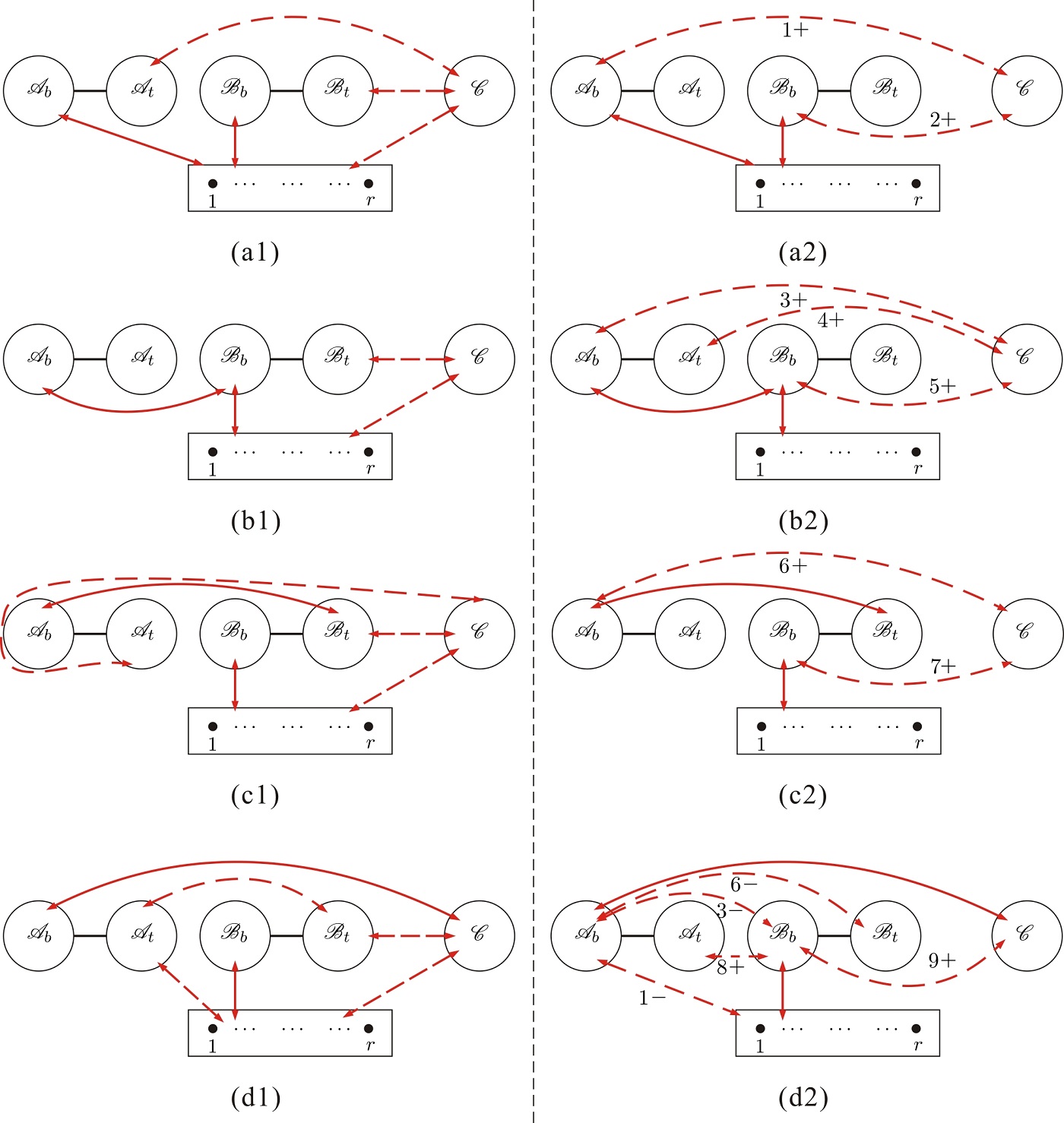}
 \caption{Physical graphs (a1), (b1), (c1), (d1) for the skeleton \figref{Fig:Figure9} by {construction-2} and the corresponding spurious graphs (a2), (b2), (c2), (d2): part 1.}\label{Fig:Figure15-1}
 \end{figure}
 \begin{figure}
 \centering
 \includegraphics[width=0.9\textwidth]{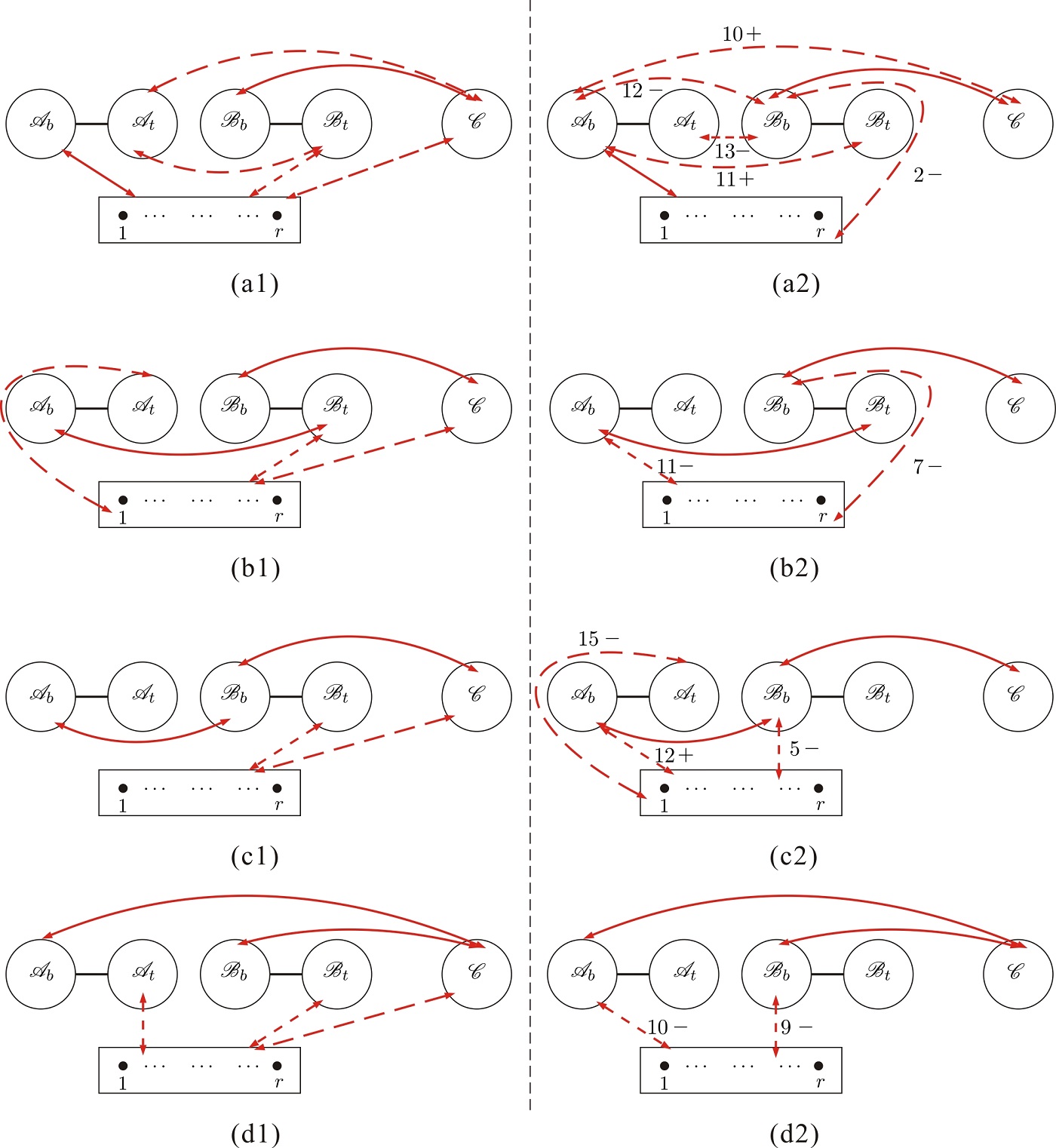}
 \caption{Physical graphs (a1), (b1), (c1), (d1) for the skeleton \figref{Fig:Figure9} by {construction-2} and the corresponding spurious graphs (a2), (b2), (c2), (d2): part 2.}\label{Fig:Figure15}
 \end{figure}

\begin{figure}
\centering
\includegraphics[width=0.9\textwidth]{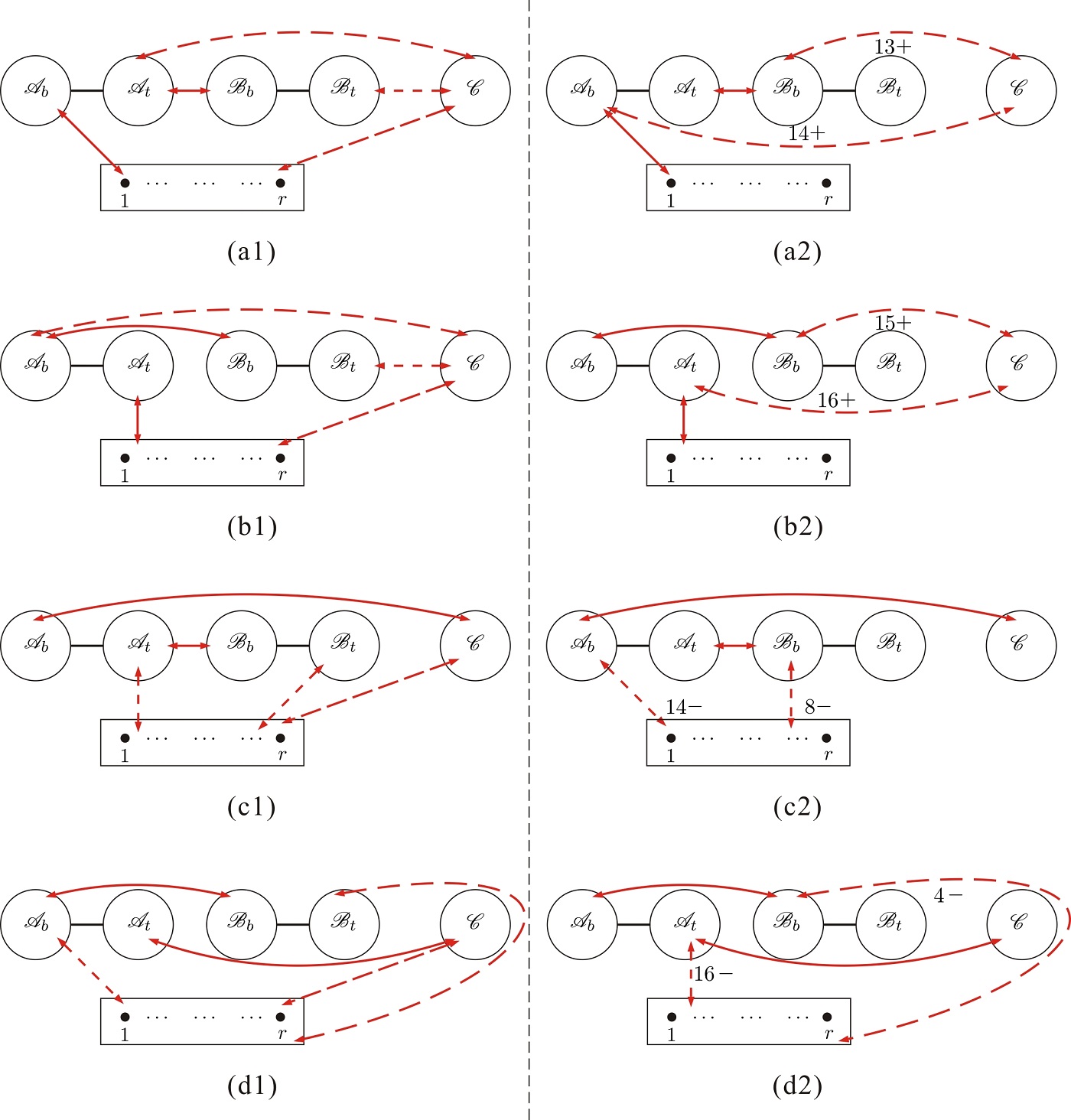}
\caption{Physical graphs (a1), (b1), (c1), (d1) for the skeleton \figref{Fig:Figure9} by {construction-2} and the corresponding spurious graphs (a2), (b2), (c2), (d2): part 3.}\label{Fig:Figure16}
\end{figure}
\begin{figure}
\centering
\includegraphics[width=0.7\textwidth]{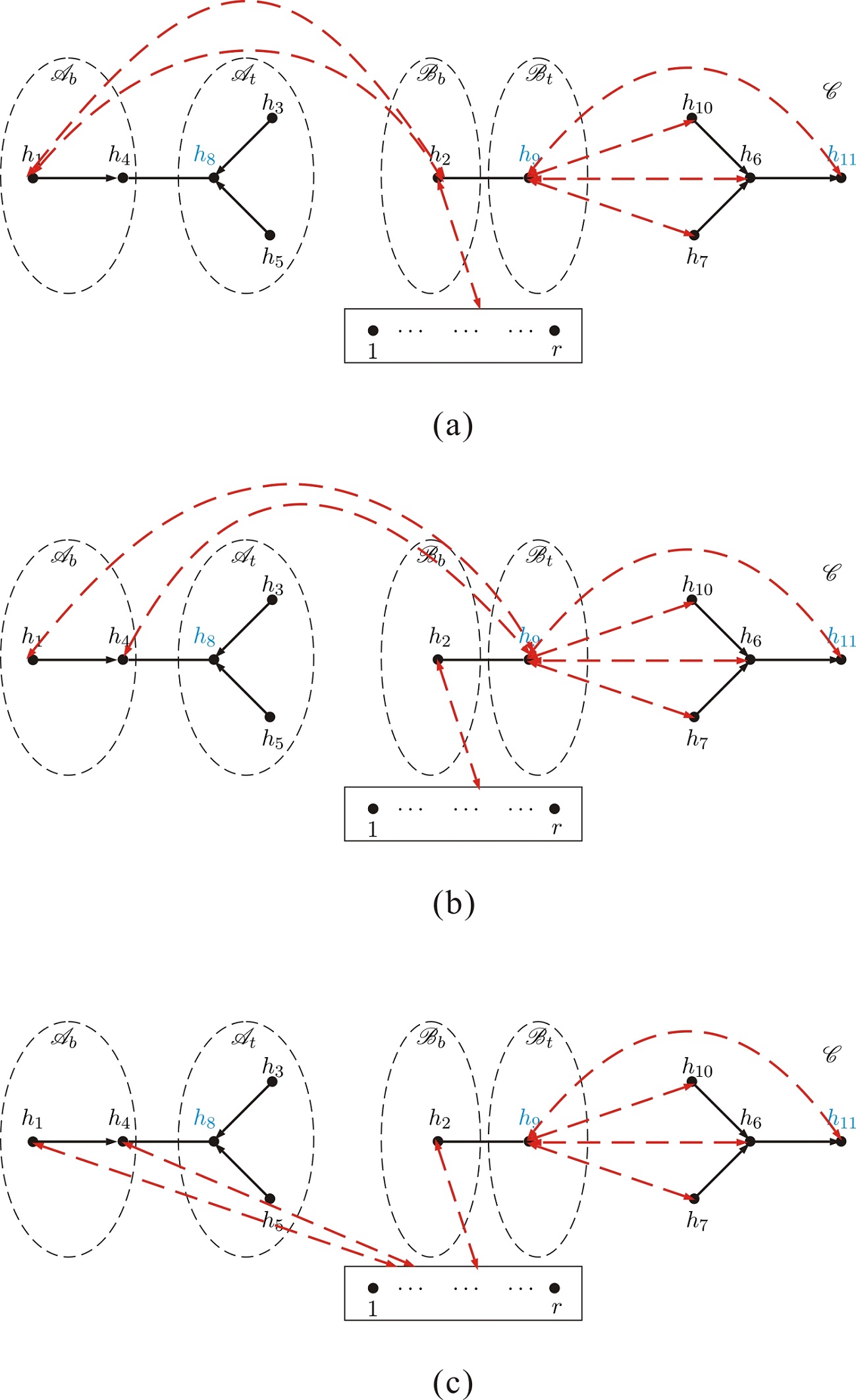}
\caption{All possible graphs for \figref{Fig:Figure12} (a). In \figref{Fig:Figure12} (a), the $\mathscr{A}$ component can be connected with all the three regions  $\mathscr{B}_b$ (see (a)), $\mathscr{B}_t$ (see (b)) and $\mathscr{C}$ (see (c)) in the dashed box.)}\label{Fig:Figure17}
\end{figure}

All physical graphs for the skeleton \figref{Fig:Figure9} obtained by construction-1 are displayed in figures (\ref{Fig:Figure12}), (\ref{Fig:Figure13}) and (\ref{Fig:Figure14}). In each graph, the inner structure in $\mathscr{C}$ and each side of $\mathscr{A}$ and  $\mathscr{B}$ are neglected for convenience. Each type-3 line can be connected to arbitrary two nodes in the corresponding region. For example in \figref{Fig:Figure12} (a), the two end nodes of the type-three line between $\mathscr{A}_b$ and the dashed box can be an arbitrary node in $\mathscr{A}_b$ and an arbitrary node in $\mathscr{B}\oplus\mathscr{R}$ (except $n$ because in the expansion of EYM amplitude $n$ cannot play as a root). All graphs corresponding to \figref{Fig:Figure12} (a) are displayed in \figref{Fig:Figure17}.

All physical graphs obtained by  construction-2 are given by \figref{Fig:Figure15-1} (a1), (b1), (c1), (d1), \figref{Fig:Figure15} (a1), (b1), (c1), (d1) and \figref{Fig:Figure16} (a1), (b1), (c1), (d1). Each graph with a given dashed double arrow line gives a possible configuration. The corresponding spurious graphs are provided by \figref{Fig:Figure15-1} (a2), (b2), (c2), (d2), \figref{Fig:Figure15} (a2), (b2), (c2), (d2) and \figref{Fig:Figure16} (a2), (b2), (c2), (d2). The two spurious graphs with the dashed lines labeled by the same number but opposite signs cancel with each other.

 \bibliographystyle{JHEP}
\bibliography{NoteONGaugeIdentity}

\end{document}